\documentclass[proceedings]{JHEP3}
\usepackage{amsfonts}
\usepackage{amsmath}
\usepackage{epsfig,multicol}
\usepackage{float}

\newbox\mybox

\newcommand\fverb{\setbox\mybox=\hbox\bgroup\verb}
\newcommand\fverbdo{\egroup\medskip\noindent\fbox{\unhbox\mybox}\ }
\newcommand\fverbit{\egroup\item[\fbox{\unhbox\mybox}]}
\conference{$\mathcal{PT}$ in complex nonlinear wave equations}
\abstract{We investigate complex versions of the Korteweg-deVries equations and an Ito type nonlinear system with two coupled nonlinear fields. We systematically construct rational, trigonometric/hyperbolic, elliptic and soliton solutions for these models and focus in particular on physically feasible systems, that is those with real energies. The reality of the energy is usually attributed to different realisations of an antilinear symmetry, as for instance $\mathcal{PT}$-symmetry. It is shown that the symmetry can be spontaneously broken in two alternative ways either by specific choices of the domain or by manipulating the parameters in the solutions of the model, thus leading to complex energies. Surprisingly the reality of the energies can be regained
in some cases by a further breaking of the symmetry on the level of the Hamiltonian.
In many examples some of the fixed points in the complex solution for the field undergo a Hopf bifurcation in the $\mathcal{PT}$-symmetry breaking process. By employing several different variants of the symmetries we propose many classes of new invariant extensions of these models and study their properties. The reduction of some of these models yields complex quantum mechanical models previously studied.}

\title{$\mathcal{PT}$-symmetry breaking in complex nonlinear wave equations
and their deformations}
\author{Andrea Cavaglia$^\bullet$, Andreas Fring$^\bullet$ and Bijan Bagchi$%
^\circ$ \\
$^\bullet$ Centre for Mathematical Science, City University London,\\
$\,\,$ Northampton Square, London EC1V 0HB, UK\\
$^\circ$ Department of Applied Mathematics, University of Calcutta,\\
$\,\,$ 92 Acharya Prafulla Chandra Road, Kolkata 700 009, India\\
E-mail: andrea.cavaglia.1@city.ac.uk, a.fring@city.ac.uk,
bbagchi123@rediffmail.com}

\input{tcilatex}

\begin{document}

\section{Introduction}

One can adopt various points of view with regard to the usefulness of the
study of complex classical and quantum mechanical systems. Being very
orthodox one may just view the complex systems as providing a larger setting
which allows a better insight from a broader framework when restricted to
the real physical system. A very successful example for this viewpoint is
the more than seventy year old proposal of the analytical S-matrix \cite%
{Heisenberg2,DIObook}, which is still pursued nowadays; especially in 1+1
dimensions. Genuinely non-Hermitian systems such as dissipative ones are
also well studied, but they are usually regarded as open and are therefore
not self-consistent \cite{Ingridrev}. In contrast, a more recent perspective
allows to regard certain complex quantum mechanical Hamiltonians also as
perfectly acceptable self-consistent descriptions of physical systems \cite%
{Bender:1998ke}, in the sense that they possess real energy eigenvalue
spectra and well defined unitary time-evolution (see \cite%
{Benderrev,Alirev,PauloPhD} for recent reviews). To consider pseudo/quasi
Hermitian systems provides a very clear conceptual view in this respect as
it makes use of a similarity transformation towards a Hermitian Hamiltonian
system for which everything is well defined in a standard conventional
sense, although such transformations do not exist for all types of such
systems. A more radical view is to give a direct meaning to the
non-Hermitian Hamiltonians without any reference to the Hermitian system.
This latter point of view has to be taken in order to explain recent
experiments in which non-Hermitian systems are studied on optical lattices,
see e.g. \cite{Experiment}, as the observed gain and loss can not be
explained in a purely Hermitian setting. A further experimental realization
has recently been proposed for graphene nanoribbons, where non-Hermitian
Hamiltonians arise as effective Hamiltonians \cite{graphene}.

Largely inspired by the study of the quantum systems, also classical complex
systems have been investigated recently. Naturally one may view them too in
various ways, either as directly meaningful or only sensible when
transformed to a real system. Many classical models have already been
investigated from these various perspectives, such as complex extensions of
standard one particle real quantum mechanical potentials \cite%
{Nana,Bender:2006tz,Bender:2008fr,Bender:2010eg}, non-Hamiltonian dynamical
systems \cite{Bender:2007pr}, chaotic systems \cite{Bender:2008qe} and
deformations of many-particle systems such as Calogero-Moser-Sutherland
models \cite{AF,AFZ,Assis:2009gt,FringSmith}. These investigations led to
new insights into the quantum theories based on the features found in these
classical models, such as tunneling behaviour \cite{Bender:2010eg,tunnel},
band structures \cite{Benderbands} and even a complex generalizations of
Bohr's correspondence principle has been formulated \cite{Bender:2009jg}.

Extensions of field theories of nonlinear wave type, such as
Korteweg-deVries (KdV) equations \cite{BBCF,AFKdV,BBAF} and closely related
models \cite{Bender:2007ij,Curtright:2007ta,comp1,CompactonsAF} have also
been investigated. These type of models will be the main subject of our
investigations in this manuscript.

Pseudo or quasi-Hermiticity are often equivalent to a simultaneous parity
and time reversal, so-called $\mathcal{PT}$-symmetry. Remarkably this
property, or more general antilinear symmetry \cite{EW}, of quantum
mechanical systems is already visible on the classical level. It is known
for more than fifty years that when this symmetry is unbroken the reality of
the spectrum of the theory is guaranteed. \emph{Unbroken} refers here to the
property that%
\begin{equation}
\lbrack \mathcal{PT},H]=0\qquad \text{and\qquad }\mathcal{PT}\Phi =\Phi 
\text{,}  \label{PT}
\end{equation}%
i.e.~the Hamiltonian $H$ is $\mathcal{PT}$-symmetric and its eigenfunctions $%
\Phi $ are also eigenstates of the $\mathcal{PT}$ or any other antilinear
operator. In case the latter property does not hold, one speaks of \emph{%
spontaneously broken} $\mathcal{PT}$-symmetry and the eigenenergies become
complex conjugate pairs. The $\mathcal{PT}$-symmetry can be broken
spontaneously in two different ways, either by tuning some of the parameters
of the models appropriately or by manipulating the domain on which the model
is defined, see e.g.~\cite{Ergun} for the latter possibility. One speaks of 
\emph{broken} $\mathcal{PT}$-symmetry when neither of the relations in (\ref%
{PT}) hold. In this case one usually expects complex energies, but we will
demonstrate in this manuscript that in some cases real energy spectra can be
produced by breaking the spontaneously broken theory further in a controlled
way.

Here we will trace these properties on the classical level and use the
reality of the classical energy as a criterium to select physically
meaningful complex models including their boundary and initial conditions.
The role of $\Phi $ in (\ref{PT}) is in this case played by the solution of
the classical field equation of motion, say $u(x,t)$. For a given
Hamiltonian density $\mathcal{H}$ depending on $u(x,t)$ the energy on the
interval $x\in \lbrack -a,a]$ is computed by 
\begin{equation}
E=\int\nolimits_{-a}^{a}\mathcal{H}\left[ u(x)\right] dx=\oint\nolimits_{%
\Gamma }\mathcal{H}\left[ u(x)\right] \frac{du}{u_{x}}.  \label{Energy}
\end{equation}%
Here and throughout the manuscript we use the standard convention $%
u_{x}\equiv du/dx$. When the system possesses periodic solutions, that is
when $u(a)=u(-a)$ along a path $\Gamma $ in the complex $u$-plane, we can
employ the alternative contour integral version in (\ref{Energy}). A simple
argument \cite{AFKdV} shows that the energy in the interval $x\in \lbrack
-a,a]$ is guaranteed to be real if the symmetry property $\mathcal{H}^{\ast }%
\left[ u(x)\right] =\mathcal{H}\left[ u(-x)\right] $ holds for the
Hamiltonian density. We will present here some unexpected examples of real
energy solutions for which neither of the relations in (\ref{PT}) hold, i.e.
the $\mathcal{PT}$-symmetry is broken for the Hamiltonian and the solutions.

Here we will investigate two different types of complexified wave equations.
First of all we study in section 2 the most immediate way to complexify wave
equations by introducing complex boundary values and initial conditions for
the KdV-system. We investigate systematically the traveling wave and soliton
solutions. We study complex trajectories in the complex plane of the
KdV-field especially with regard to their properties under $\mathcal{PT}$%
-symmetry breaking. In section 3 we employ the different types of $\mathcal{%
PT}$-symmetry as a construction principle to propose new extended versions
of the KdV-system. We investigate the new models in a similar fashion as
their undeformed counterparts. Particular attention is paid to the question
of what kind of conditions will lead to physical models, in the sense that
they possess real energies. In section 4 we study the effect of complex
boundary conditions and initial values on a nonlinear system with two
fundamental fields coupled to each other, which is referred to as Ito system
for some specific parameter choice. We also investigate the $\mathcal{PT}$%
-symmetry properties of these models. Due to the presence of an additional
field when compared with the KdV-system we can identify four different
versions of $\mathcal{PT}$-symmetry being realised in these systems, which
we exploit in section 5 to construct new models. We study them from similar
points of view as the previous ones. We draw our conclusions in section 6.

\section{Complex Korteweg-deVries equation}

Complex extensions of the KdV equation have been investigated already some
time ago for instance in~\cite{CKdV1,CKdV2,CKdV3,FSA}. However, $\mathcal{PT}
$-symmetry has only been utilized recently in \cite{BBCF,AFKdV,BBAF} in
order to understand some of their properties and in particular to construct
new models. The standard KdV system is known to be a Hamiltonian system with
density%
\begin{equation}
\mathcal{H}_{\text{KdV}}=-\frac{\beta }{6}u^{3}+\frac{\gamma }{2}%
u_{x}^{2}\quad \ \ \ \ \ \ \ \ \ \beta ,\gamma \in \mathbb{C},  \label{Hkdv}
\end{equation}%
leading to the Korteweg-deVries equation in the form 
\begin{equation}
u_{t}+\beta uu_{x}+\gamma u_{xxx}=0.\quad \ \ \ \ \ \ \   \label{KdV}
\end{equation}%
Usually the constants $\beta $ and $\gamma $ are chosen to be real, but here
we allow them to take complex values, thus including the possibility for $%
u(x,t)$ to be complex. $\mathcal{PT}$-symmetry may then be realised in two
alternative ways as 
\begin{eqnarray}
\mathcal{PT}_{+} &:&x\mapsto -x,t\mapsto -t,i\mapsto -i,u\mapsto u\quad \ \ 
\text{for }\beta ,\gamma \in \mathbb{R}, \\
\mathcal{PT}_{-} &:&x\mapsto -x,t\mapsto -t,i\mapsto -i,u\mapsto -u\quad 
\text{for }i\beta ,\gamma \in \mathbb{R},
\end{eqnarray}%
both possibilities guaranteeing that $\mathcal{PT}_{\pm }:\mathcal{H}_{\text{%
KdV}}\mapsto $ $\mathcal{H}_{\text{KdV}}$ holds. The underlying models
respect one of the two symmetries and are therefore different as they
correspond to two distinct choices of the coupling constant which may,
however, be related by a simple rotation in $u$. Nonetheless, with regard to
possible deformations to be discussed below this second symmetry allows to
construct different types of new models.

A crucial feature of the model to be acceptable as physically consistent is
the reality of the energy (\ref{Energy}). In case of the Hamiltonian density 
$\mathcal{H}_{\text{KdV}}$ the reality of the energy (\ref{Energy}) is
guaranteed for the two possibilities $u^{\ast }(x)=u(-x)$ when $\beta
,\gamma \in \mathbb{R}$ or $u^{\ast }(x)=-u(-x)$ when $i\beta ,\gamma \in 
\mathbb{R}$, resulting from $\mathcal{PT}_{+}$ or $\mathcal{PT}_{-}$,
respectively.

\subsection{$\mathcal{PT}$-symmetric, spontaneously broken and broken
solutions}

Let us now see which complex boundary conditions and initial values are
physically permissible. In order to establish this, we first briefly recall
how the traveling wave solutions of the KdV-equations of motion may be
constructed systematically. Integrating (\ref{KdV}) twice leads to the
equation 
\begin{equation}
u_{\zeta }^{2}=\frac{2}{\gamma }\left( \kappa _{2}+\kappa _{1}u+\frac{c}{2}%
u^{2}-\frac{\beta }{6}u^{3}\right) =:\lambda P(u)  \label{P}
\end{equation}%
with integration constants $\kappa _{1},\kappa _{2}\in \mathbb{C}$ and $P(u)$
denoting a third order polynomial in $u$ multiplied by an overall constant $%
\lambda $. A further integration yields%
\begin{equation}
\pm \sqrt{\lambda }\left( \zeta -\zeta _{0}\right) =\int du\frac{1}{\sqrt{%
P(u)}},  \label{int}
\end{equation}%
where we made the usual assumption that the field $u(x,t)$ acquires the form
of a traveling wave $u(x,t)=u(\zeta )$ with $\zeta =x-ct$ and $c$ denoting
the wave speed. In complete generality this is an elliptic integral, but it
is instructive to generate simpler solutions by systematically making some
specific assumptions on the factorization of the polynomial $P(u)$. This
solution method may then be extended to the deformed cases.

Demanding specific boundary conditions will impose further restrictions or
might be entirely incompatible with certain factorizations of $P(u)$. For
instance, in case we wish to implement vanishing asymptotic boundary
conditions for $u$ and its derivatives, the once integrated version of
equation (\ref{KdV}) and (\ref{P}) implies%
\begin{equation}
\lim_{\zeta \rightarrow \pm \infty }u,u_{\zeta }=0\quad \Rightarrow \quad
\kappa _{1}=\kappa _{2}=0.  \label{u0}
\end{equation}

In the following we will study the solutions in the complex $u$-plane. For
this purpose it is useful to separate $u$ into its real and imaginary part $%
u^{R}$ and $u^{I}$, respectively, and decouple (\ref{P}) into two first
order differential equations in these variables%
\begin{equation}
u_{\zeta }^{R}=\pm \func{Re}\left[ \sqrt{\lambda }\sqrt{P(u^{R}+iu^{I})}%
\right] \qquad \text{and\qquad }u_{\zeta }^{I}=\pm \func{Im}\left[ \sqrt{%
\lambda }\sqrt{P(u^{R}+iu^{I})}\right] .  \label{121}
\end{equation}%
In this set up we may then apply many of the techniques which have been
developed for two dimensional dynamical systems, see for instance \cite%
{arrow}. Most immediate is the application of the linearisation theorem at
some fixed point $u_{f}$, converting the nonlinear system into%
\begin{equation}
\left( 
\begin{array}{r}
u_{\zeta }^{R} \\ 
u_{\zeta }^{I}%
\end{array}%
\right) =\left. J(u^{R},u^{I})\right\vert _{u=u_{f}}\left( 
\begin{array}{r}
u_{\zeta }^{R} \\ 
u_{\zeta }^{I}%
\end{array}%
\right) ,  \label{lin}
\end{equation}%
with Jacobian matrix 
\begin{equation}
\left. J(u^{R},u^{I})\right\vert _{u=u_{f}}=\left. \left( 
\begin{array}{ll}
\pm \frac{\partial \func{Re}[\sqrt{\lambda }\sqrt{P(u)}]}{\partial u^{R}} & 
\pm \frac{\partial \func{Re}[\sqrt{\lambda }\sqrt{P(u)}]}{\partial u^{I}} \\ 
\pm \frac{\partial \func{Im}[\sqrt{\lambda }\sqrt{P(u)}]}{\partial u^{R}} & 
\pm \frac{\partial \func{Im}[\sqrt{\lambda }\sqrt{P(u)}]}{\partial u^{I}}%
\end{array}%
\right) \right\vert _{u=u_{f}}.  \label{Jac}
\end{equation}%
We denote the eigenvalues of $J(u=u_{f})$ by $j_{1}$, $j_{2}$. Provided the
system (\ref{lin}) is simple and $\func{Re}j_{i}\neq 0$ the linearisation
theorem applies, stating that the phase portraits of the systems (\ref{121})
and (\ref{lin}) are qualitatively the same in some neighbourhood of the
fixed point $u_{f}$. The ten similarity classes for $2\times 2$-matrices
fully characterizing all possible behaviours for the fixed points of (\ref%
{121}) are reported for reference in appendix A.

\subsubsection{Rational solutions}

\begin{figure}[h!]
\centering  \includegraphics[width=7.0cm]{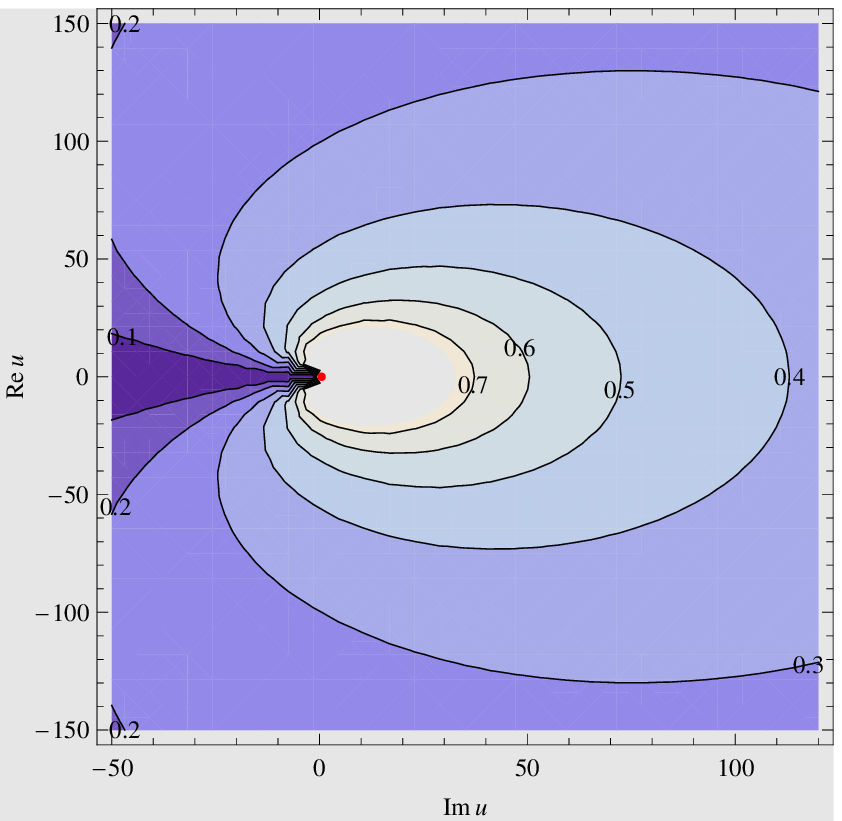} %
\includegraphics[width=7.0cm]{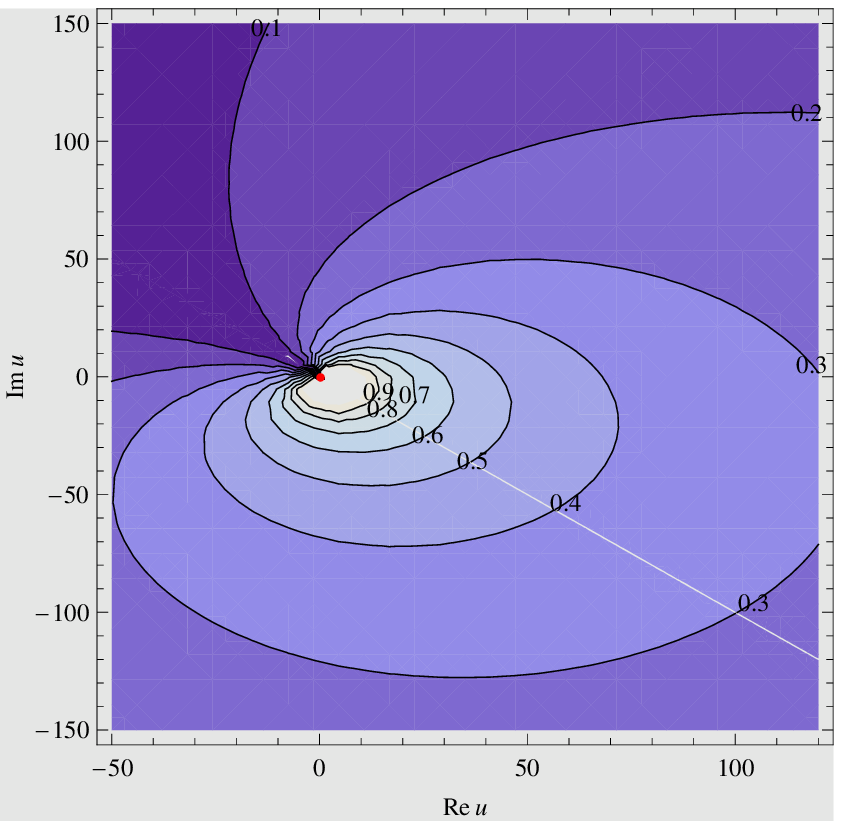}
\caption{Complex rational solutions of the KdV equation for different values
of purely complex initial conditions $\protect\zeta _{0}$: (a) $\mathcal{PT}$%
-symmetric solutions for $c=1$, $\protect\beta =2$, $\protect\gamma =3$ and $%
A=1/2$; (b) Broken $\mathcal{PT}$-symmetric solutions for $c=1$, $\protect%
\beta =2+i2$, $\protect\gamma =3$ and $A=(1-i)/4$. Different trajectories
are characterized by different initial conditions $\protect\zeta _{0}$. Some
values for the imaginary part of $\func{Im}\protect\zeta _{0}<1$ are
indicated on the trajectories. The unresolved white region corresponds to
values for $\func{Im}\protect\zeta _{0}>1$. Throughout this manuscript we
label panels from the left to the right as (a), (b), (c), etc.}
\label{fig1}
\end{figure}

Factorizing $P(u)$ at first in the simplest way as $P(u)=(u-A)^{3}$ with one
constant $A$ leaves no freedom when solving (\ref{P}), as all constants are
fixed%
\begin{equation}
\lambda =-\frac{\beta }{3\gamma },\quad \kappa _{1}=-\frac{c^{2}}{2\beta }%
,\quad \kappa _{2}=\frac{c^{3}}{6\beta ^{2}}\quad \text{and\quad }A=\frac{c}{%
\beta }.  \label{rat}
\end{equation}%
Clearly asymptotically vanishing boundary conditions (\ref{u0}) are only
possible for a static solution with $c=0$. The evaluation of the expression (%
\ref{int}) produces for this factorization the rational solution%
\begin{equation}
u\left( \zeta \right) =\frac{c}{\beta }-\frac{12\gamma }{\beta \left( \zeta
-\zeta _{0}\right) ^{2}},  \label{28}
\end{equation}%
with additional integration constant $\zeta _{0}\in \mathbb{C}$. Taking $%
\zeta _{0}$ as purely imaginary maintains the $\mathcal{PT}$-symmetry of the
solution, whereas any real part may be compensated by a shift in $\zeta $,
which is kept to be real. Independently of the parameter choice,
asymptotically all rational solutions of the type (\ref{28}) end up at $%
\lim\nolimits_{\zeta \rightarrow \pm \infty }u(\zeta )=c/\beta $, as we can
also observe in figure \ref{fig1}.

Focussing at first on the $\mathcal{PT}$-symmetric scenario, we notice that
while keeping the speed of the wave $c$ real there are two possible choices
for the conjugation leading to physical solutions, namely $u^{\ast }(\zeta
)=u(-\zeta )$ when $i\zeta _{0},\beta ,\gamma \in \mathbb{R}$ or $u^{\ast
}(\zeta )=-u(-\zeta )$ when $i\zeta _{0},i\beta ,\gamma \in \mathbb{R}$. We
depict some complex trajectories in the $u$-plane in figure \ref{fig1}a for
several different initial conditions $\zeta _{0}$. We have taken the plus
and minus sign in (\ref{int}) for the upper and lower half plane,
respectively. We observe that for a specific branch the point $A$ in the $u$%
-plane appears to be either a stable or an unstable improper asymptotic
fixed point. We adopt here the notion for the characterizations of fixed
points from the linearisation (see appendix A for a classification), despite
the fact that $u_{\zeta }$ is not a meromorphic function and the system is
not easily linearized. This implies that only the choice with different
branches for the upper and lower half will give rise to closed orbits as
depicted. For increasing $\zeta $, they run either out of the fixed point in
the upper half plane and into it in the lower half or vice versa. The
crossing of the trajectories with the real line is easily computed to be at $%
u(0)=c/\beta +12\gamma /\beta (\func{Im}\zeta _{0})^{2}$. This makes it
evident that only complex trajectories may be closed, whereas the real
solution is the only trajectory drifting off to minus infinity.

Since the rational solution (\ref{28}) does not have any free parameter
left, as specified in (\ref{rat}), there is no possibility to break the $%
\mathcal{PT}$-symmetry spontaneously for this type of solution. We may,
however, break the\ $\mathcal{PT}$-symmetry completely directly on the level
of the Hamiltonian by fully complexifying $\beta $ or $\gamma $, an example
of which is depicted in figure \ref{fig1}b. As expected we observe that the
symmetry $u^{\ast }(\zeta )=u(-\zeta )$ has been lost, but instead the
trajectories are almost symmetric about the line passing the two points $A$
and $u(0)$, which are now both located away from the real axis. However, the
nature of the fixed point, being either an unstable or stable improper node,
has not changed.

Considering the expression for the energy (\ref{Energy}) it is evident that
it will be real when we have the symmetry $u^{\ast }(x)=u(-x)$. We also
compute the expression explicitly for the solution (\ref{28}) by
substituting it into (\ref{Energy}). Then the energy in the interval $[-a,a]$
is computed to%
\begin{equation}
E_{a}=-\frac{ac^{2}}{3\beta ^{2}}\left( c+\frac{36\gamma }{a^{2}-\zeta
_{0}^{2}}\right) +\frac{72\gamma ^{2}}{15\beta ^{2}}\left[ \frac{10c\left(
a^{3}+3a\zeta _{0}^{2}\right) }{\left( a^{2}-\zeta _{0}^{2}\right) ^{3}}-%
\frac{48\gamma \left( a^{5}+10a^{3}\zeta _{0}^{2}+5a\zeta _{0}^{4}\right) }{%
\left( a^{2}-\zeta _{0}^{2}\right) ^{5}}\right] .
\end{equation}%
Evidently $E_{a}$ is real even when $i\zeta _{0},\beta ,\gamma \in \mathbb{R}
$ or $i\zeta _{0},i\beta ,\gamma \in \mathbb{R}$ and complex otherwise, that
is for the $\mathcal{PT}$-symmetric and broken $\mathcal{PT}$-symmetric
case, respectively.

\subsubsection{Trigonometric/hyperbolic solutions}

As the next possibility for the factorization we specify $%
P(u)=(u-A)^{2}(u-B) $ involving now two constants $A$ and $B$, thus leaving
one of them at our disposal when solving (\ref{P})%
\begin{equation}
\lambda =-\frac{\beta }{3\gamma },\quad \kappa _{1}=\frac{A}{2}(\beta
A-2c),\quad \kappa _{2}=\frac{A^{2}}{6}(3c-2\beta A)\quad \text{and\quad }B=%
\frac{3c}{\beta }-2A.  \label{ak}
\end{equation}%
Having now some freedom in the choice of the constants one may ask which
ones are the most natural to use for the symmetry breaking. As we will see
the constants $A$ and $B$ have a direct physical meaning and it appears
therefore natural to view them as the free parameters to tune for a concrete
model with fixed coupling constants $\beta ,\gamma $, rather than the
integration constants $\kappa _{1}$ or $\kappa _{2}$ emerging more
indirectly without immediate interpretation. By (\ref{u0}) vanishing
asymptotic boundary conditions require the choice $A=0$.

\begin{figure}[h!]
\centering  \includegraphics[width=7.0cm]{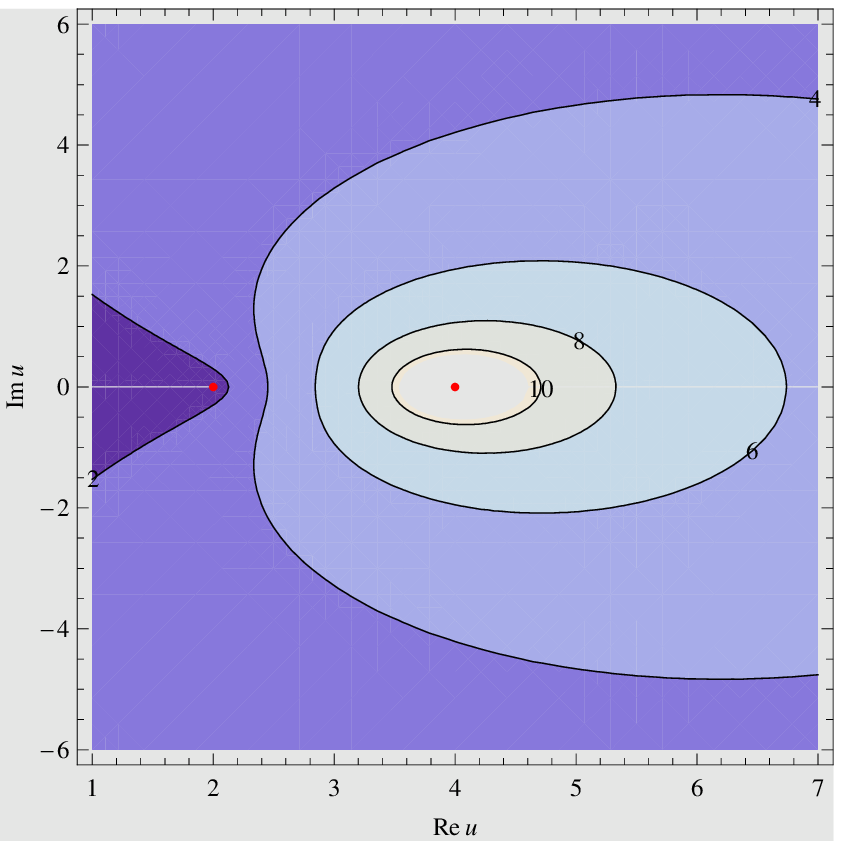} %
\includegraphics[width=7.0cm]{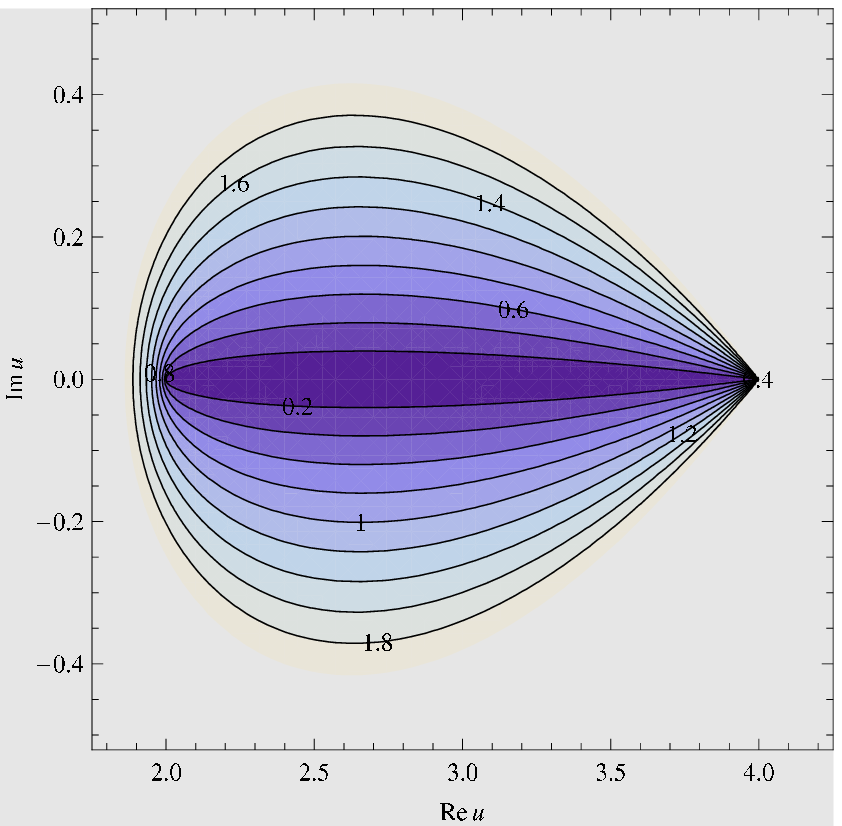}
\caption{Complex trigonometric/hyperbolic solutions of the KdV equation: (a) 
$\mathcal{PT}$-symmetric periodic solutions with $c=1$, $\protect\beta =3/10$%
, $\protect\gamma =3$, $A=4$, $B=2$ and $T=2\protect\sqrt{15}\protect\pi $;
(b) $\mathcal{PT}$-symmetric asymptotically constant solutions with $c=1$, $%
\protect\beta =3/10$, $\protect\gamma =-3,$ $A=4$ and $B=2$.}
\label{fig2}
\end{figure}

In general, the solution to (\ref{int}) produces in this case the
trigonometric/hyperbolic solution%
\begin{equation}
u\left( \zeta \right) =B+(A-B)\tanh ^{2}\left[ \frac{1}{2}\sqrt{A-B}\sqrt{%
\lambda }\left( \zeta -\zeta _{0}\right) \right] .  \label{uu}
\end{equation}%
Let us first discuss the $\mathcal{PT}$-symmetric scenario for which all
constants are taken to be real except for $\zeta _{0}$, which we still allow
to be complex. When in that case either $A<B$, $\lambda >0$ or $A>B$, $%
\lambda <0$ we obtain a periodic solution with period $T=2\pi /(\sqrt{%
\left\vert A-B\right\vert }\sqrt{\left\vert \lambda \right\vert })$ as
depicted in figure \ref{fig2}a. The closed trajectories surround the point $%
A $, whereas the point $B$ is situated on its outside, following from the
fact that on the real axis we always have $u\left( \zeta \right) <B$ or $%
u\left( \zeta \right) >B$, in the respective cases. This behaviour is also
confirmed by the linearisation (\ref{lin}) at the fixed point $A$.
Parameterizing $A-B=r_{AB}e^{i\theta _{AB}}$ and $\lambda =r_{\lambda
}e^{i\theta _{\lambda }}$ the Jacobian (\ref{Jac}) is easily computed to%
\begin{equation}
\left. J(u)\right\vert _{u=A}=\left( 
\begin{array}{ll}
\pm \sqrt{r_{AB}r_{\lambda }}\cos \left[ \frac{1}{2}(\theta _{AB}+\theta
_{\lambda })\right] & \mp \sqrt{r_{AB}r}_{\lambda }\sin \left[ \frac{1}{2}%
(\theta _{AB}+\theta _{\lambda })\right] \\ 
\pm \sqrt{r_{AB}r_{\lambda }}\sin \left[ \frac{1}{2}(\theta _{AB}+\theta
_{\lambda })\right] & \pm \sqrt{r_{AB}r_{\lambda }}\cos \left[ \frac{1}{2}%
(\theta _{AB}+\theta _{\lambda })\right]%
\end{array}%
\right) ,  \label{Jaco}
\end{equation}%
with eigenvalues%
\begin{equation}
j_{1}=\pm \sqrt{r_{AB}r_{\lambda }}\exp \left[ \frac{i}{2}(\theta
_{AB}+\theta _{\lambda })\right] \quad \text{and\quad }j_{2}=\pm \sqrt{%
r_{AB}r_{\lambda }}\exp \left[ -\frac{i}{2}(\theta _{AB}+\theta _{\lambda })%
\right] .  \label{eigen}
\end{equation}%
Clearly for the two cases here, that is $A<B$, $\lambda >0$ or $A>B$, $%
\lambda <0$, the eigenvalues of the Jacobian are purely complex, i.e. $%
j_{1/2}=\pm i\sqrt{\left\vert A-B\right\vert }\sqrt{\left\vert \lambda
\right\vert }$, indicating that the point $A$ is a centre. The result is
independent of the sign in (\ref{121}). The linearisation at the fixed point 
$B$ does not exist, due to the occurrence of the square root.

For this periodic case we may also compute the energy $E_{T}$ for one period 
$T$ from $-\pi /\sqrt{\left\vert A-B\right\vert }\sqrt{\left\vert \lambda
\right\vert }$ to $\pi /\sqrt{\left\vert A-B\right\vert }\sqrt{\left\vert
\lambda \right\vert }$ analytically to%
\begin{equation}
E_{T}=\oint\nolimits_{\Gamma }\mathcal{H}\left[ u(\zeta )\right] \frac{du}{%
u_{\zeta }}=\oint\nolimits_{\Gamma }\frac{\mathcal{H}\left[ u\right] }{\sqrt{%
\lambda }\sqrt{u-B}(u-A)}du=-\pi \sqrt{\frac{\beta \gamma }{3}}\frac{A^{3}}{%
\sqrt{A-B}}.  \label{ET}
\end{equation}%
Here the contour $\Gamma $ is taken to be any complex trajectory resulting
for $i\zeta _{0}\in \mathbb{R}$. Then (\ref{ET}) follows from Cauchy's
residue theorem and the fact that $B$ will always be outside the contour $%
\Gamma $. We find that the energy is real and takes the same value for all
trajectories independently of the concrete choice for the initial condition $%
\zeta _{0}$. Notice that the energy for the real solution, i.e.~$\zeta
_{0}\in \mathbb{R}$, can not be computed in such an easy way as in that case 
$\Gamma $ does not form a closed contour. Thus demanding $\mathcal{PT}$%
-symmetry leads inevitably to purely complex initial value conditions and
constitutes a natural "$\epsilon $-prescription" to deform the real solution
with the purpose to compute the energy for one period.

For the remaining possibilities $A>B$, $\lambda >0$ or $A<B$, $\lambda <0$
all solutions tend asymptotically to $A$, which we depict for some examples
in figure \ref{fig2}b. The linearisation (\ref{lin}) at the fixed point $A$
yields in these cases the two degenerate real eigenvalues $j_{1}=j_{2}=\sqrt{%
\left\vert A-B\right\vert }\sqrt{\left\vert \lambda \right\vert }$ or $%
j_{1}=j_{2}=-\sqrt{\left\vert A-B\right\vert }\sqrt{\left\vert \lambda
\right\vert }$ for $J$ depending on the plus or minus sign in (\ref{121}),
respectively. The Jacobian (\ref{Jaco}) is diagonalisable in these cases and
therefore $A$ is an unstable or stable star node. As for the rational
solutions, this implies that only the choice with different branches for the
upper and lower half will give rise to closed orbits as seen in figure \ref%
{fig2}b. Now the energies can not be computed in a simple manner as for the
periodic case since the singularity at $A$ is situated on the contour.

Let us next embark on the case in which the $\mathcal{PT}$-symmetry is
spontaneously broken, which unlike as for the rational solutions, is
possible for (\ref{uu}) since we have additional parameters at our disposal.
From (\ref{ak}) and (\ref{uu}) it is clear that when $2\beta \kappa
_{1}+c^{2}<0$ the constant $A$ becomes complex and we no longer have the
property $u^{\ast }(\zeta )=u(-\zeta )$ ensuring the reality of the
spectrum. The nature of the fixed point is now changed again to an unstable
or stable focus depending on whether $\func{Re}j_{i}>0$ or $\func{Re}j_{i}<0$%
, respectively. Thus once again closed orbits are obtained with the choice
of different branches for the upper and lower half plane. This means of
course that the periodic solution ceases to be periodic.

\begin{figure}[h!]
\centering  \includegraphics[width=7.0cm]{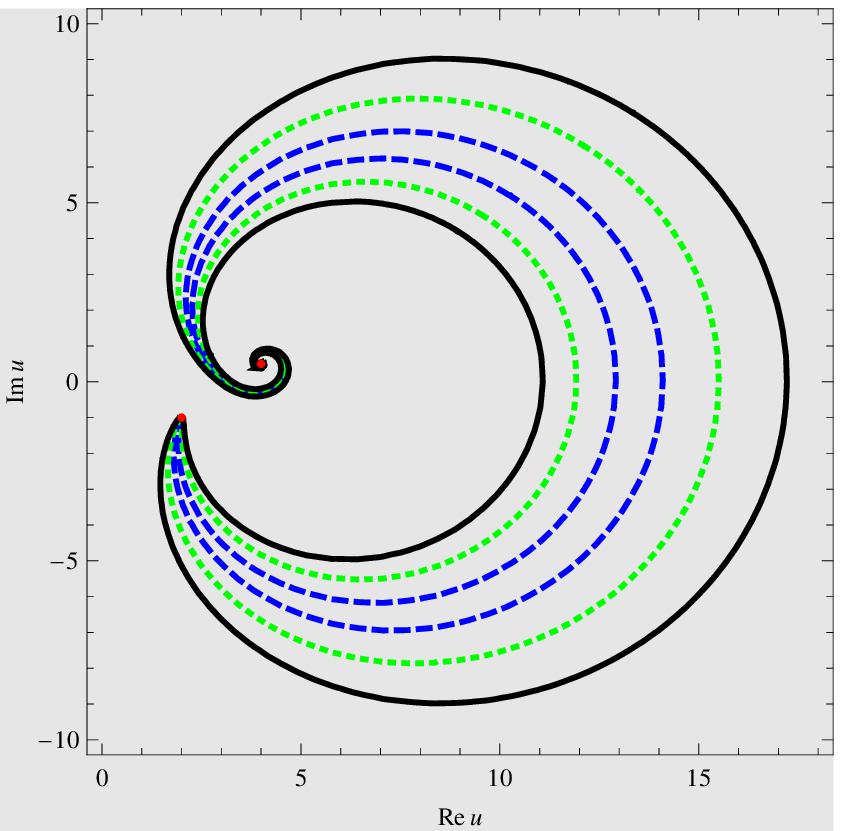} %
\includegraphics[width=7.0cm]{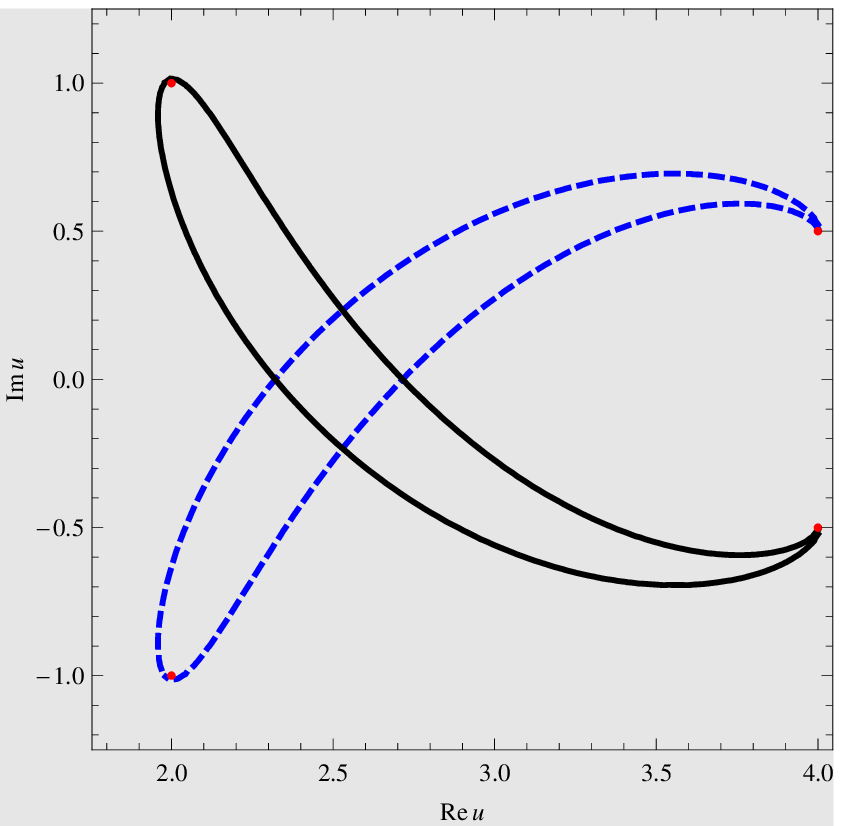}
\caption{Complex trigonometric/hyperbolic solutions of the KdV equation: (a)
Spontaneously broken $\mathcal{PT}$-symmetry of the periodic solution with $%
c=1$, $\protect\beta =3/10$, $\protect\gamma =3$, $A=4+i/2$ and $B=2-i$ for $%
\func{Im}\protect\zeta _{0}=0.5$ solid (black), $\func{Im}\protect\zeta %
_{0}=0.3$ dotted (green) $\func{Im}\protect\zeta _{0}=0.1$ dashed (blue);
(b) Spontaneously broken $\mathcal{PT}$-symmetry of the asymptotically
constant solution with $c=1$, $\protect\beta =3/10$, $\protect\gamma =-3$
for $A=4-i/2$, $B=2+i$ $\func{Im}\protect\zeta _{0}=-0.5$ solid (black)and $%
A=4+i/2$, $B=2-i$, $\func{Im}\protect\zeta _{0}=0.5$ dashed (blue).}
\label{fig3}
\end{figure}

The energies $E_{T}$ for one period in the periodic solution corresponding
to the two choices $A,B,\zeta _{0}$ and $A^{\ast },B^{\ast },\zeta
_{0}^{\ast }$ occur in complex conjugate pairs. This is the typical scenario
for spontaneously broken $\mathcal{PT}$-symmetry, i.e.~the Hamiltonian is
still $\mathcal{PT}$-symmetric but the solutions to the Schr\"{o}dinger
equation in the quantum case and in the classical case to the equation of
motion, are not. Formula (\ref{ET}) may still be used in this case for the
computation of the energy even though the singularities are moved away from
the real axis as they still lie within the contour. Once again the result
does not depend on the choice of\ $\zeta _{0}$. We depict some trajectories
for this type of spontaneous symmetry breaking in figure \ref{fig3}. The
conjugate solution is simply obtained by a reflection about the real axis,
as shown explicitly for the asymptotically constant solution in panel (b).

\begin{figure}[h!]
\centering  \includegraphics[width=7.0cm]{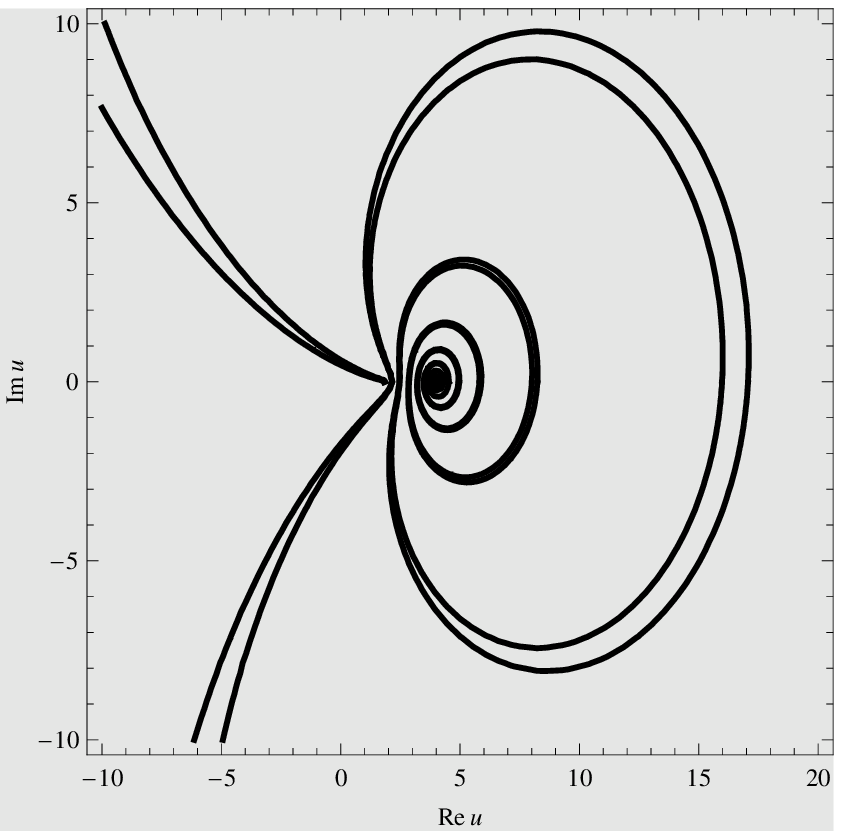} %
\includegraphics[width=7.0cm]{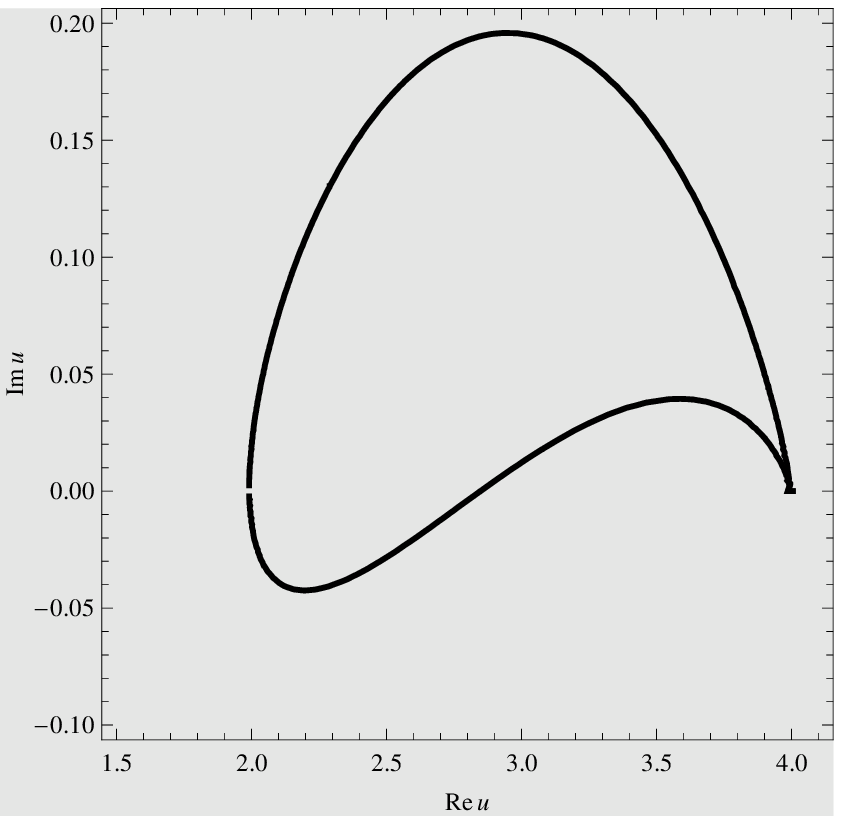}
\caption{Complex trigonometric/hyperbolic solutions of the KdV equation: (a)
Broken $\mathcal{PT}$-symmetry of the periodic solution with $A=4$, $B=2$, $%
c=1$, $\protect\beta =3/10$, $\protect\gamma =3+i/2$ and $\func{Im}\protect%
\zeta _{0}=6$; (b) Broken $\mathcal{PT}$-symmetry of the asymptotically
constant solution with $A=4$, $B=2$, $c=1$, $\protect\beta =3/10$, $\protect%
\gamma =3+i/2$ and $\func{Im}\protect\zeta _{0}=1/2$.}
\label{fig4}
\end{figure}

Finally, there is of course also the possibility to break $\mathcal{PT}$%
-symmetry directly for the Hamiltonian itself. For instance when $\beta
\notin \mathbb{R}$ and/or $\gamma \notin \mathbb{R}$ we expect the energies
to be complex. In that case the complex conjugate energy would be obtained
by considering a new type of Hamiltonian with $\beta ^{\ast }$, $\gamma
^{\ast }$ and thus it does not arise from within one specific model. We
depict an example trajectory for this scenario in figure \ref{fig4}.

We observe from figure \ref{fig4}a that the periodic broken solution near
the fixed point is qualitatively the same as the one for the spontaneously
broken case, namely a stable or unstable focus. This behaviour follows from
the eigenvalues (\ref{eigen}), which indicate that there is no distinction
at this point in whether the complexification of the $j_{i}$ result from a
spontaneous or a complete breaking of the $\mathcal{PT}$-symmetry. Note that
the trajectories still close even though this is not shown in figure \ref%
{fig4}a, but this may be seen on a larger scaled plot. We find a similar
behaviour for the fixed point of the broken asymptotic solution as depicted
in figure \ref{fig4}b.

\begin{figure}[h]
\centering  \includegraphics[width=7.0cm]{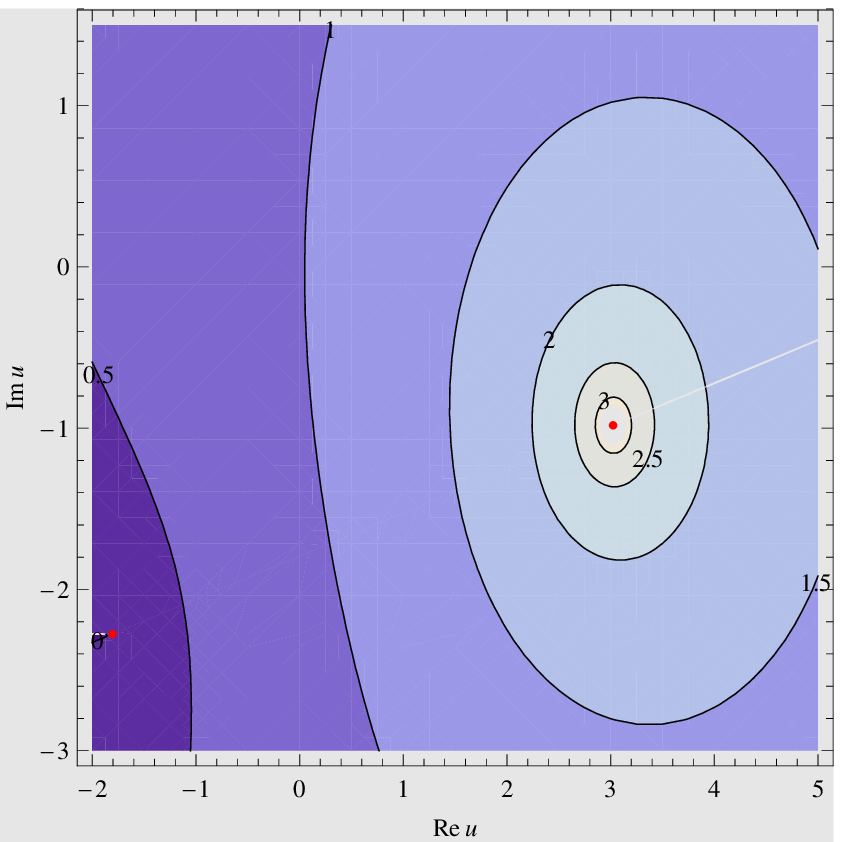} %
\includegraphics[width=7.0cm]{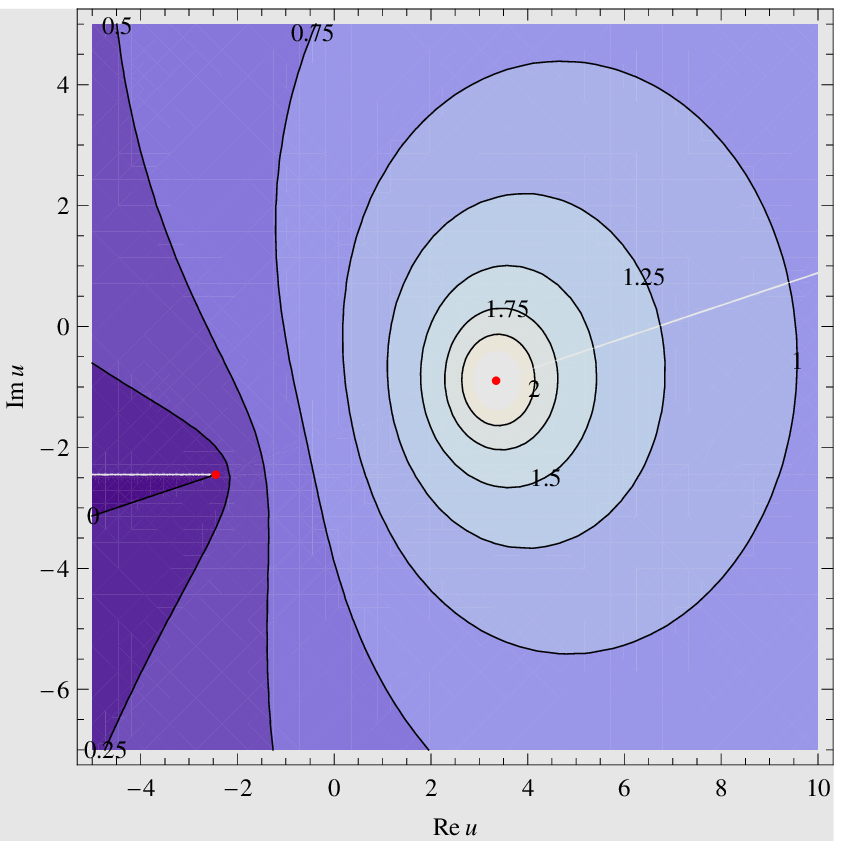}
\caption{Complex trigonometric/hyperbolic solutions of the KdV equation with
completely broken $\mathcal{PT}$-symmetry: (a) Periodic solution with
complex energy $E_{T}=-10.518+i1.666$ for $c=1$, $\protect\beta =1/2\exp (i%
\protect\pi /4)$, $\protect\gamma =1/3\exp (i\protect\pi /3)$, $A=\protect%
\sqrt{3}\sin (11\protect\pi /60)\exp (-i\protect\pi /10)\approx 3.025-i0.983$
and $B=6\exp (-i\protect\pi /4)-2\protect\sqrt{3}\sin (11\protect\pi %
/60)\exp (-i\protect\pi /10)\approx -1.806-i2.277$; (b) Periodic solution
with real energy $E_{T}=-4\protect\pi $ for $c=1$, $\protect\beta =1/2\exp (i%
\protect\pi /4)$, $\protect\gamma =1/3\exp i\protect\pi /3$, $A=2\protect%
\sqrt{3}\exp (-i\protect\pi /12)\approx 3.346-i0.896$ and $B=-\protect\sqrt{6%
}(1+i)\approx -2.449-i2.449$.}
\label{figreal}
\end{figure}

So far we have simply broken the symmetry by choosing some more or less
random complex value for $\gamma $. However, we can also carry out this
process in a more controlled fashion producing periodic motion and even some
non-Hermitian non-$\mathcal{PT}$-symmetry cases with real energies. First of
all we observe from eigenvalues of the Jacobian (\ref{Jac}) that the fixed
point $A$ becomes a centre when $\theta _{AB}+\theta _{\lambda }=\pi $,
which is also compatible with the solution (\ref{uu}) from which we notice
that a periodic motion occurs when $\lambda (A-B)<0$. This allows for
complex values of the parameters $\beta ,\gamma ,A$ and $B$. Combining this
constraint with the last equation in (\ref{ak}) leads to 
\begin{equation}
A=\frac{\sin \theta _{\gamma }}{\left\vert \beta \right\vert \sin \left(
\theta _{\gamma }-\theta _{\beta }-\theta _{A}\right) }\exp \left( i\theta
_{A}\right) .  \label{A}
\end{equation}%
Thus, for a given model, that is for \emph{any }fixed values of $\beta $ and 
$\gamma $ we obtain a periodic motion around the point $A$, given by the
expression in (\ref{A}) for any value of $\theta _{A}$. Indeed this is the
case as we observe for instance in figure \ref{figreal}a. When using the
constraint $\theta _{AB}+\theta _{\lambda }=\pi $ for the periodic motion in
the expression for the energy (\ref{ET}) we find%
\begin{equation}
E_{T}=-\frac{\pi }{3}\frac{\beta A^{3}}{\sqrt{\left\vert \lambda \right\vert
\left\vert A-B\right\vert }}.
\end{equation}%
Hence, demanding that the energy is to be real leads to the further
constraint $3\theta _{A}+\theta _{\beta }=0,\pi $. Implementing this in (\ref%
{A}) yields%
\begin{equation}
E_{T}\in \mathbb{R\quad }\text{for }A=\frac{\sin \theta _{\gamma }}{%
\left\vert \beta \right\vert \sin \left( \theta _{\gamma }-2\theta _{\beta
}/3\right) }\exp \left( -i\frac{\theta _{\beta }}{3}\right) .
\label{Ilsebil}
\end{equation}%
This means for a given model, that is for \emph{any} fixed values of $\beta $
and $\gamma $ we can find a point $A$, given by the expression in (\ref%
{Ilsebil}), around which the trajectory is periodic and the corresponding
energy is real. This holds irrespective of whether $\beta $ and $\gamma $
are real or complex, or in other words whether the $\mathcal{PT}$-symmetry
is intact or completely broken. In figure \ref{figreal}b we depict an
example solution of (\ref{Ilsebil}) corresponding to a periodic motion with
completely broken $\mathcal{PT}$-symmetry but real energy. An obvious
question to ask at this point is whether by breaking this symmetry the
system has acquired a new kind of anti-linear symmetry, which could serve to
explain the reality of the energy. A systematic study of this more general
issue will be presented elsewhere.

In general we observe from the various cases studied in this subsection that
when we let the parameters vary one of the fixed points for the periodic
solution undergoes a Hopf bifurcation, i.e. it changes its behaviour from a
stable focus to a centre and then to an unstable focus. As we can see from (%
\ref{eigen}) this behaviour is governed by the sign of $\pm \cos \left[ 
\frac{1}{2}(\theta _{AB}+\theta _{\lambda })\right] $ this bifurcation can
be achieved either by a spontaneous symmetry breaking, that is by varying
the free parameter $A$, or by a complete breaking of the symmetry, by
varying $\beta $ or $\gamma $. Both cases pass through the $\mathcal{PT}$%
-symmetric case as realised by the centre.

\subsubsection{Elliptic solutions}

Next we specify $P(u)=(u-A)(u-B)(u-C)$ with three constants $A$, $B$ and $C$
leaving two constants at our disposal when solving (\ref{P}) 
\begin{eqnarray}
\lambda &=&-\frac{\beta }{3\gamma },\quad \kappa _{1}=\frac{1}{6}\left[
\beta (A^{2}+AC+C^{2})-3c(A-C)\right] ,\quad  \label{ell1} \\
\kappa _{2} &=&\frac{AC}{6}[3c-\beta (A+C)]\quad \text{and\quad }B=\frac{3c}{%
\beta }-(A+C).  \label{ell2}
\end{eqnarray}%
Vanishing asymptotic boundary conditions reduce this case to the previous
one as by (\ref{u0}) they require either $A=B=0$, $A=C=0$ or $B=C=0$. The
evaluation of (\ref{int}) yields in this case the elliptic solution%
\begin{equation}
u\left( \zeta \right) =A+(B-A)\limfunc{ns}{}^{2}\left[ \frac{1}{2}\sqrt{B-A}%
\sqrt{\lambda }\left( \zeta -\zeta _{0}\right) \right. \left\vert \frac{A-C}{%
A-B}\right] ,  \label{uell}
\end{equation}%
with $\limfunc{ns}(z|m)$ being one of the Jacobi elliptic functions.
Therefore $u\left( \zeta \right) $ is a double periodic function%
\begin{equation}
u\left( \zeta \right) =u\left( \zeta +\omega _{1}+\omega _{2}\right)
\end{equation}%
with periods%
\begin{equation}
\omega _{1}=\frac{8}{\sqrt{B-A}\sqrt{\lambda }}K\left( \frac{A-C}{A-B}%
\right) \qquad \text{and\qquad }\omega _{2}=i\frac{16}{\sqrt{B-A}\sqrt{%
\lambda }}K\left( \frac{C-B}{A-B}\right) .
\end{equation}%
Here $K(m)$ denotes a complete elliptic integral of the first kind. As
expected we recover the previous case when two of the constants $A,B,C$
coincide as in these cases the periods become dependent on each other, i.e.~$%
\lim\nolimits_{A\rightarrow B}\omega _{1}/\omega _{2}=1/2\limfunc{sign}(C-A)$%
, $\lim\nolimits_{A\rightarrow C}\omega _{1}/\omega _{2}=-i\infty $ and $%
\lim\nolimits_{B\rightarrow C}\omega _{1}/\omega _{2}=0$.

Studying at first the $\mathcal{PT}$-symmetric solutions, we find periodic
solutions encircling some of the fixed points. For instance in figure \ref%
{fig5} we depict an example in which the fixed points $A$ and $B$ are
encircled, whereas $C$ is situated outside of the trajectories. Different
types of scenarios are also expected, changing for instance the sign in $%
\gamma $ will lead to a periodic orbit surrounding the points $B$ and $C$.

\begin{figure}[h!]
\centering  \includegraphics[width=7.0cm]{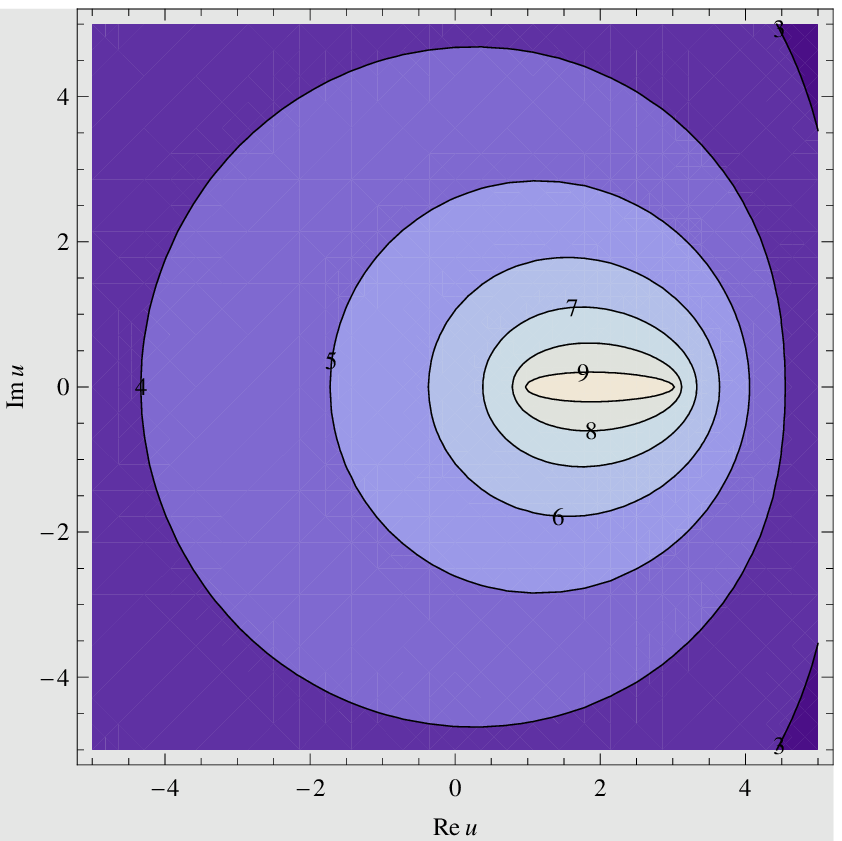} %
\includegraphics[width=7.0cm]{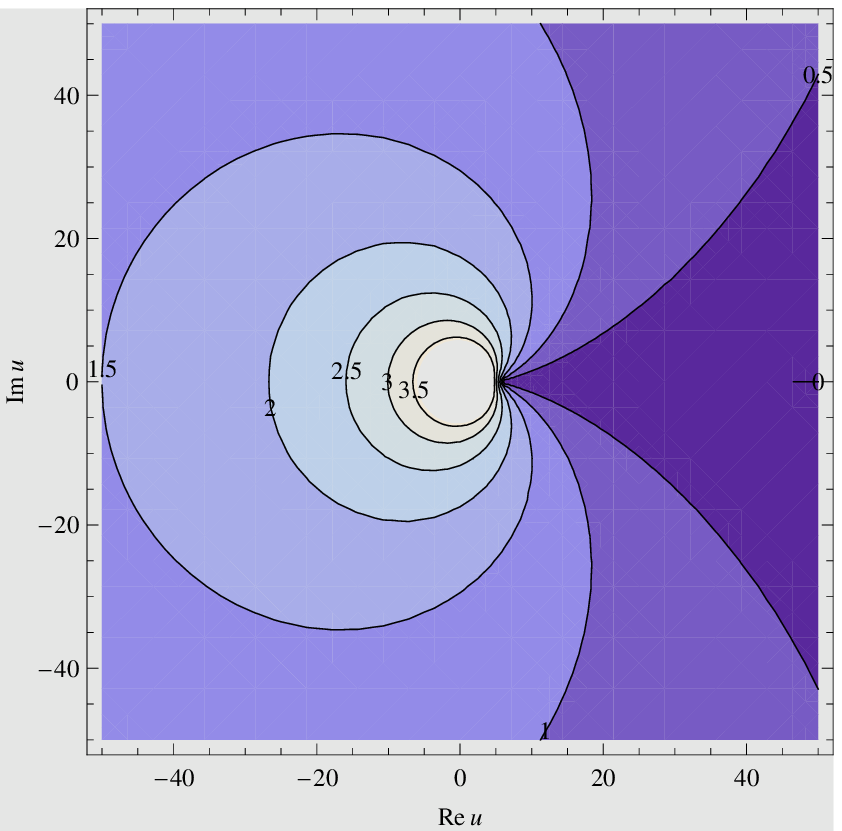}
\caption{$\mathcal{PT}$-symmetric complex elliptic solutions of the KdV
equation with $A=1$, $B=3$, $C=6$, $c=1$, $\protect\beta =3/10$, $\protect%
\gamma =-3$ for different values of $\func{Im}\protect\zeta _{0}$.}
\label{fig5}
\end{figure}

\begin{figure}[h]
\centering  \includegraphics[width=7.0cm]{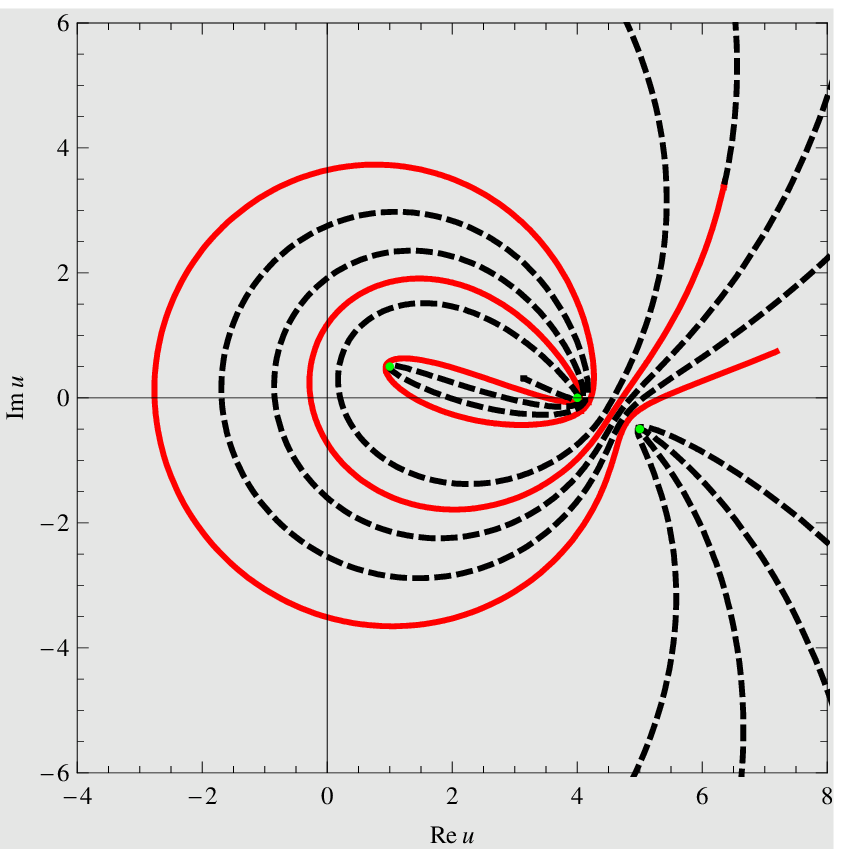} %
\includegraphics[width=7.0cm]{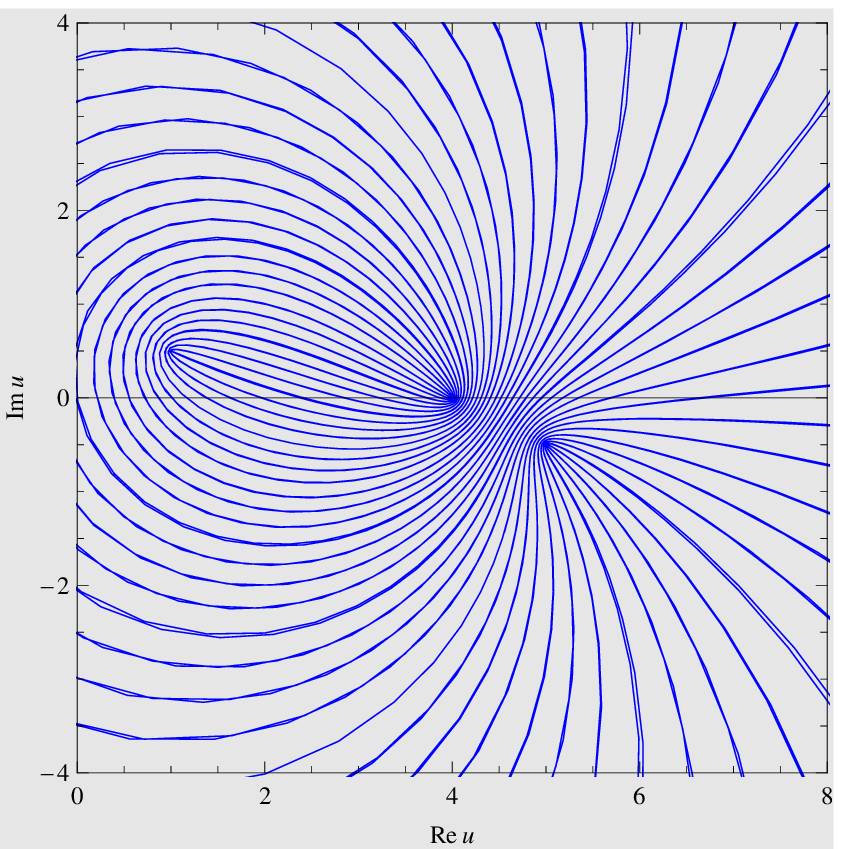}
\caption{Spontaneously broken $\mathcal{PT}$-symmetric complex elliptic
solutions of the KdV equation for $\func{Im}\protect\zeta _{0}=6$ with $A=4$%
, $B=5-i/2$, $C=1+i/2$, $c=1$, $\protect\beta =3/10$ and $\protect\gamma =3$%
: (a) $-64\leq \protect\zeta \leq 18$ solid (red) and $18<\protect\zeta \leq
200$ dashed (black); (b) $-200<\protect\zeta <1400$. }
\label{fig5x}
\end{figure}

Next we break the $\mathcal{PT}$-symmetry spontaneously by complexfying the
free parameters of the solutions. We find that the periodic trajectories
open up and cease to be periodic. In figure \ref{fig5x}a we trace part of
the trajectory to illustrate the behaviour. Starting at the point $%
u(-64)\approx 7.19+i0.74$ the trajectory passes down between the points $A$
and $B$ and moves then up again surrounding once the points $C$ and $A$ in a
clockwise sense. Thereafter it encircles $C$ once more but instead of moving
around $A$ it passes inbetween $C$ and $A$ surrounding $A$ in an
anti-clockwise sense. It keeps progressing in an anti-clockwise manner
encircling $C$ before passing from below between the points $A$ and $B$
reaching the point $u(18)\approx 6.36+i3.45$. It appears that this type of
movement is repeated indefinitely. In figure \ref{fig5x}b we depict a wider
range for $\zeta $ indicating that the region of the phase space depicted
will eventually be filled densely by the trajectory, hence suggesting a
chaotic behaviour. However, we do not observe a sensitivity towards the
initial condition, which would be typical for a proper chaotic system.

\begin{figure}[h]
\centering  \includegraphics[width=7.0cm]{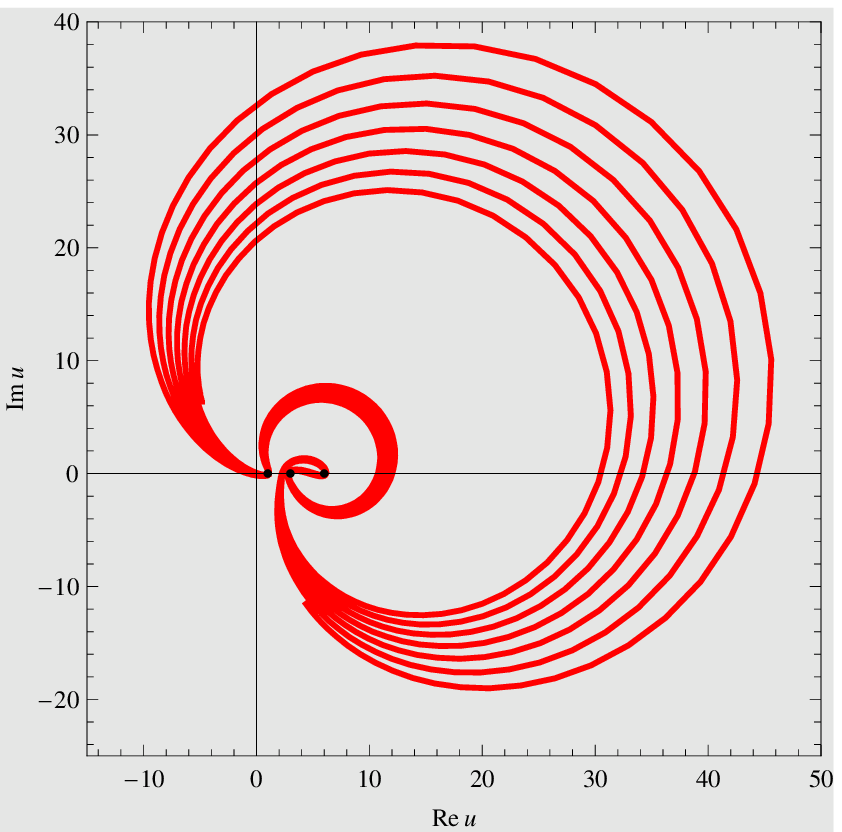} %
\includegraphics[width=7.0cm]{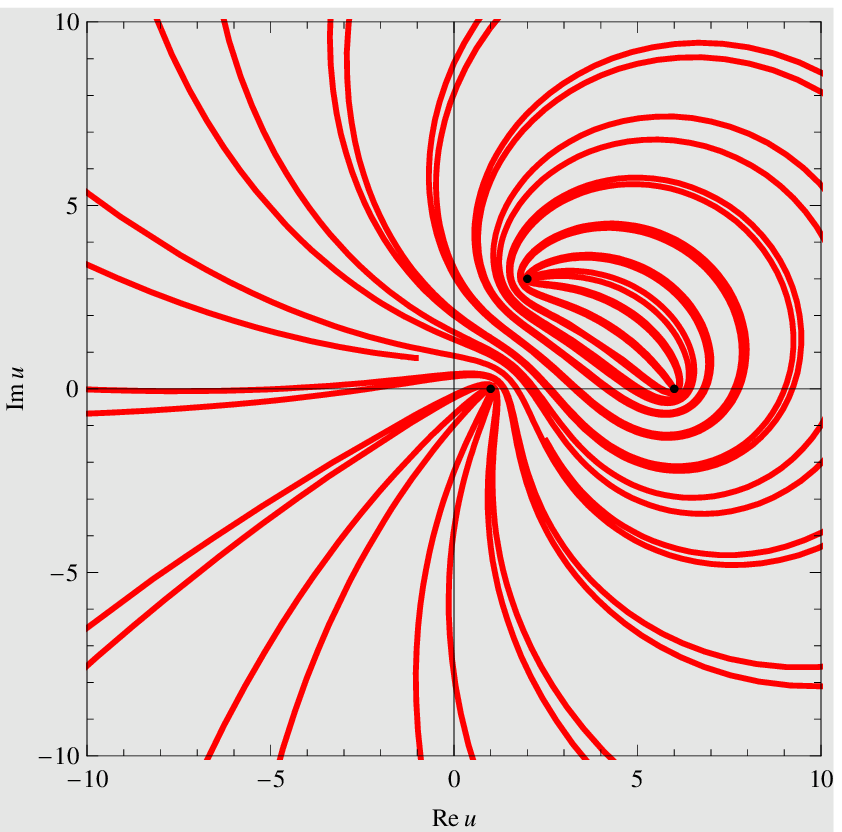}
\caption{Broken $\mathcal{PT}$-symmetric complex elliptic solutions of the
KdV equation for $\func{Im}\protect\zeta _{0}=6$: (a) $A=1$, $B=3$, $C=6$, $%
c=1$, $\protect\beta =3/10$ and $\protect\gamma =3+2i$ for $-200\leq \protect%
\zeta \leq 200$;(b) $A=1$, $B=2+3i$, $C=6$, $c=1$, $\protect\beta =3/10-i/10$
and $\protect\gamma =3$ for $-200\leq \protect\zeta \leq 200$.}
\label{fig5xx}
\end{figure}

Finally we may also break the $\mathcal{PT}$-symmetry completely by
complexfying the parameters of the model $\beta $ or/and $\gamma $. Examples
for such a scenario are depicted in figure \ref{fig5xx}. The behaviour is
similar to the one of the spontaneously broken case, i.e. the periodic
motion has turned into open trajectories with a noncompact limit set.
Increasing the range for $\zeta $ will fill the depicted part of the phase
space similarly as in the spontaneously broken $\mathcal{PT}$-symmetry case,
thus suggesting a chaotic behaviour, albeit once again we do not observe the
typical sensitivity towards the initial conditions.

Figures \ref{fig5x} and \ref{fig5xx} are very reminiscent of the plots which
may be found in section 5 of reference \cite{Bender:2008fr}. This is not
surprising as formally the differential equations solved in there for the
quantum mechanical setting for the potential $V\sim x^{3}$ are special cases
of our more general treatment when making the identification $u\rightarrow x$
and $\zeta \rightarrow t$ for our traveling wave equation. With the further
identifications 
\begin{equation}
\kappa _{1}=0,\quad \kappa _{2}=\gamma E,\quad \beta =6cg\quad \text{%
and\quad }\gamma =-c  \label{match}
\end{equation}%
equation (\ref{P}) converts precisely into the quantum mechanical Hamiltonian%
\begin{equation}
H=E=\frac{1}{2}p^{2}+\frac{1}{2}x^{2}-gx^{3},
\end{equation}%
treated in \cite{Bender:2008fr}. The identification (\ref{match}) explains
why no analogue to our rational solution was found in \cite{Bender:2008fr},
since $\kappa _{1}=0$ implies $c=0$ and therefore the vanishing of $\kappa
_{2}$, i.e. the confining potential, and $A$. However, as can be seen from (%
\ref{ak}) there should be an analogue to our trigonometric solution with
energy $E=-4c^{3}/\gamma \beta ^{2}$ when setting $A=2c/\beta $. This energy
depends on the coupling constant and is of course not freely choosable as in
the elliptic case presented in this section where $\kappa _{2}$ is a free
parameter. More potentials of the type treated in \cite{Bender:2008fr} can
be obtained systematically from the deformations discussed below.

The new elliptic solutions recently found for generalized shallow wave
equations \cite{BijanDas} are expected to exhibit a similar behaviour as the
solution discussed in this subsection.

\subsubsection{Soliton solutions}

\begin{figure}[h]
\centering  \includegraphics[width=7.0cm]{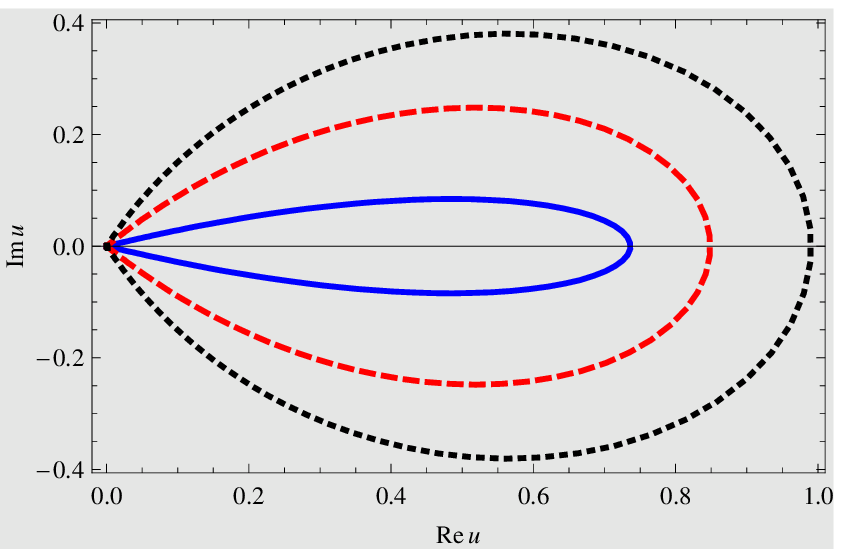} %
\includegraphics[width=7.0cm]{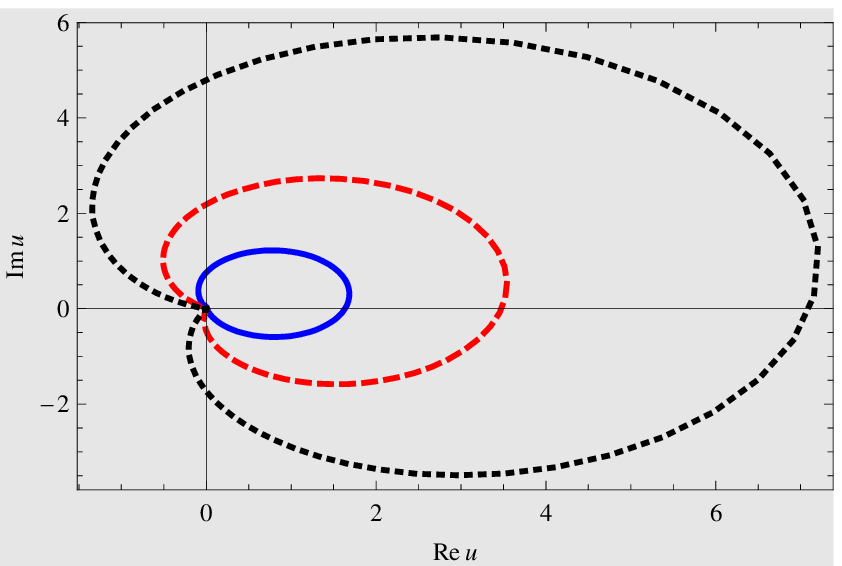}
\caption{Complex one-soliton solution of the KdV-equation at fixed time $%
t=-2 $: (a) $\mathcal{PT}$-symmetric solution with $\protect\beta =6$, $%
\protect\gamma =1$, $p_{1}=1.2$ for $\protect\phi =i0.3$ solid (blue), $%
\protect\phi =i0.8$ dashed (red) and $\protect\phi =i1.1$ dotted (black);
(b) Broken $\mathcal{PT}$-symmetric solution with $\protect\beta =6$, $%
\protect\gamma =1+i0.4$, $p_{1}=1.2$ for $\protect\phi =i0.3$ solid (blue), $%
\protect\phi =i0.8$ dashed (red) and $\protect\phi =i1.1$ dotted (black).}
\label{fig6}
\end{figure}

There exist various techniques to construct soliton solutions. Here we
briefly recall and apply Hirota's direct method \cite{Hirotabook} for that
purpose. The starting point of this approach is Hirota's bilinear form,
which reads in general 
\begin{equation}
p(D)\tau \cdot \tau =0,  \label{Hirota}
\end{equation}%
with $p(D)$ being a polynomial in the Hirota derivatives, i.e.~$D_{x}f\cdot
g=$ $f_{x}\cdot g-f\cdot g_{x}$ where the dot "$\cdot $" indicates the
noncommutative nature of the product $f\cdot g$. The soliton solutions to (%
\ref{Hirota}) may then be constructed perturbatively order by order in $%
\varepsilon $ from the expansion 
\begin{equation}
\tau =\sum\limits_{n=0}^{\infty }\varepsilon ^{n}\tau _{n}.
\end{equation}%
Defining the function $\tau (x,t)$ via the relation 
\begin{equation}
u\left( x,t\right) =\frac{12\gamma }{\beta }(\ln \tau )_{xx},
\end{equation}%
the KdV-equation (\ref{KdV}) can be brought into Hirota's bilinear form%
\begin{equation}
\frac{6\gamma }{\beta }\left( \gamma D_{x}^{4}+D_{x}D_{t}\right) \tau \cdot
\tau =0.
\end{equation}%
We have set here the integration constant to zero. Taking the zeroth order
solution to be $\tau _{0}=1$ the first order tau-function results to $\tau
_{1}(x,t)=\exp (p_{1}x-\gamma p_{1}^{3}t+\phi _{1})$ and we obtain the
well-known one-soliton solution up to this order%
\begin{equation}
u\left( x,t\right) =\frac{3\gamma p_{1}^{2}}{\beta \cosh ^{2}\left[ \frac{1}{%
2}(p_{1}x-\gamma p_{1}^{3}t+\phi _{1})\right] }.  \label{solisol}
\end{equation}%
A $\mathcal{PT}$-symmetric variant of this solution is depicted in figure %
\ref{fig6}a. As expected the origin in the $u$-plane is an asymptotic fixed
point, reached for $x\rightarrow \pm \infty $ or $t\rightarrow \pm \infty $.

Breaking the $\mathcal{PT}$-symmetry in various ways yields more unexpected
results. Regarding $p_{1}$, $\phi _{1}$, $\zeta _{0}$ as parameters of the
solution and $\beta $, $\gamma $ as model defining constants, their
complexification constitutes a spontaneous or complete $\mathcal{PT}$%
-symmetry breaking, respectively. The corresponding solutions in the $u$%
-plane appear as distorted versions of the symmetric case, as can be seen in
figure \ref{fig6}b. An interesting feature of the broken complex solution is
that, unlike its real counterpart, the soliton does not maintain its overall
shape while traveling. Instead, it becomes a breather and regains its shape
after a certain distance $\Delta _{x}$ and certain amount of time $\Delta
_{t}$ governed by the equation%
\begin{equation}
u\left( x+\Delta _{x},t\right) =u\left( x,t+\Delta _{t}\right) .  \label{ud}
\end{equation}%
Separating $\gamma $, $p_{1}$ into their real and imaginary part as $\gamma
=\gamma _{r}+i\gamma _{i}$ and $p_{1}=p_{r}+ip_{i}$, we can solve (\ref{ud})
for the one-soliton solution (\ref{solisol}) to

\begin{eqnarray}
\Delta _{t} &=&\frac{2\pi p_{r}}{\left( p_{i}^{4}-p_{r}^{4}\right) \gamma
_{i}-2p_{i}p_{r}\left( p_{i}^{2}+p_{r}^{2}\right) \gamma _{r}},\text{\qquad }
\label{d1} \\
\Delta _{x} &=&2\pi \frac{p_{i}\left( 3p_{r}^{2}-p_{i}^{2}\right) \gamma
_{i}+2\pi p_{r}\left( 3p_{i}^{2}-p_{r}^{2}\right) \gamma _{r}}{\left(
p_{i}^{4}-p_{r}^{4}\right) \gamma _{i}-2p_{i}p_{r}\left(
p_{i}^{2}+p_{r}^{2}\right) \gamma _{r}}.  \label{d2}
\end{eqnarray}%
The speed of the soliton is therefore given by%
\begin{equation}
v=-\frac{\Delta _{x}}{\Delta _{t}}=\left( 3p_{i}^{2}-p_{r}^{2}\right) \gamma
_{r}-\frac{p_{i}\left( p_{i}^{2}-3p_{r}^{2}\right) \gamma _{i}}{p_{r}}.
\end{equation}

\begin{figure}[h]
\centering  \includegraphics[width=7.0cm]{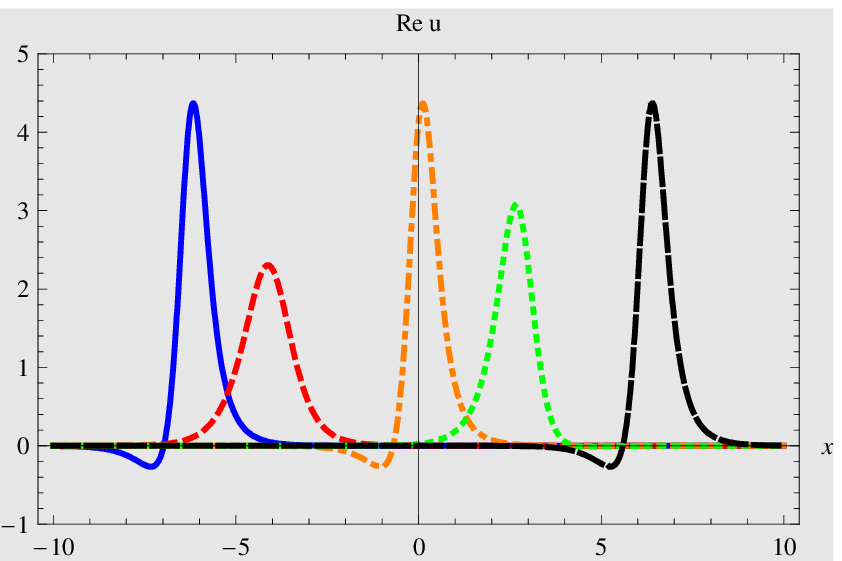} %
\includegraphics[width=7.0cm]{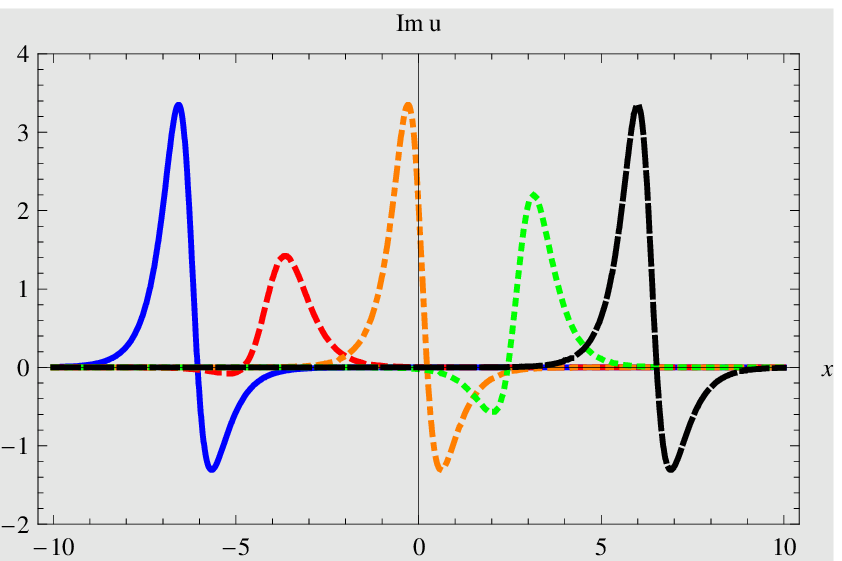}
\caption{Complex spontaneously broken one-soliton solution of the
KdV-equation with $\protect\beta =6$, $\protect\gamma =1+i/2$, $p_{1}=2$, $%
\protect\phi =i0.8$ and $\Delta _{t}=-\protect\pi /2$ for different times $%
t=-\protect\pi /2$ solid (blue), $t=-1$ dashed (red), $t=0$ dasheddot
(orange), $t=0.7$ dotted (green), and $t=\protect\pi /2$ dasheddotdot
(black) (a) real part; (b) imaginary part.}
\label{figsol}
\end{figure}

Clearly for the $\mathcal{PT}$-symmetric solution the shape will be the same
in the $u$-plane even at different times, whereas the complex solution will
change its shape as we observe in figure \ref{fig6}. In figure \ref{figsol}
we observe that the one-soliton changes its shape while progressing in time
but regains it after $\Delta _{t}=-\pi /2$ for the chosen values.

\begin{figure}[h!]
\centering  \includegraphics[width=7.0cm]{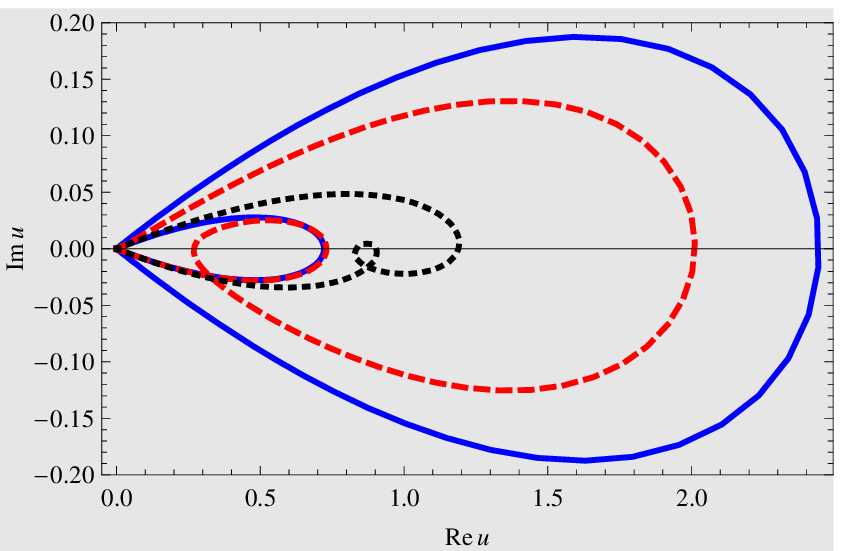} %
\includegraphics[width=7.0cm]{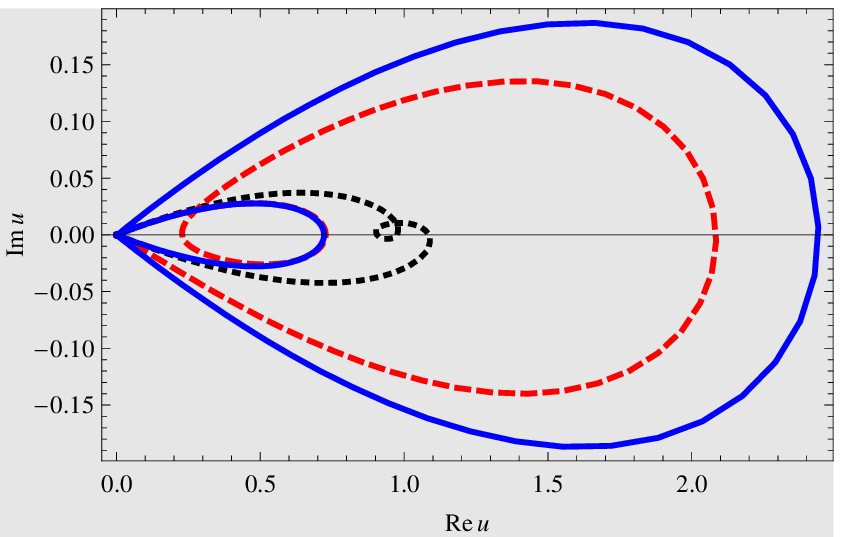}
\caption{ $\mathcal{PT}$-symmetric two-soliton solutions of the KdV equation
for $\protect\beta =6$, $\protect\gamma =1$, $p_{1}=1.2$, $p_{2}=2.2$, $%
\protect\phi _{1}=i0.1$ and $\protect\phi _{2}=i0.2$. (a) $t=-2$ solid
(blue), $t=-0.2$ dashed (red), $t=0.2$ dotted (black); (b) $t=0.3$ dotted
(black), $t=0.8$ dashed (red), $t=2.0$ solid (blue).}
\label{fig7}
\end{figure}

\begin{figure}[h]
\centering  \includegraphics[width=7.0cm]{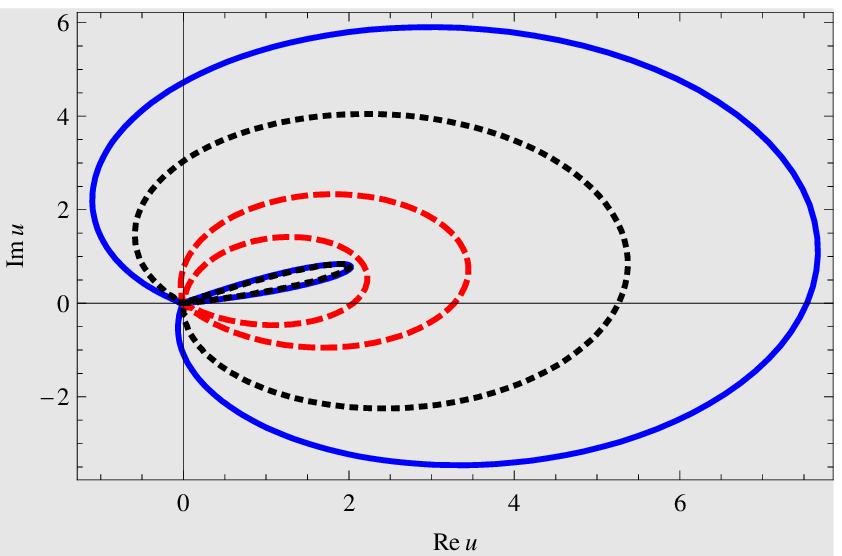} %
\includegraphics[width=7.0cm]{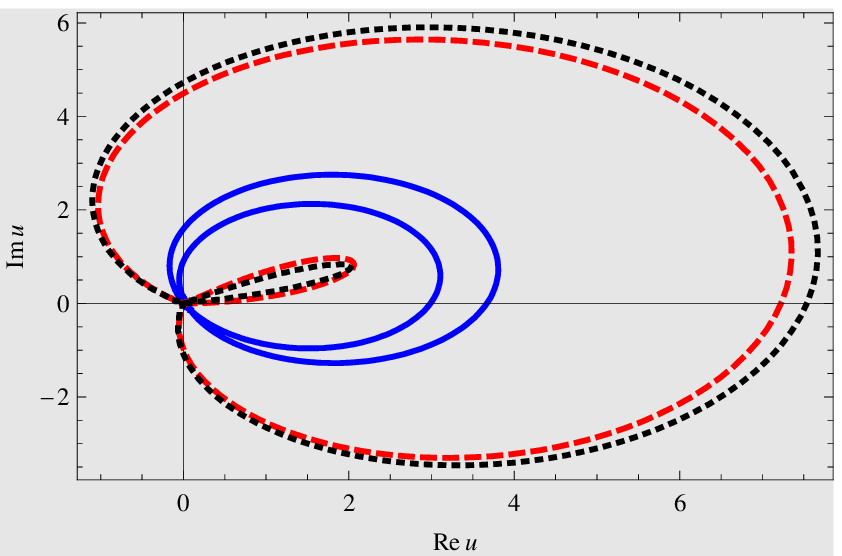}
\caption{Broken $\mathcal{PT}$-symmetric two-soliton solutions of the KdV
equation for $\protect\beta =6$, $\protect\gamma =1+i\protect\pi /8$, $%
p_{1}=2(2/3)^{1/3}$, $p_{2}=2$, $\protect\phi _{1}=i0.1$ and $\protect\phi %
_{2}=i0.2$. (a) $t=-4$ solid (blue), $t=-3.5$ dashed (red), $t=-2.$ dotted
(black); (b) $t=0.7$ solid (blue), $t=2$ dashed (red), $t=8$ dotted (black).}
\label{figxx}
\end{figure}

Proceeding to the next order in $\varepsilon $ we compute the two soliton
solution to%
\begin{equation}
u\left( x,t\right) =\frac{24\gamma
\sum\nolimits_{k=0}^{6}c_{k}(-1)^{k}p_{2}^{k}p_{1}^{6-k}}{\beta \left(
p_{1}+p_{2}\right) {}^{4}\left[ 2\cosh \left( \frac{1}{2}\left( \eta
_{1}-\eta _{2}\right) \right) +e^{-\frac{\eta _{1}}{2}-\frac{\eta _{2}}{2}%
}\left( \frac{e^{\eta _{1}+\eta _{2}}\left( p_{1}-p_{2}\right) {}^{4}}{%
\left( p_{1}+p_{2}\right) {}^{4}}+1\right) \right] ^{2}{}}  \label{twos}
\end{equation}%
where we abbreviated $\eta _{i}=p_{i}x-\gamma p_{i}^{3}t+\phi _{i}$ for $%
i=1,2$ with%
\begin{equation}
\begin{array}{r}
c_{0}=1+\cosh \eta _{2},\quad c_{1}=4\sinh \eta _{2},\quad c_{2}=\cosh \eta
_{1}+6\cosh \eta _{2}-1,\quad c_{3}=4\left( \sinh \eta _{1}+\sinh \eta
_{2}\right)%
\end{array}%
\end{equation}%
and used the symmetry relations $c_{i}(\eta _{1},\eta _{2})=c_{6-i}(\eta
_{2},\eta _{1})$. We depict the time evolution of this solution in figures %
\ref{fig7} and \ref{figxx}.

We observe in figure \ref{fig7} the typical scenario for soliton scattering,
albeit in the complex plane. For large negative time the two-solitons are
separated, indicated here by two individual one-soliton solutions resembling
the type depicted in figure \ref{fig6} touching each other only in the
asymptotic point at the origin. In the scattering regime the two solutions
merge in a non $\mathcal{PT}$-symmetric manner until they separate again for
large positive time, acquiring once again their individual one-soliton shape.

For the values in figure \ref{figxx} the periods for the one-soliton
breather to regain its shape with $p_{1}$ and $p_{2}$ are computed by (\ref%
{d1}) to be $\Delta _{t}^{1}=-3$ and $\Delta _{t}^{2}=-2$, respectively.
Indeed we observe in figure \ref{figxx}a that one of the solitons has
recovered its shape comparing the solutions at $t=-4$ and $t=-2$. In figure %
\ref{figxx}b we observed that both solitons have almost regained their
original shape passing from $t=2$ to $t=8$, that is after $2\Delta
_{t}^{1}=3\Delta _{t}^{2}$. The slight discrepancy visible in the figure is
due to the well known time delay occurring in a multi-soliton scattering,
see e.g. \cite{Fring:1994mz}.

The energy for the one-soliton solution on the interval $(-\infty ,\infty )$
is easily computed with formula (\ref{Energy})%
\begin{equation}
E_{1s}=-\frac{36\gamma ^{3}p_{1}^{5}}{5\beta ^{2}},
\end{equation}%
exhibiting the typical behaviour, that is being real for the $\mathcal{PT}$%
-symmetric case and complex for the (spontaneously) broken scenario. The
integral (\ref{Energy}) with $u(x,t)$ given by two-soliton solution (\ref%
{twos}) is not easily computed analytically. For large values of time we
compute numerically real energies for $\mathcal{PT}$-symmetric solutions,
for instance for the values used in figure \ref{fig7} we obtain $%
E_{2s}\approx -10.8049=E_{1s}(p_{1})+E_{1s}(p_{2})$. On the other hand we
compute complex energies for broken $\mathcal{PT}$-symmetric solutions, such
as for the values used in figure \ref{figxx} for which we evaluate $%
E_{2s}\approx -7.8876-i9.4327=E_{1s}(p_{1})+E_{1s}(p_{2})$. This means that
in all scenarios the energy of the two-soliton equals the sum of the two
separate one-solitons.

\section{$\mathcal{PT}$-symmetric deformations of the KdV equation}

Employing the standard arguments used in the study of non-Hermitian
Hamiltonian systems with real eigenvalues, we may now $\mathcal{PT}$%
-symmetrically deform the Hamiltonian and maintain the possibility to have
well defined physical systems, e.g.~we still obtain real energies resulting
from the new models despite their non-Hermitian nature. The general
principle is simply to deform $\mathcal{PT}$-anti-symmetric quantities, that
means if we have the property $\mathcal{PT}:\phi (x,t)\mapsto -\phi (x,t)$
for some field $\phi (x,t)$, we define a deformation map $\delta
_{\varepsilon }:$ $\phi (x,t)\mapsto -i[i\phi (x,t)]^{\varepsilon }$. The
undeformed case is recovered in the limit $\varepsilon =1$. The new quantity
will remain anti-$\mathcal{PT}$-symmetric with the crucial difference that
the overall minus sign is generated from the antilinear nature of the $%
\mathcal{PT}$-operator, i.e.~$i\mapsto -i,$ rather than from $\phi
(x,t)\mapsto -\phi (x,t)$. This means for the Hamiltonian resulting from the
density (\ref{Hkdv}) we can make use of either of the two possibilities%
\begin{equation}
\delta _{\varepsilon }^{+}:u_{x}\mapsto u_{x,\varepsilon
}:=-i(iu_{x})^{\varepsilon }\qquad \text{or\qquad }\delta _{\varepsilon
}^{-}:u\mapsto u_{\varepsilon }:=-i(iu)^{\varepsilon },~~~~\ \ ~~
\label{delta}
\end{equation}%
depending on whether we choose $u(x,t)$ to be $\mathcal{PT}$-symmetric or $%
\mathcal{PT}$-anti-symmetric, respectively. Accordingly we define the
deformed models with some suitable normalisation by the following
Hamiltonian densities%
\begin{equation}
\mathcal{H}_{\varepsilon }^{+}=-\frac{\beta }{6}u^{3}-\frac{\gamma }{%
1+\varepsilon }(iu_{x})^{\varepsilon +1},\quad \qquad \text{and\qquad }%
\mathcal{H}_{\varepsilon }^{-}=\frac{\beta }{(1+\varepsilon )(2+\varepsilon )%
}(iu)^{\varepsilon +2}+\frac{\gamma }{2}u_{x}^{2},\quad
\end{equation}%
with corresponding equations of motion%
\begin{equation}
u_{t}+\beta uu_{x}+\gamma u_{xxx,\varepsilon }=0\qquad \text{and\qquad }%
u_{t}+i\beta u_{\varepsilon }u_{x}+\gamma u_{xxx}=0,  \label{dIto2}
\end{equation}%
respectively. The Hamiltonian $\mathcal{H}_{\varepsilon }^{+}$ was proposed
in \cite{AFKdV} whereas $\mathcal{H}_{\varepsilon }^{-}$ corresponds to a
complex version of the generalized KdV equation. For the higher deformed
derivatives we use here the notation $u_{xx,\varepsilon }:=\partial
_{x}u_{x,\varepsilon }$, $u_{xxx,\varepsilon }:=\partial
_{x}^{2}u_{x,\varepsilon }$,$\ldots ,u_{nx,\varepsilon }:=\partial
_{x}^{n-1}u_{x,\varepsilon }$, which means we only deform the first
derivative and keep acting on it with $\partial _{x}$ to define the higher
order derivatives. On the level of the equation of motion we could also
deform the dispersion term proportional to $\beta $, as investigated in \cite%
{BBCF} for the KdV-equation. However, such deformations do not lead to
Hamiltonian systems and the question of how $\mathcal{PT}$-symmetry may be
utilized in this scenario remains an open issue.

Let us now present some solutions to these equations.

\subsection{The $\mathcal{H}_{\protect\varepsilon }^{+}$-models}

Proceeding similarly as in the previous section, we integrate the first
equation in (\ref{dIto2}) twice and obtain the deformed version of equation (%
\ref{P})%
\begin{equation}
u_{\zeta }^{(n)}=\exp \left[ \frac{i\pi }{2(\varepsilon +1)}(1-\varepsilon
+4n)\right] \left[ \lambda _{\varepsilon }P(u)\right] ^{\frac{1}{%
1+\varepsilon }},  \label{uzeta}
\end{equation}%
where we abbreviated $\lambda _{\varepsilon }=-\beta (1+\varepsilon
)/(6\gamma \varepsilon )$ and denote different branches by $n$. The
polynomial $P(u)$ is identical to the one introduced in (\ref{P}), which
means we can employ the same factorization as in the previous section.
Integrating once more yields 
\begin{equation}
\zeta ^{(n)}-\zeta _{0}=\exp \left[ \frac{i\pi }{2(\varepsilon +1)}%
(\varepsilon -1+4n)\right] \int du\frac{1}{\left[ \lambda _{\varepsilon }P(u)%
\right] ^{\frac{1}{1+\varepsilon }}}.  \label{jj}
\end{equation}%
We may now proceed as before and specify further the form of $P(u)$. As for
the case $\varepsilon =1$, in some specific cases we succeed to compute the
remaining integral and subsequently solve the resulting equation for $u$,
thus obtaining $u(\zeta )$. However, even when this is not possible
analytically we can still investigate (\ref{jj}) numerically for \emph{all}
cases by viewing $\zeta $ as a function of the complex variable $u$ and
plotting the contours of $\func{Im}[\zeta (u)]=\zeta _{0}$ in the $u$-plane,
while taking special care about the different Riemann sheets labeled by $n$.
One should distinguish here these type of Riemann sheets from those arising
due to the technique employed in our solution procedure. Considering $\zeta $
as a function of $u$, as we do in some intermediate steps, will introduce
new branch cuts which are sometimes seen in our figures, e.g. figure \ref%
{fig22}b. However, we do not attribute any meaning to them in the $u$-plane.
Genuine branch cuts can always be distinguished from the \textquotedblleft
technical\textquotedblright\ ones by the fact that they have to be connected
to the fixed points which are branch points.

\subsubsection{Rational solutions}

We start with the same assumption as in the previous section, namely $%
P(u)=(u-A)^{3}$, which will impose the same constraints (\ref{rat}) for the
factorization. The case $\varepsilon =2$ is special since we may take the
root $1/3$ in this case. Integrating (\ref{jj}) and solving the result for $%
u $ we find 
\begin{equation}
u^{(n)}\left( \zeta \right) =A+\exp \left[ -\frac{ie^{-\frac{2}{3}in\pi }}{%
2^{2/3}}\left( \frac{\beta }{\gamma }\right) ^{1/3}\left( \zeta -\zeta
_{0}\right) \right] .
\end{equation}%
For the remaining integer values $\varepsilon \in \mathbb{Z}\backslash \{2\}$
we can compute also a particular solution%
\begin{equation}
u\left( \zeta \right) =\frac{c}{\beta }+\exp \left( \frac{i\pi }{2}\frac{%
5+\varepsilon }{2-\varepsilon }\right) \left( \frac{6\gamma \varepsilon }{%
\beta }\right) ^{\varepsilon -2}(1+\varepsilon )^{\frac{\varepsilon }{%
\varepsilon -2}}\left( \varepsilon -2\right) ^{\frac{\varepsilon +1}{%
\varepsilon -2}}\left( \zeta -\zeta _{0}\right) ^{\frac{\varepsilon +1}{%
\varepsilon -2}}.  \label{ratsol}
\end{equation}%
However, in general we have to take care about the different branches
present in (\ref{jj}). Our numerical findings for some specific cases are
depicted in the figures below.

We supplement our numerical analysis with the prediction of some analytical
features. Of special interest are the lines approaching or leaving the fixed
points radially. When the symmetry is unbroken they can be defined
equivalently as the lines for which $\zeta _{0}=0$. We compute them by
noting first that on one hand any point $\tilde{u}$ on the line radially
crossing the point $u=A$ is characterized by a constant value for $\arg (%
\tilde{u}-A)=:\theta _{0}$. On the other hand the change of $u$ with respect
to $\zeta $ has to point into the same direction, i.e.~$\arg (\pm \tilde{u}%
_{\zeta })=\theta _{0}$. Hence the lines of real initial conditions are
determined by solving%
\begin{equation}
\arg (\pm \tilde{u}_{\zeta }^{(n)})=\arg (\tilde{u}^{(n)}-A)+2\pi m\text{.}
\label{00}
\end{equation}%
Employing (\ref{uzeta}) and the factorized form of $P(u)$ we can solve (\ref%
{00}), obtaining%
\begin{eqnarray}
\theta _{0}^{(n,m,\pm )} &:&=\arg (\tilde{u}^{(n)}-A)  \label{37} \\
&=&\frac{\pi }{2(2-\varepsilon )}\left[ \varepsilon -1+4(m-n+m\varepsilon
)-(1+\varepsilon )(1\pm 1)\right] +\frac{1}{\varepsilon -2}\arg \lambda
_{\varepsilon }  \notag
\end{eqnarray}%
for the angle in which the trajectories with $\zeta _{0}=0$ enter the point $%
A$.

\begin{figure}[h!]
\centering \includegraphics[width=7.0cm]{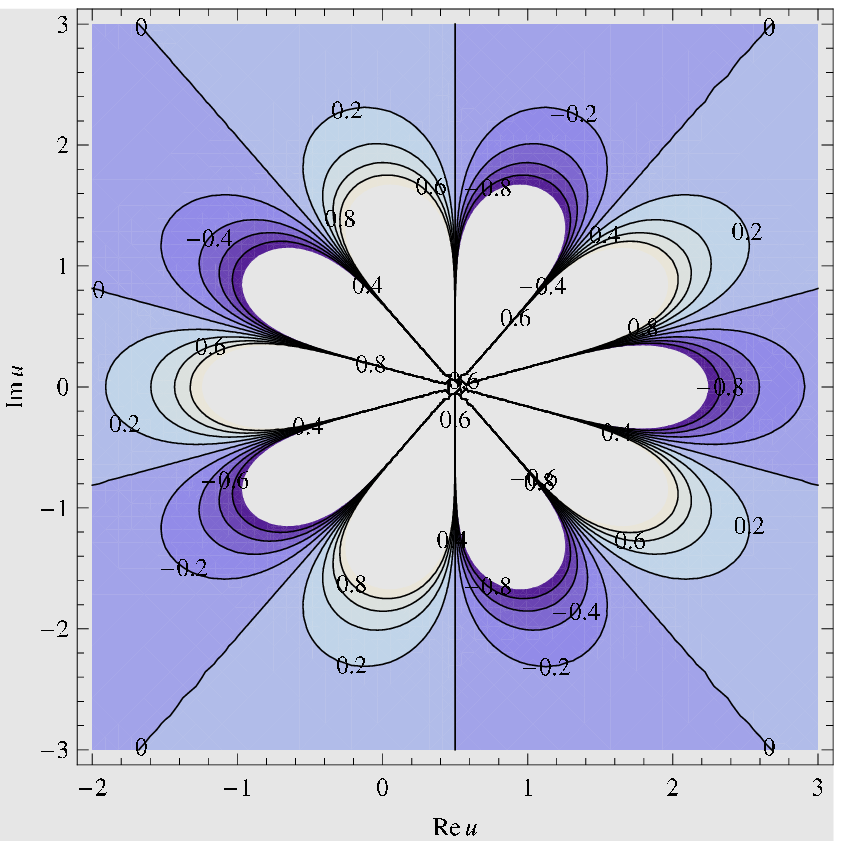} %
\includegraphics[width=7.0cm]{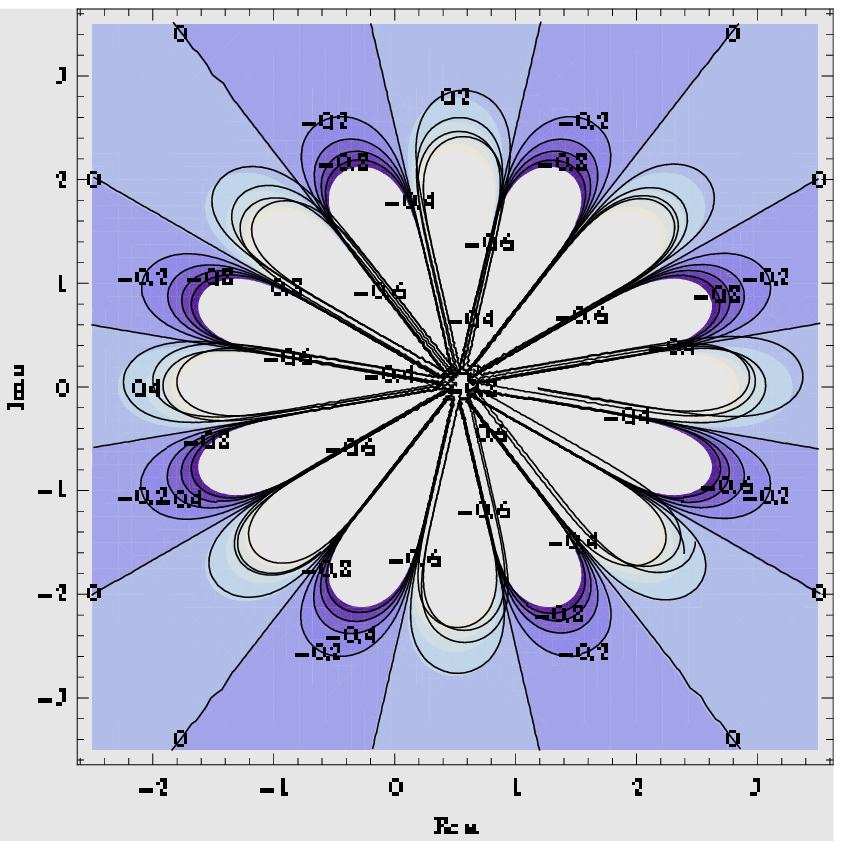}
\caption{Complex $\mathcal{PT}$-symmetric rational solutions of the deformed
KdV equation with $A=1/2,c=1$, $\protect\beta =2$ and $\protect\gamma =3$
for (a) $\mathcal{H}_{-1/2}^{+}$ and (b) $\mathcal{H}_{-2/3}^{+}$.}
\label{fig9}
\end{figure}

In figure \ref{fig9} we show the first two members of the sequel
parameterizing the deformation parameter as $\varepsilon =-n/(n+1)$ with $%
n=1,2,\ldots $ We observe that the pictures resemble flowers with $4+6n$
petals. We may predict the number of petals analytically as they are equal
to the number of $\zeta _{0}=0$ solutions computable by (\ref{37}). For
instance for the case $n=3$ the formula (\ref{37}) predicts the $16$ values $%
\theta _{0}=(2\ell -1)\pi /16$ for the parameter choice $c=1$, $\beta =2$
and $\gamma =3$ with $\ell =1,\ldots ,16$. We recognize these values in
figure \ref{fig9}b. Note that the wedge regions separate different solutions
from each other as the point $A$ is always an asymptotic fixed point.

\begin{figure}[h!]
\centering  \includegraphics[width=7.0cm]{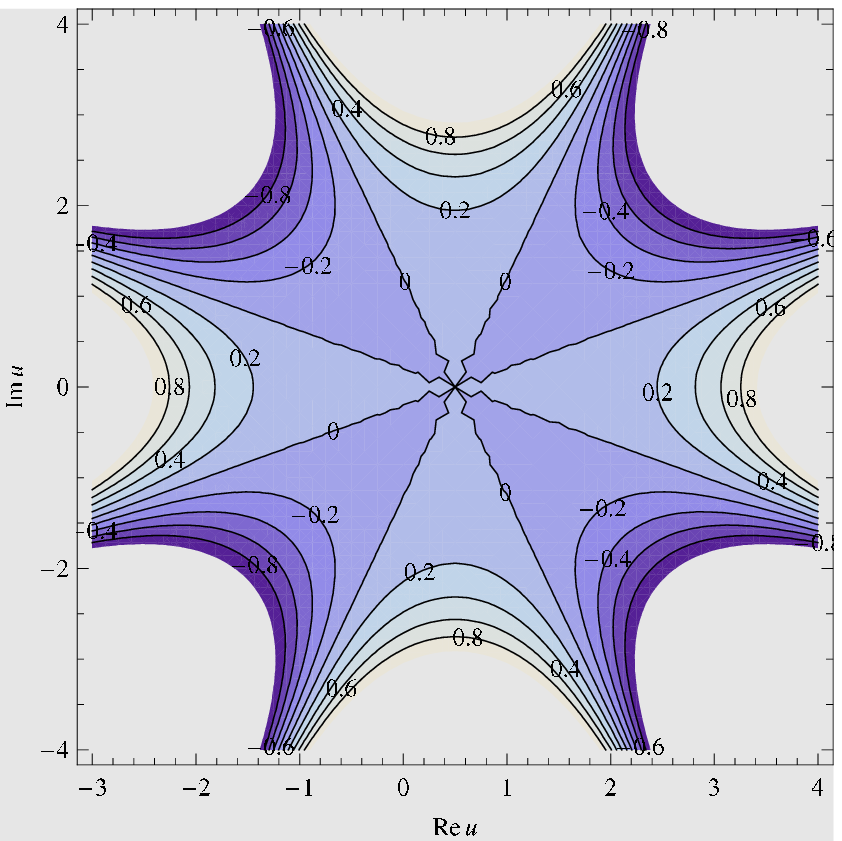} %
\includegraphics[width=7.0cm]{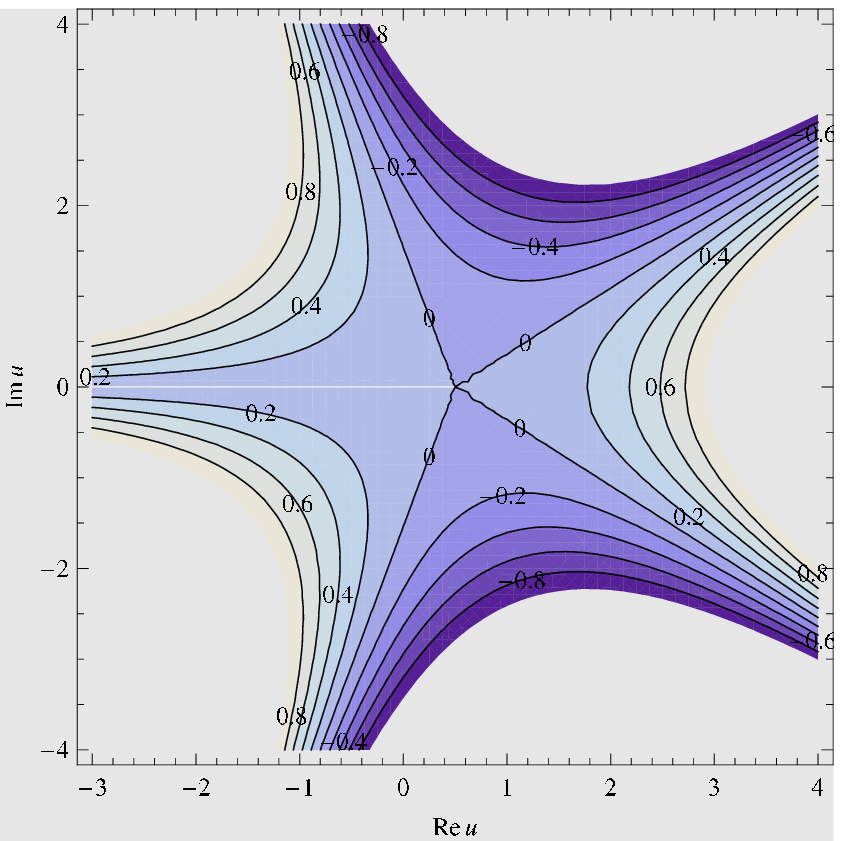}
\caption{Complex $\mathcal{PT}$-symmetric rational solutions of the deformed
KdV equation with $A=1/2$, $c=1$, $\protect\beta =2$ and $\protect\gamma =3$
for (a) $\mathcal{H}_{-2}^{+}$ and (b) $\mathcal{H}_{-3}^{+}$.}
\label{fig10}
\end{figure}

The trajectories appear to be qualitatively quite different for negative
integer values of $\varepsilon $. For instance in figure \ref{fig10}a and %
\ref{fig10}b we depict some trajectories for the models with $\varepsilon
=-2 $ and $\varepsilon =-3$, respectively. It appears that the fixed point $%
A $ is more like a saddle point, i.e. we exhibit trajectories more of a
hyperbolic nature running away to infinity rather than converging
asymptotically to $A$ as in figure \ref{fig9}. The $\zeta _{0}=0$ solutions
are predicted once again correctly by (\ref{37}) to be at $\theta
_{0}=(2\ell -1)\pi /8$ and $\theta _{0}=(2\ell -1)\pi /5$.

\begin{figure}[h!]
\centering  \includegraphics[width=7.0cm]{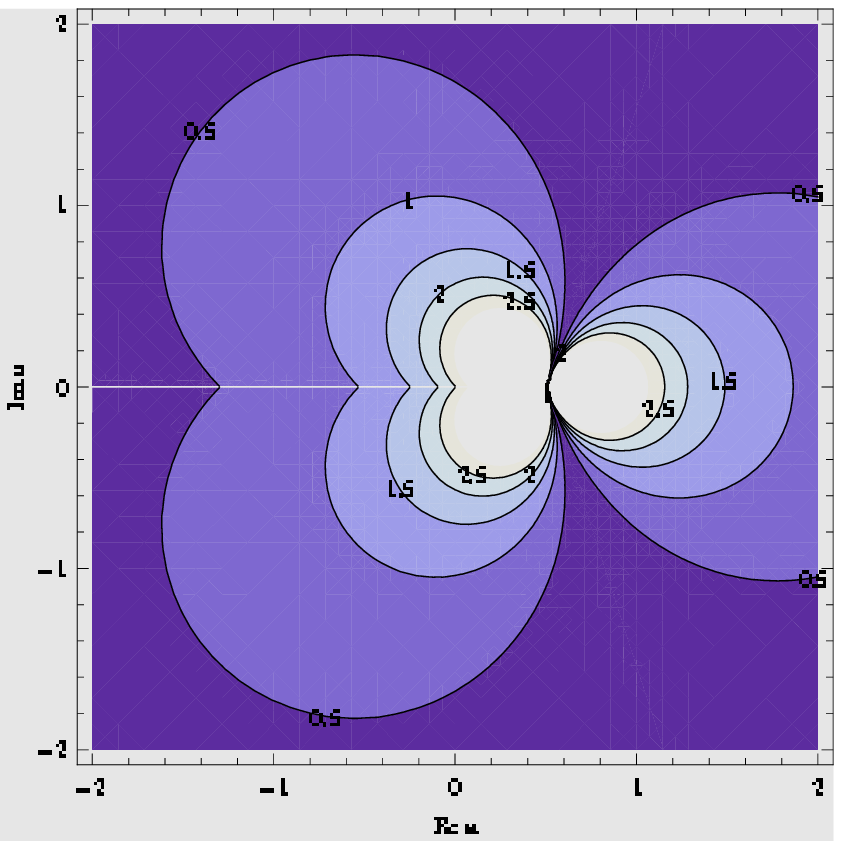} %
\includegraphics[width=7.0cm]{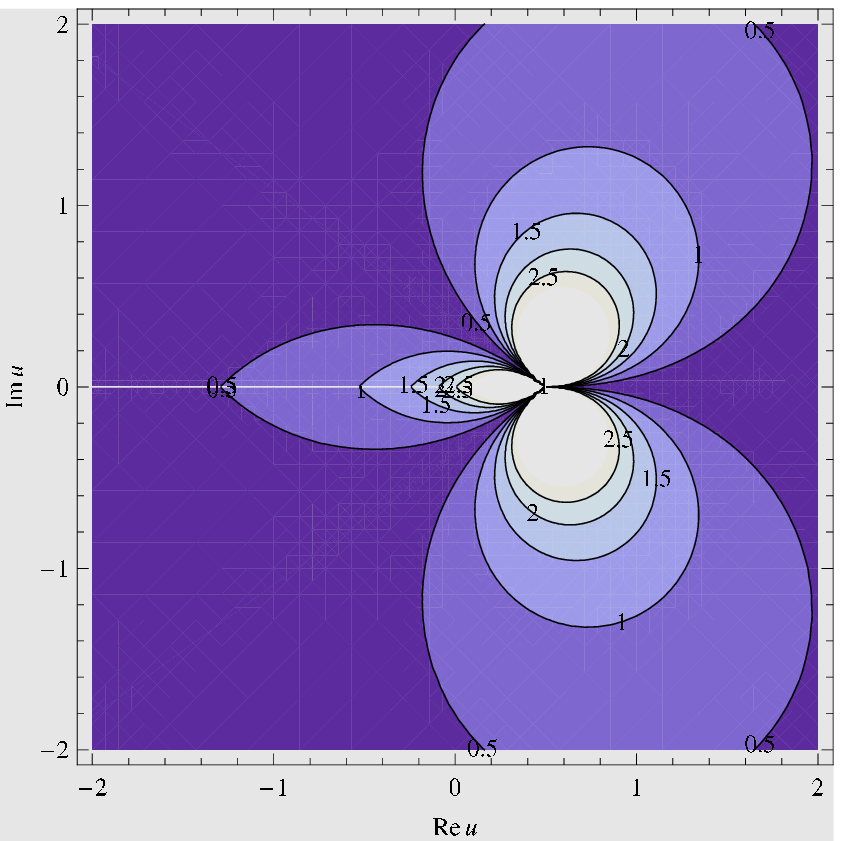}
\caption{$\mathcal{PT}$-symmetric rational solutions for $\mathcal{H}%
_{1/3}^{+}:$ Different Riemann sheets for $A=1/2$, $c=1$, $\protect\beta =2$
and $\protect\gamma =3$ (a) $\protect\zeta ^{(1)}$ and (b) $\protect\zeta %
^{(2)}$.}
\label{fig11}
\end{figure}

For positive rational values of $\varepsilon $ a complete trajectory extends
over several different Riemann sheets. Figure \ref{fig11}a and \ref{fig11}b
show the solutions $\zeta ^{(1)}$ and $\zeta ^{(2)}$, respectively. The
angles for the $\zeta _{0}=0$ solutions are predicted correctly by (\ref{37}%
) to be $\theta _{0}=4\ell \pi /5$ for $\ell =1,\ldots ,5$. A closed
trajectory is obtained when passing the branch cut at $-\infty $ to $1/2$
from the upper half plane in panel (a) to the lower half plane in panel (b).
In figure \ref{fig13}a a single trajectory is depicted for $\func{Im}\zeta
_{0}=1$.

For the broken $\mathcal{PT}$-symmetry we present the solution in figure \ref%
{fig12}. The branch cut at $-\infty -i/4$ to $(1-i)/4$ is passed from above
in panel (a) to below in panel (b). This is illustrated in figure \ref{fig13}%
b, where we depict just one single trajectory for $\func{Im}\zeta _{0}=1$.
The trajectories for the $\mathcal{PT}$-symmetric and broken $\mathcal{PT}$%
-symmetric case look qualitatively very similar, the major difference being
that the fixed point has moved away from the real axis, thus leading to a
loss of the symmetry.

\begin{figure}[h!]
\centering \includegraphics[width=7.0cm]{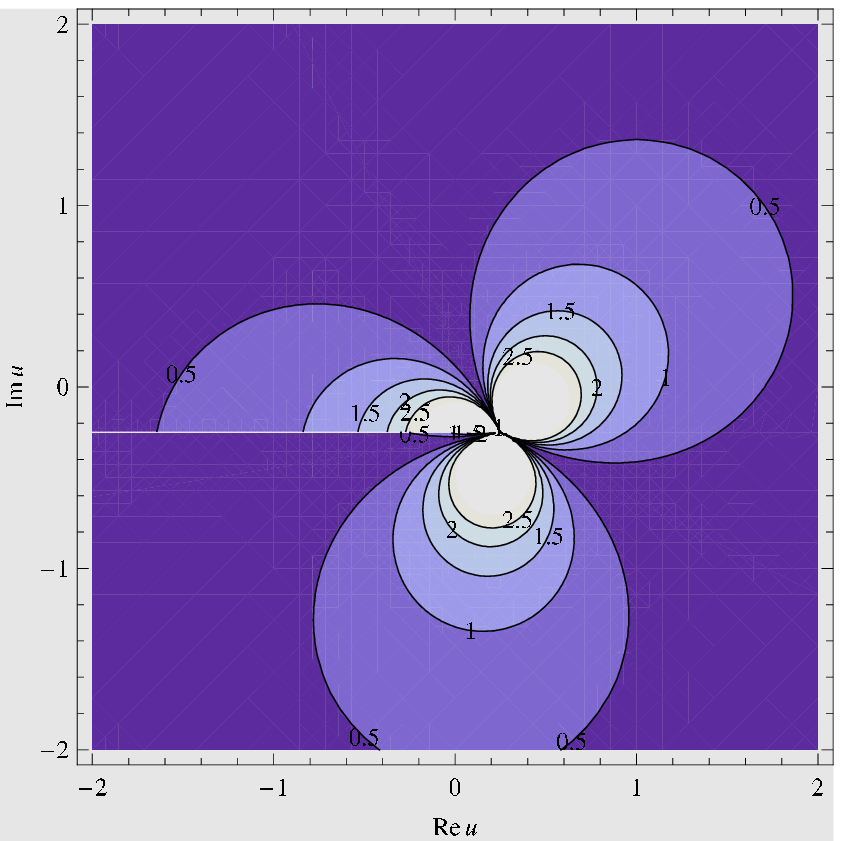} %
\includegraphics[width=7.0cm]{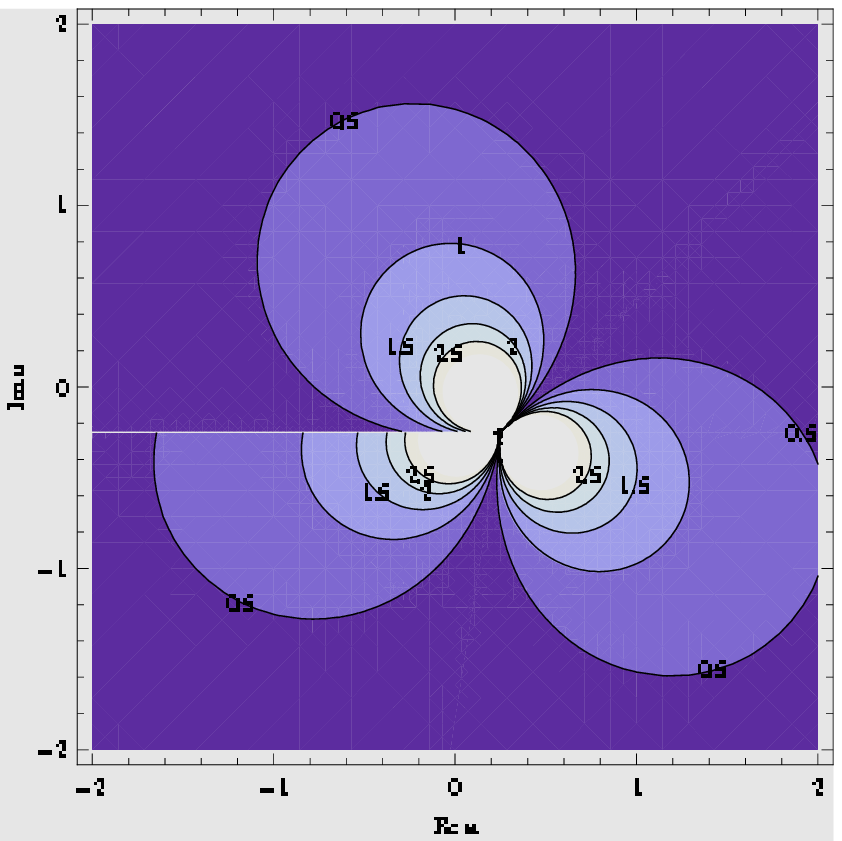}
\caption{Broken $\mathcal{PT}$-symmetric rational solutions for $\mathcal{H}%
_{1/3}^{+}:$ Different Riemann sheets for $A=(1-i)/4$, $c=1$, $\protect\beta %
=2+2i$ and $\protect\gamma =3$ (a) $\protect\zeta ^{(1)}$ and (b) $\protect%
\zeta ^{(2)}$.}
\label{fig12}
\end{figure}
\begin{figure}[h!]
\centering  \includegraphics[width=7.0cm]{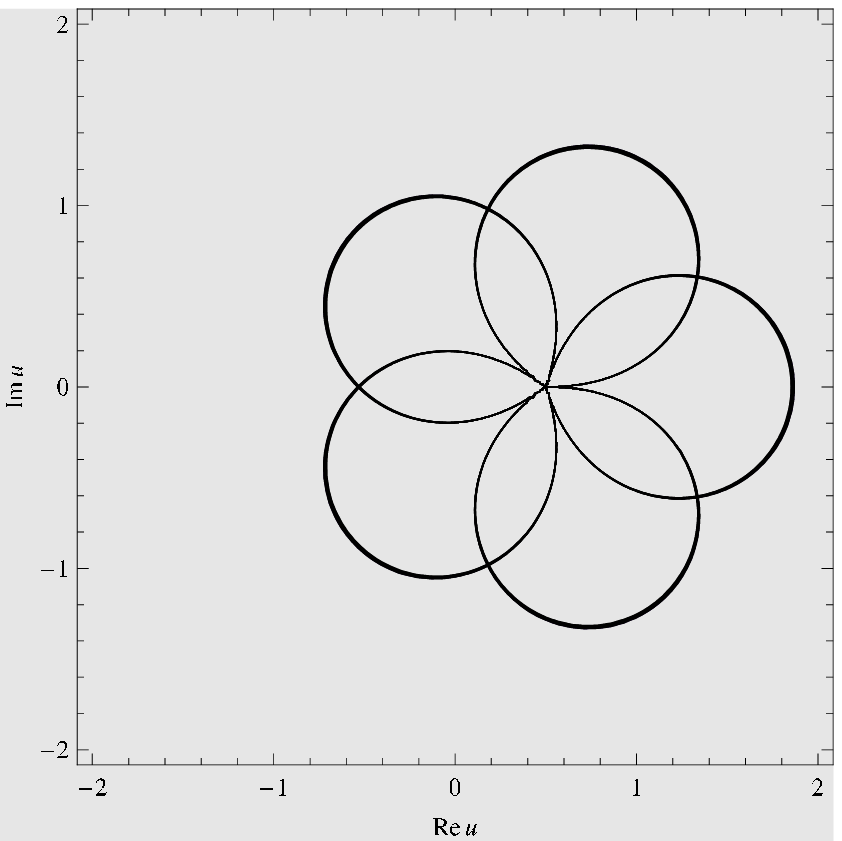} %
\includegraphics[width=7.0cm]{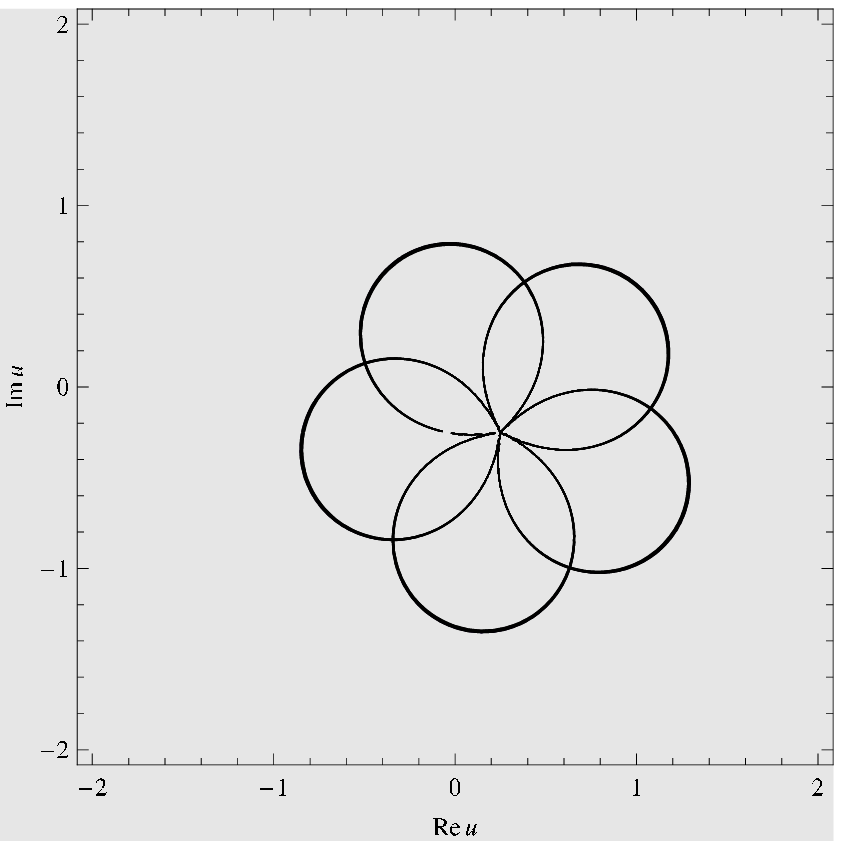}
\caption{Single trajectory for the rational solution of $\mathcal{H}%
_{1/3}^{+}$ with $\func{Im}\protect\zeta _{0}=1$ $:$ (a) $\mathcal{PT}$%
-symmetric solutions for the values as specified in figure \protect\ref%
{fig11}, (b) Broken $\mathcal{PT}$-symmetric solutions for the values as
specified in figure \protect\ref{fig12}.}
\label{fig13}
\end{figure}
\begin{figure}[h!]
\centering  \includegraphics[width=7.0cm]{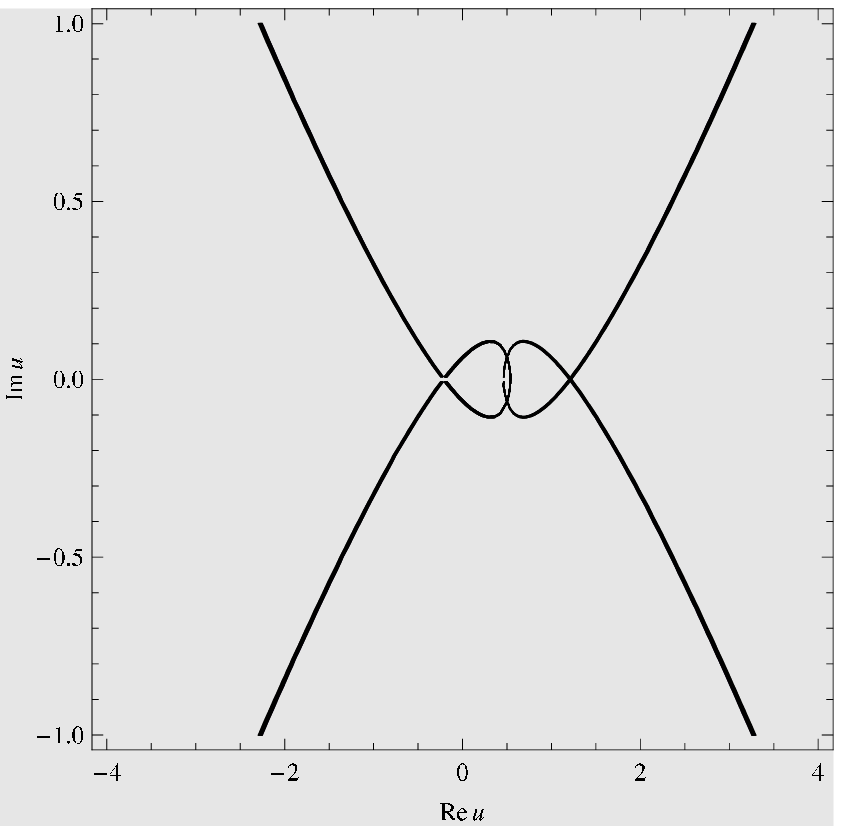} %
\includegraphics[width=7.0cm]{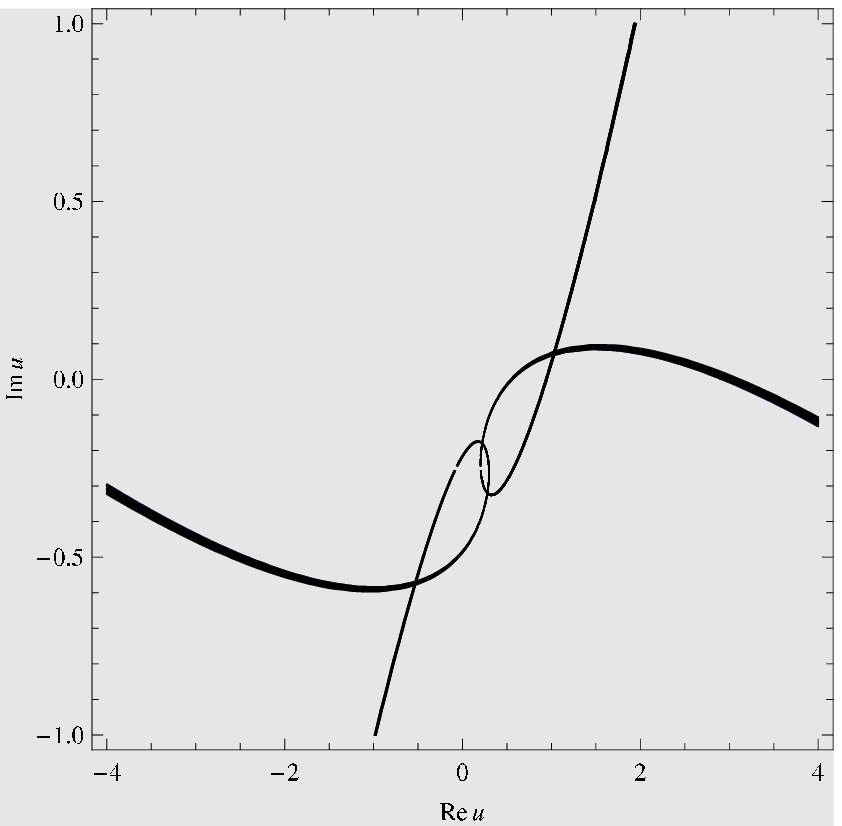}
\caption{Single trajectory for the rational solution for of $\mathcal{H}%
_{3}^{+}$ with $\func{Im}\protect\zeta _{0}=1$: (a) $\mathcal{PT}$-symmetric
solutions for $A=1/2$, $c=1$, $\protect\beta =2$ and $\protect\gamma =3$,
(b) Broken $\mathcal{PT}$-symmetric solutions for $A=1/4(1-i)$, $c=1$, $%
\protect\beta =2+2i$ and $\protect\gamma =3$.}
\label{fig14}
\end{figure}

\begin{figure}[h!]
\centering  \includegraphics[width=7.0cm]{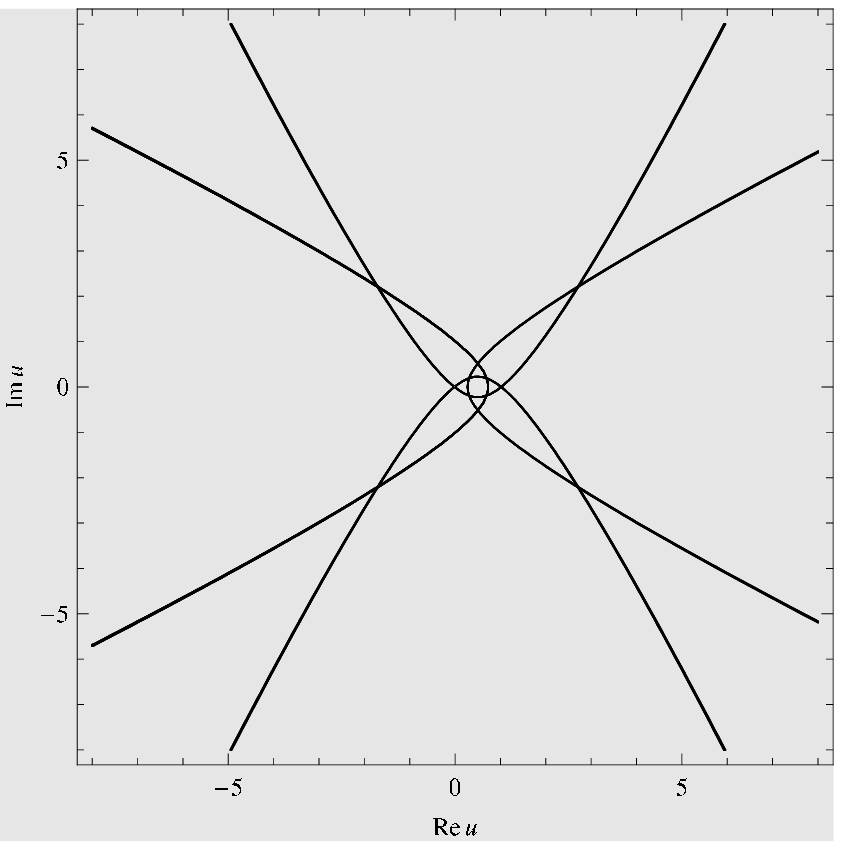} %
\includegraphics[width=7.0cm]{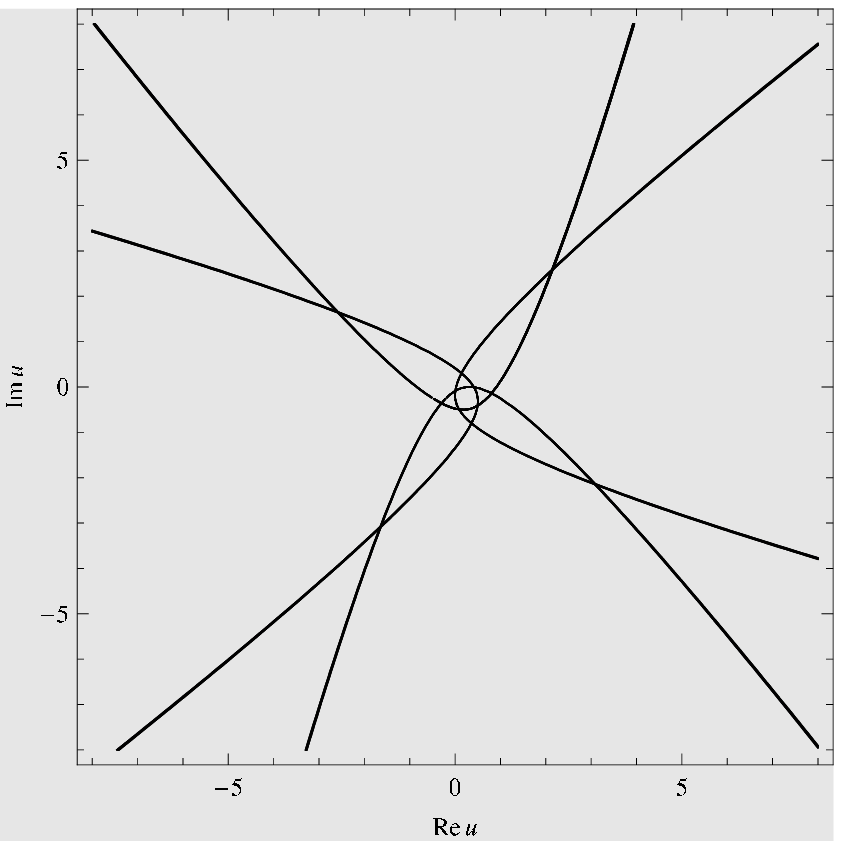}
\caption{Single trajectory for the rational solution for of $\mathcal{H}%
_{6}^{+}$ with $\func{Im}\protect\zeta _{0}=1$: (a) $\mathcal{PT}$-symmetric
solutions for $A=1/2$, $c=1$, $\protect\beta =2$ and $\protect\gamma =3$,
(b) Broken $\mathcal{PT}$-symmetric solutions for $A=1/4(1-i)$, $c=1$, $%
\protect\beta =2+2i$ and $\protect\gamma =3$.}
\label{fig15}
\end{figure}

Potentially there are of course many more possible values for $\varepsilon $
to be considered. We conclude here just by presenting two more examples with 
$\varepsilon $ being an integer, as these are the most common deformations
usually considered. For these values we obtain yet another type of
characteristics as more and more branches have to be taken into account for
increasing $\varepsilon $. In figure \ref{fig14} and \ref{fig15} we depict
all branches for the trajectory with $\func{Im}\zeta _{0}=1$ for $%
\varepsilon =3$ and $\varepsilon =6$, respectively.

We observe some intricate winding behaviour near the fixed point, which,
however, is not asymptotic in these cases. As is apparent from the solution (%
\ref{ratsol}) the trajectories diverge to infinity into various directions
depending on the chosen Riemann sheet. Evidently the number of these
asymptotes grows with increasing $\varepsilon $. Breaking the $\mathcal{PT}$%
-symmetry will only distort the trajectories, giving rise to new directions
of the asymptotes, but not changing their numbers.

\subsubsection{Trigonometric/hyperbolic solutions}

As in the non-deformed case we assume next the factorization $%
P(u)=(u-A)^{2}(u-B)$, which will impose once more the constraints (\ref{ak}%
). Also in this case we can compute (\ref{jj}) numerically.

Again we supplement our numerical findings by some analytical results. We
can predict once more the lines for which $\zeta _{0}=0$ explicitly
following the arguments of the previous subsection, with the difference that
we have to take the limit to either the point $A$ or $B$. For the angle of
lines near the point $A$ we find%
\begin{equation}
\theta _{A}^{(n,m,\pm )}=\frac{\pi }{2(\varepsilon -1)}\left[ 1-\varepsilon
+4(n-m)-4m\varepsilon \right] +\frac{1}{\varepsilon -1}\arg (A-B)+\frac{1}{%
\varepsilon -2}\arg (\pm \lambda _{\varepsilon })
\end{equation}%
whereas for the angle of lines near the point $B$ we compute%
\begin{equation}
\theta _{B}^{(n,m,\pm )}=\frac{\pi }{2}\left( \varepsilon -1-4n\right)
+(1+\varepsilon )2\pi m+(\varepsilon -1)\arg (B-A)-\arg (\pm \lambda
_{\varepsilon }).
\end{equation}

\begin{figure}[h!]
\centering  \includegraphics[width=7.0cm]{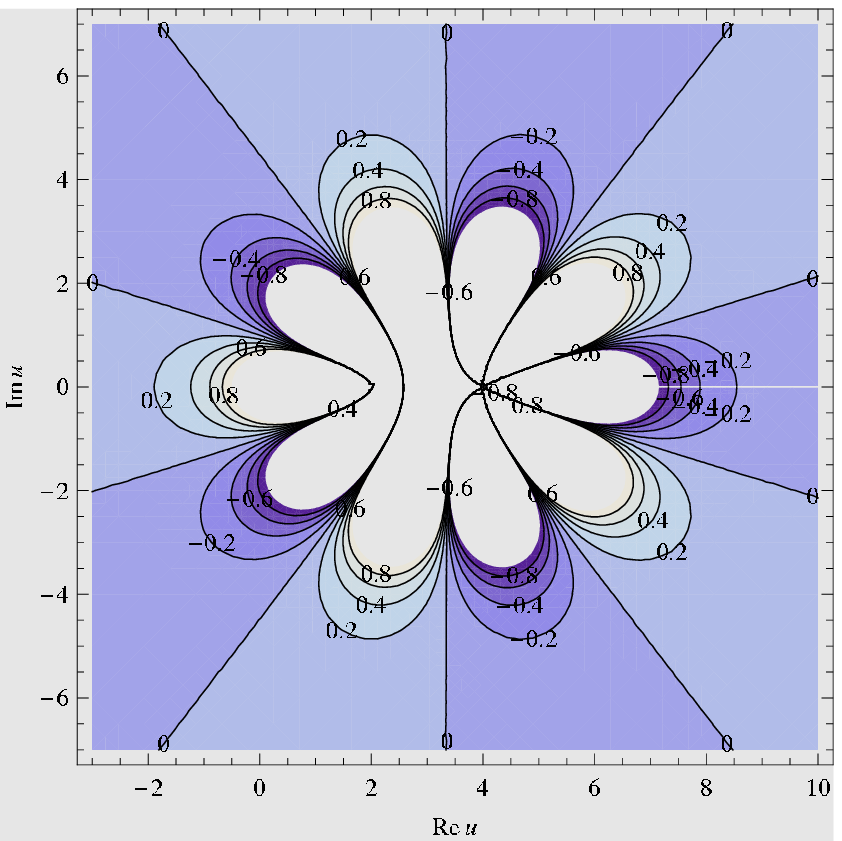} %
\includegraphics[width=7.0cm]{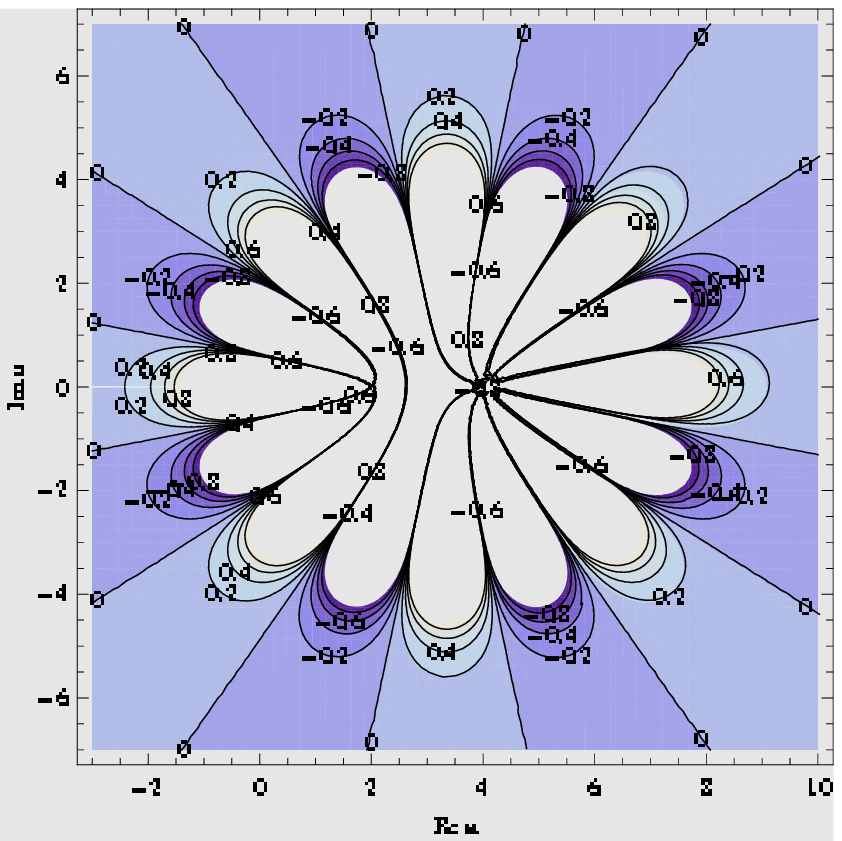}
\caption{Complex $\mathcal{PT}$-symmetric trigonometric/hyperbolic solutions
of the deformed KdV equation with $A=4,B=2,c=1$, $\protect\beta =2$ and $%
\protect\gamma =3$ for (a) $\mathcal{H}_{-1/2}^{+}$ and (b) $\mathcal{H}%
_{-2/3}^{+}$.}
\label{fig16}
\end{figure}

\begin{figure}[h!]
\centering  \includegraphics[width=7.0cm]{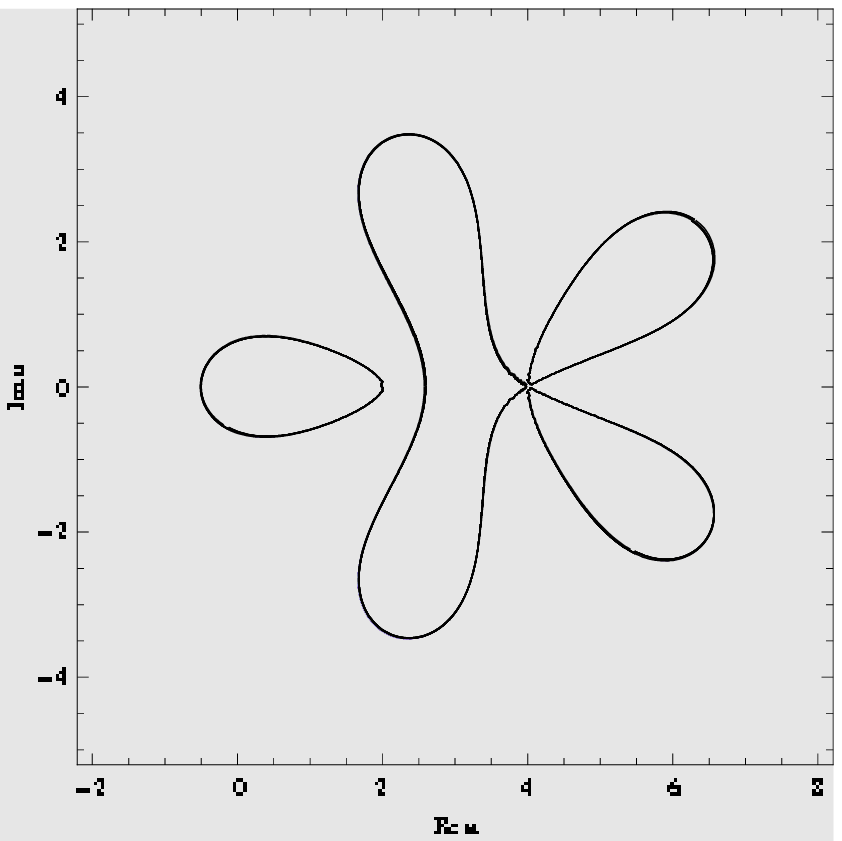} %
\includegraphics[width=7.0cm]{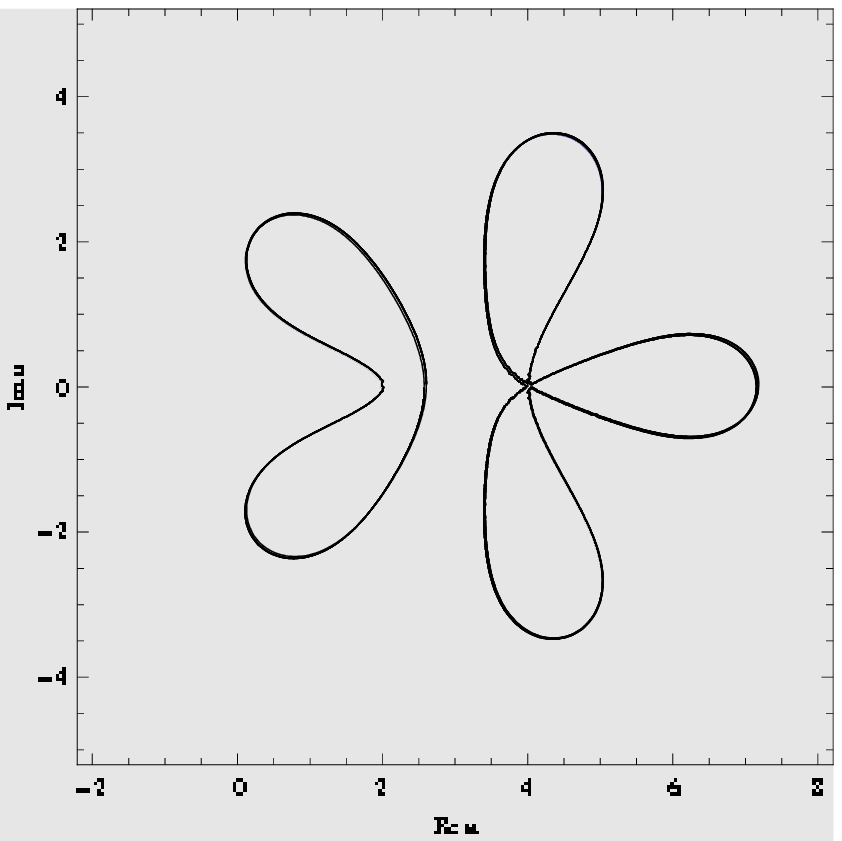}
\caption{Single trajectory for the complex $\mathcal{PT}$-symmetric
trigonometric/hyperbolic solutions of the deformed KdV equation for $%
\mathcal{H}_{-1/2}^{+}$ with the same values as specified in figure  \protect
\ref{fig16} for (a) $\func{Im}\protect\zeta _{0}=1$ and (b) $\func{Im}%
\protect\zeta _{0}=-1$.}
\label{fig17}
\end{figure}

\begin{figure}[h!]
\centering  \includegraphics[width=7.0cm]{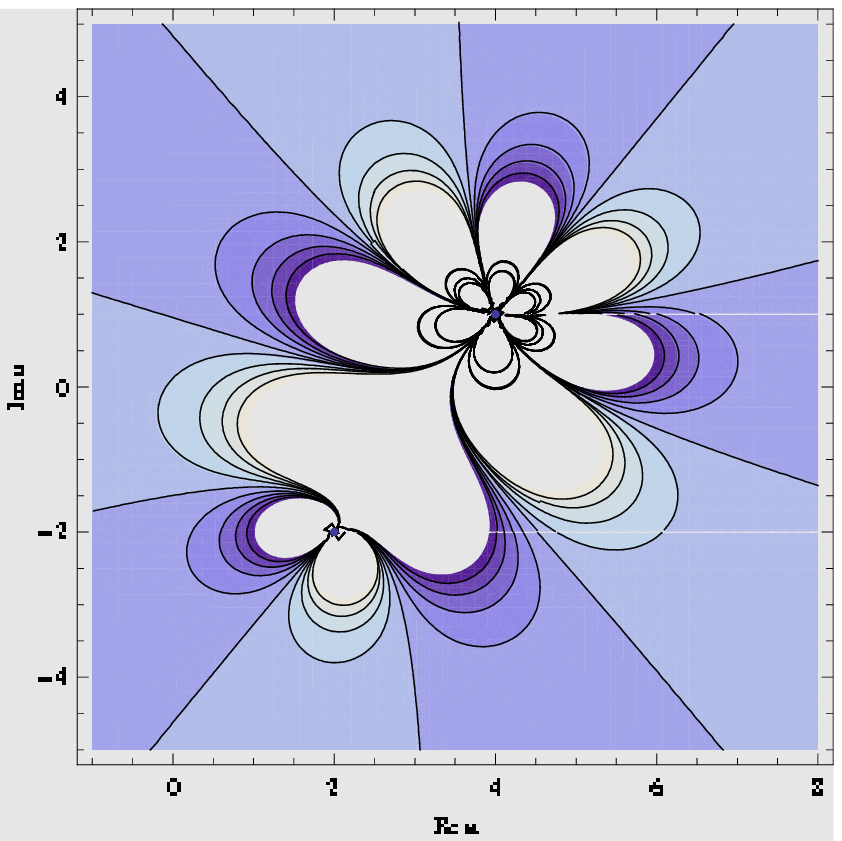} %
\includegraphics[width=7.0cm]{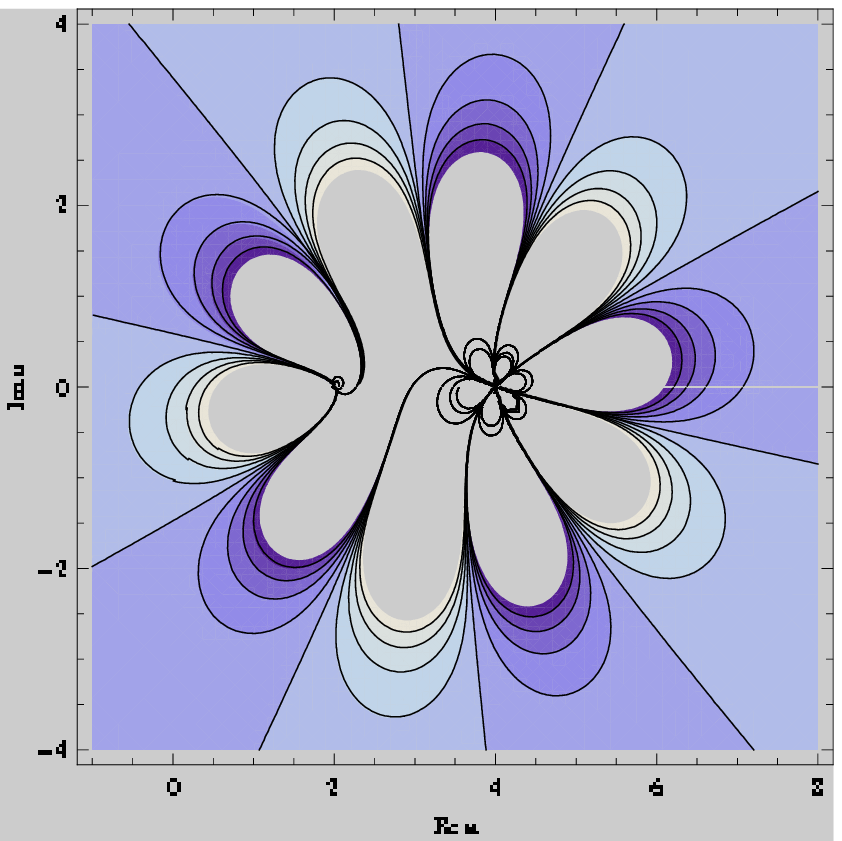}
\caption{Broken $\mathcal{PT}$-symmetric trigonometric/hyperbolic solutions
of the deformed KdV equation $\mathcal{H}_{-1/2}^{+}$: (a) Spontaneously
broken $\mathcal{PT}$-symmetry with $A=4+i$, $B=2-2i$, $c=1$, $\protect\beta %
=3/10$ and $\protect\gamma =3 $; (b) broken $\mathcal{PT}$-symmetry with $%
A=4 $, $B=2$, $c=1$, $\protect\beta =3/10$ and $\protect\gamma =3+i $.}
\label{fig20xx}
\end{figure}

\begin{figure}[h!]
\centering  \includegraphics[width=7.0cm]{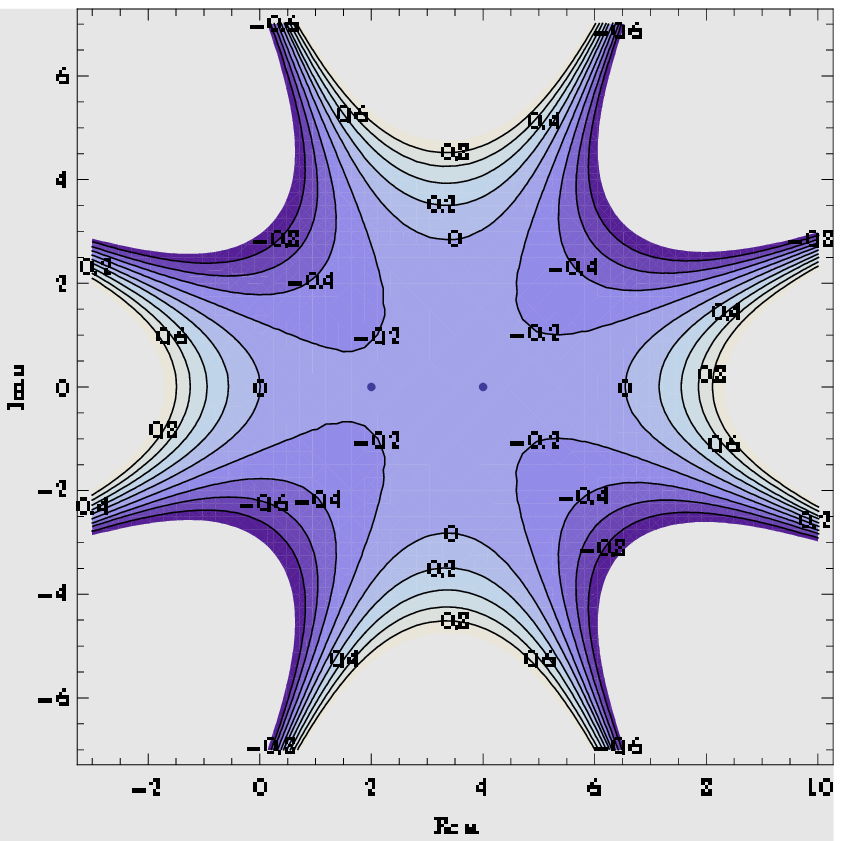} %
\includegraphics[width=7.0cm]{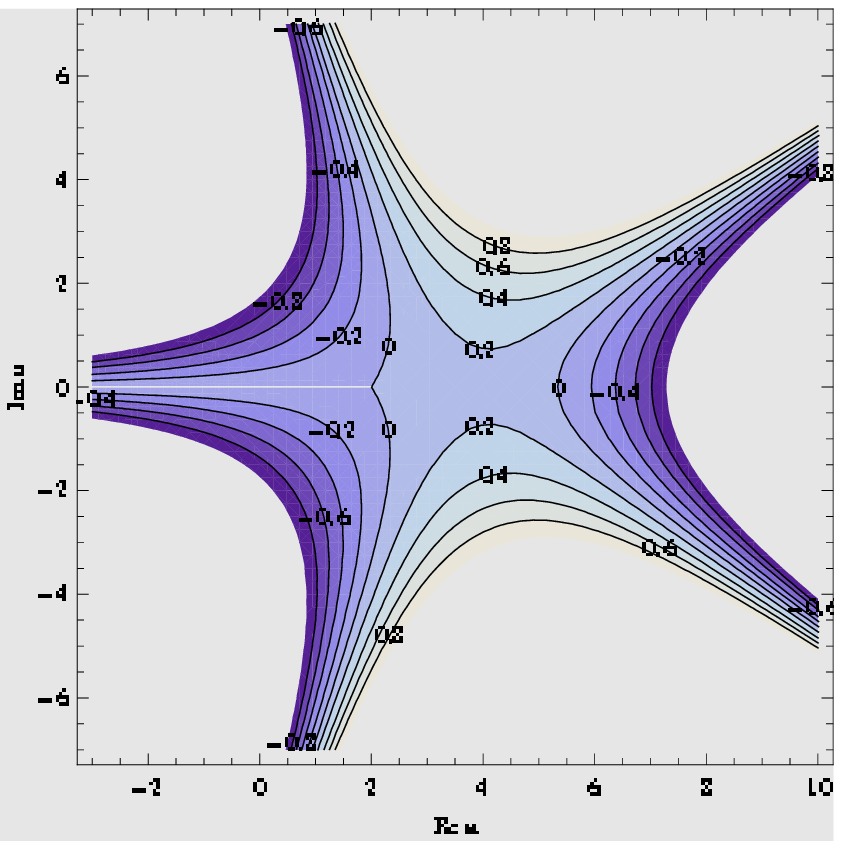}
\caption{Complex $\mathcal{PT}$-symmetric trigonometric/hyperbolic solutions
of the deformed KdV equation with $A=4$, $c=1$, $\protect\beta =2$ and $%
\protect\gamma =3$ for (a) $\mathcal{H}_{-2}^{+}$ and (b) $\mathcal{H}%
_{-3}^{+}$.}
\label{fig18}
\end{figure}

\begin{figure}[h!]
\centering  \includegraphics[width=7.0cm]{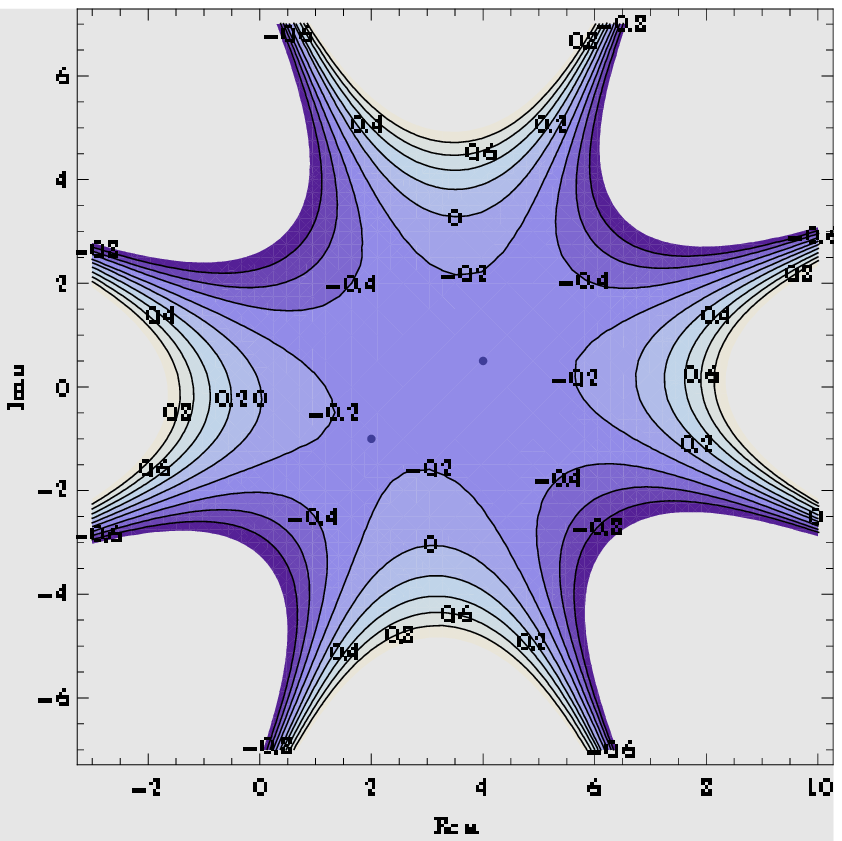} %
\includegraphics[width=7.0cm]{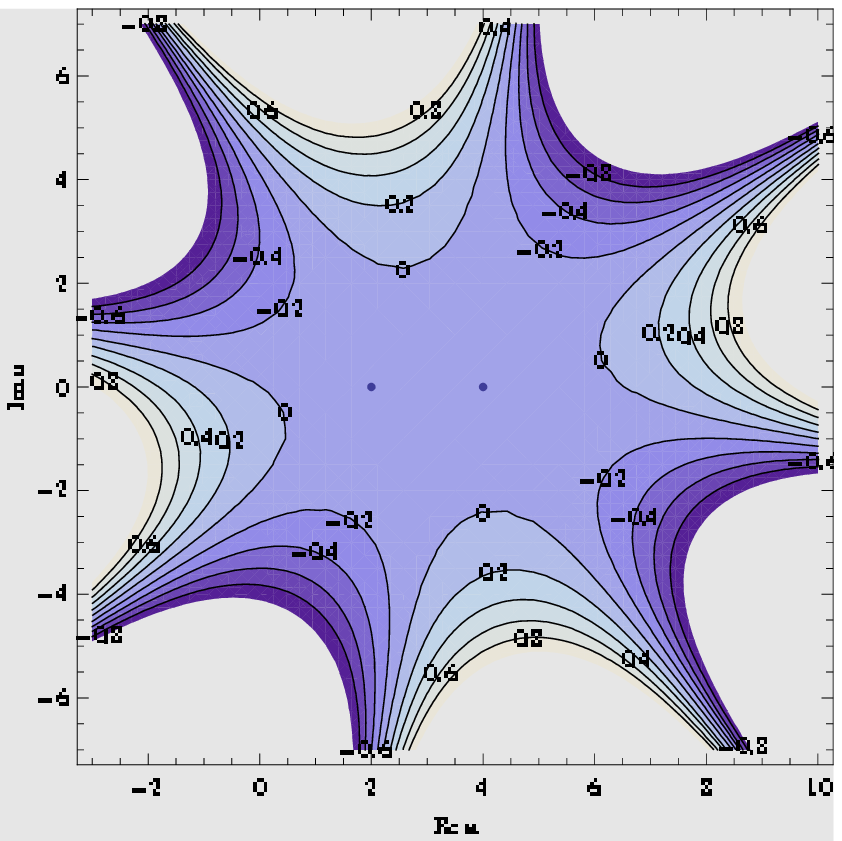}
\caption{Broken $\mathcal{PT}$-symmetric trigonometric/hyperbolic solutions
of the deformed KdV equation for $\mathcal{H}_{-2}^{+}$: (a) Spontaneously
broken $\mathcal{PT}$-symmetry with $A=4+i/2,B=2-i,c=1$, $\protect\beta %
=3/10 $ and $\protect\gamma =3$; (b) broken $\mathcal{PT}$-symmetry with $%
A=4,B=2,c=1$, $\protect\beta =3/10$ and $\protect\gamma =3+4i$.}
\label{fig19}
\end{figure}

We depict some specific examples with different characteristics in the
figures \ref{fig16} to \ref{fig19}.

In figure \ref{fig16} we recognize a similar characteristic behaviour as for
the rational solution of the same model depicted in figure \ref{fig9}.
Roughly the portrait \ref{fig16} corresponds to \ref{fig9} with the
difference that the single flower centre in the rational case situated as $A$
has been pulled apart to the two points $A$ and $B$ in the trigonometric
one. The overall effect is that some of trajectories separated in the
rational case in different wedges are joint in the trigonometric/hyperbolic
case. To illustrate this we have extracted one single trajectory in figure %
\ref{fig17}. We observe a similar behaviour for the entire sequel
parameterized by $\varepsilon =-n/(n+1)$ with $n=1,2,\ldots $ Notice also
that unlike as in the case $\varepsilon =1$ there is no distinction between
a periodic and an asymptotically constant solution.

Having enough free parameters available we can now break the symmetry for
this solution also spontaneously or completely as illustrated in figure \ref%
{fig20xx}. In both cases we observe that the amount of wedges remains
unchanged, the symmetry about the real axis is lost and the winding around
the fixed points becomes more intricate. In addition, trajectories from
certain wedge regions connect in different ways, e.g. in figure \ref{fig16}
we observe a connected trajectory in the two light shaded wedge region to
the left of the vertical as can also be seen in figure \ref{fig17}. Such
type of trajectories do not exist in the broken case as we can observe in
figure \ref{fig20xx}. We note that instead trajectories from the light
shaded region in the upper half plane connect to a dark shaded region in the
lower half plane to the left of the corresponding light shaded region in the
unbroken case.

Comparing next the solutions for $\varepsilon \in \mathbb{R}^{-}$ we observe
in figure \ref{fig18}a a similar transformation from the rational to the
trigonometric/hyperbolic case as for the previous example. Again we find
that the rational solution resembles this solution for the same values of
the deformation parameters, with the difference that the characteristic
behaviour around the point $A$ in the rational case has been distorted to
the points $A$ and $B$.

The broken $\mathcal{PT}$-symmetric case for this model is presented in
figure \ref{fig19}. In panel (a) we depict the spontaneously broken $%
\mathcal{PT}$-symmetric scenario for $\mathcal{H}_{-2}^{+}$, observing that
the symmetry is lost since the trajectories rotate together with the two
fixed points. The $\mathcal{PT}$-symmetry is completely broken in the
presentation in \ref{fig19}b for the same model. Having kept the values for
the fixed points, we note that in this case the orbits rotate around them.

We have investigated also other models observing similar patterns, which we
will however not present here.

\subsubsection{Elliptic solutions}

\begin{figure}[h!]
\centering  \includegraphics[width=4.9cm]{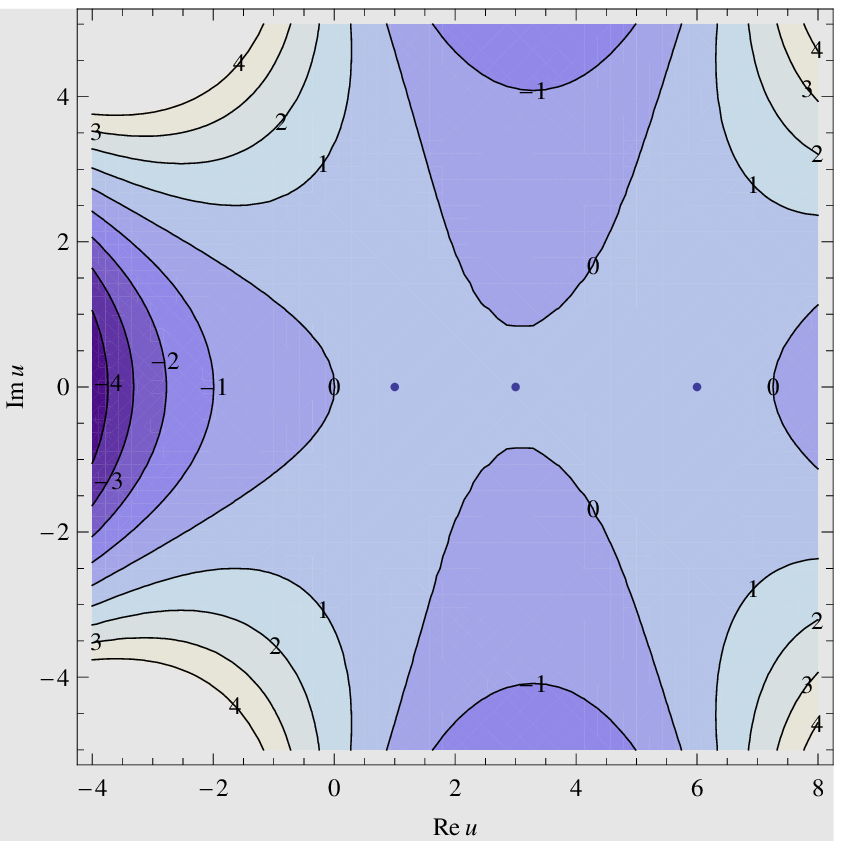} %
\includegraphics[width=4.9cm]{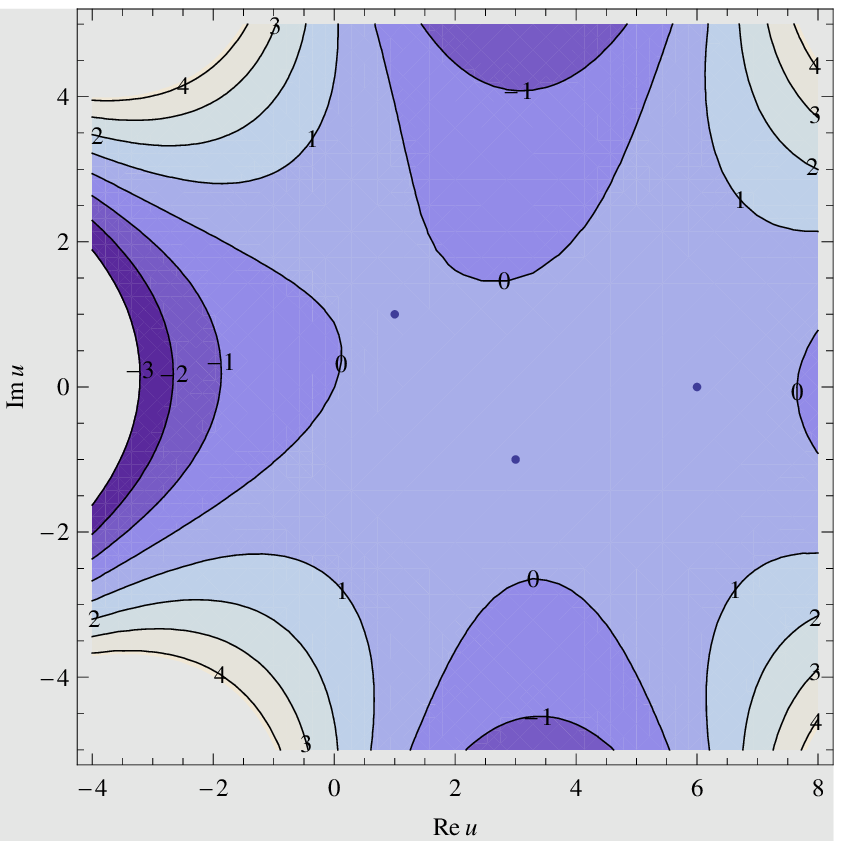} %
\includegraphics[width=4.9cm]{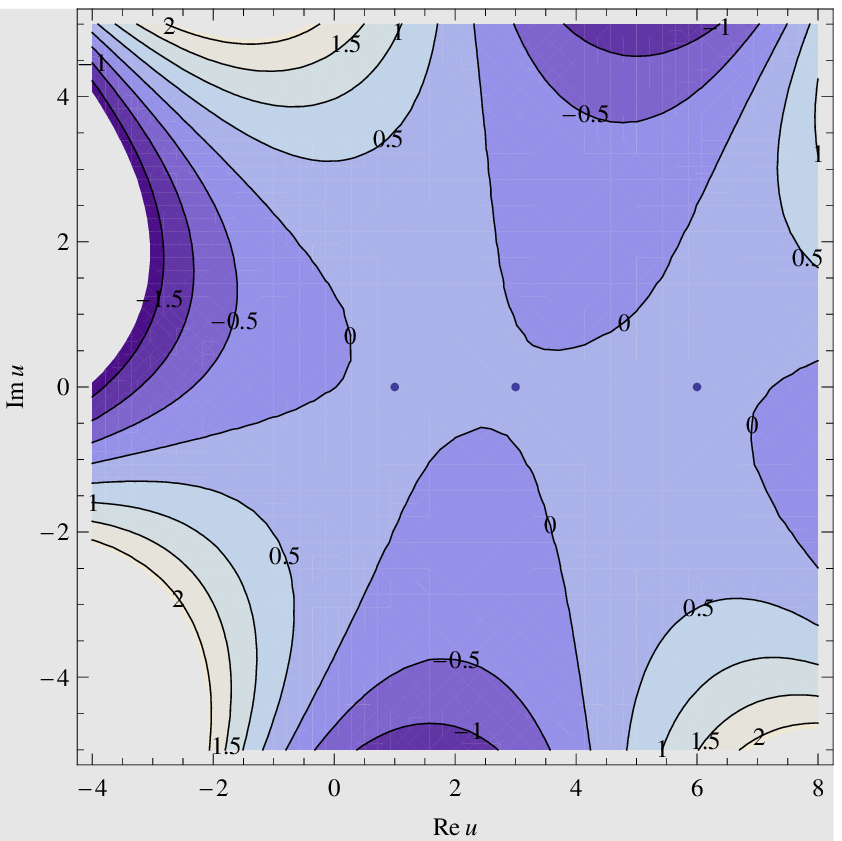}
\caption{Elliptic solutions for $\mathcal{H}_{-2}^{+}$: (a) $\mathcal{PT}$%
-symmetric with $A=1$, $B=3$, $C=6$, $\protect\beta =3/10$, $\protect\gamma %
=-3$ and $c=1$; (b) spontaneously broken $\mathcal{PT}$-symmetry with $A=1+i$%
, $B=3-i$, $C=6$, $\protect\beta =3/10$, $\protect\gamma =-3$ and $c=1$; (c)
broken $\mathcal{PT}$-symmetry with $A=1$, $B=3$, $C=6$, $\protect\beta %
=3/10 $, $\protect\gamma =-3+4i$ and $c=1$.}
\label{figell1}
\end{figure}

\begin{figure}[h!]
\centering  \includegraphics[width=6.0cm]{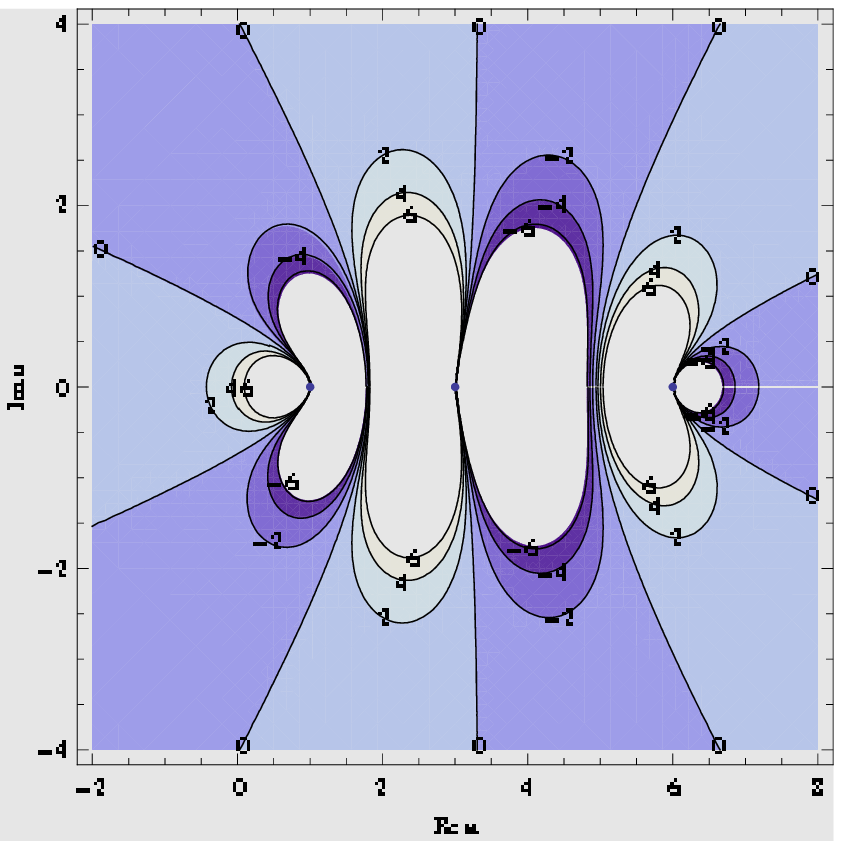} %
\includegraphics[width=6.0cm]{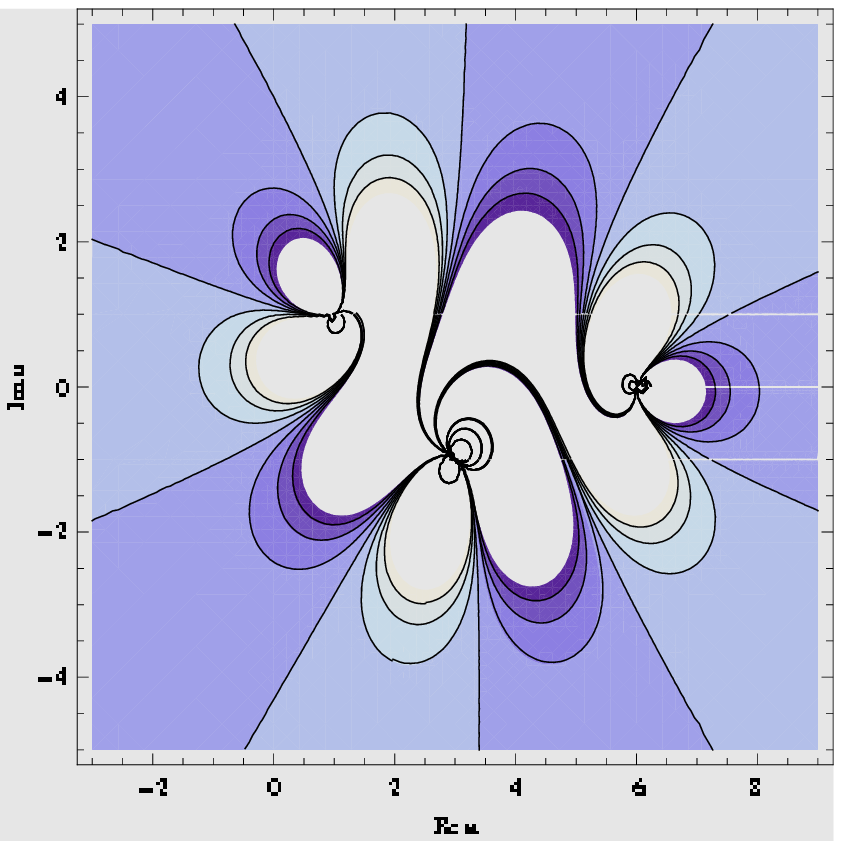}
\caption{Elliptic solutions for $\mathcal{H}_{-1/2}^{+}$: (a) $\mathcal{PT}$%
-symmetric with $A=1$, $B=3$, $C=6$, $\protect\beta =3/10$, $\protect\gamma %
=-3$ and $c=1$; (b) spontaneously broken $\mathcal{PT}$-symmetry with $A=1+i$%
, $B=3-i$, $C=6$, $\protect\beta =3/10$, $\protect\gamma =-3$ and $c=1$.}
\label{figell2}
\end{figure}

Proceeding just as in the undeformed case we specify next $%
P(u)=(u-A)(u-B)(u-C)$, which imposes the constraints (\ref{ell1}) and (\ref%
{ell2}), thus leaving two constants free. We present here only the results
for the cases $\varepsilon =-2$ and $\varepsilon =-1/2$ in the figures \ref%
{figell1} and \ref{figell2}, respectively.

In all three possible scenarios we observe a similar qualitative behaviour
as for the trigonometric/hyperbolic solutions of the previous subsection
with the difference that we have three instead of two fixed points.

\subsection{The $\mathcal{H}_{\protect\varepsilon }^{-}$-models}

Let us now turn to the second type of deformation. In order to construct the
solutions we proceed in a similar manner as in the previous case.
Integrating now twice the second equation in equation (\ref{dIto2}) we
obtain 
\begin{equation}
u_{\zeta }^{2}=\frac{2}{\gamma }\left( \kappa _{2}+\kappa _{1}u+\frac{c}{2}%
u^{2}-\beta \frac{i^{\varepsilon }}{(1+\varepsilon )(2+\varepsilon )}%
u^{2+\varepsilon }\right) =:\lambda Q(u),  \label{po}
\end{equation}%
where 
\begin{equation}
\lambda =-\frac{2\beta i^{\varepsilon }}{\gamma (1+\varepsilon
)(2+\varepsilon )}.
\end{equation}%
A crucial difference between $\mathcal{H}_{\varepsilon }^{+}$ and $\mathcal{H%
}_{\varepsilon }^{-}$ is that unlike as the polynomial $P(u)$, which was of
fixed order $3$, the order of $Q(u)$ depends on $\varepsilon $, meaning that
we have more and more possibilities to factorize it for growing $\varepsilon 
$. For instance, for a given integer value $n\in \mathbb{N}_{0}$ the
factorization of $Q(u)$ as%
\begin{equation}
Q(u)=(u-A_{1})^{\varepsilon +2-n}\prod\nolimits_{i=1}^{n}(u-A_{i+1}),
\label{Q}
\end{equation}%
admits solutions provided $n-2\leq \varepsilon \leq n+1$ and $\varepsilon
\in \mathbb{N}$. This allows of course for yet another infinity of
possibilities.

When $\kappa _{1}=\kappa _{2}=0$ we can find a closed solution valid for all 
$\varepsilon $ by integrating (\ref{po}) and solving it for $u$ 
\begin{equation}
u\left( \zeta \right) =\left( \frac{c(\varepsilon +1)(\varepsilon +2)}{%
i^{\varepsilon }\beta \left[ \cosh \left( \frac{\sqrt{c}\varepsilon (\zeta
-\zeta _{0})}{\sqrt{\gamma }}\right) +1\right] }\right) ^{1/\varepsilon }.
\label{closed}
\end{equation}%
The generic scenario does not yield such a simple answer.

\subsubsection{$\mathcal{H}_{2}^{-}$}

This case is especially interesting as it corresponds to a complex version
of the modified KdV-equation. Specifying (\ref{Q}) for instance as $%
Q(u)=(u-A)^{3}(u-B)$ we can factorize the polynomial in (\ref{po}) with the
choice%
\begin{equation}
\kappa _{1}=-\frac{2c^{3/2}}{3\sqrt{-\beta }},\quad \kappa _{2}=-\frac{c^{2}%
}{4\beta },\quad A=-\frac{B}{3}\quad \text{and\quad }B=-\frac{3\sqrt{c}}{%
\sqrt{-\beta }}.
\end{equation}%
Notice that this fixes all the boundary conditions for a given model.
Solving (\ref{po}) then yields a rational solution for the second equation
in (\ref{dIto2})%
\begin{equation}
u\left( \zeta \right) =\sqrt{-\frac{c}{\beta }}\frac{2c\zeta ^{2}-9\gamma }{%
3\gamma +2c\zeta ^{2}}.  \label{umod}
\end{equation}%
As is well know for the real case one may construct solutions for the
KdV-equation from those of the modified KdV-equation by means of a Miura
transformation. We expect this also to hold for their complex versions.
Indeed, using the transformation of the form%
\begin{equation}
u_{\text{KdV}}\left( \zeta \right) =\sqrt{\frac{6\gamma }{\beta }}u_{\zeta
}-u^{2}  \label{Miura}
\end{equation}%
yields the rational solution of the KdV-equations (\ref{28}) from (\ref{umod}%
) when we identify $\zeta _{0}=i\sqrt{3\gamma /2c}$ therein.

Assuming in (\ref{Q}) instead $Q(u)=(u-A)^{2}(u-B)(u-C)$ we can factorize
the polynomial in (\ref{po}) with the constraints%
\begin{eqnarray}
\kappa _{1} &=&\frac{\beta C^{2}\left( \vartheta -5C\beta \right) +9c\left(
\vartheta -3C\beta \right) }{81\beta },\quad \kappa _{2}=\frac{C\left(
2\vartheta -C\beta \right) \left( C\beta +\vartheta \right) ^{2}}{324\beta
^{2}},\quad \\
A &=&-\frac{1}{2}(B+C)\quad \text{and\quad }B=\frac{2\vartheta -C\beta }{%
3\beta }
\end{eqnarray}%
where we abbreviated $\vartheta :=\sqrt{2}\sqrt{-\beta \left( \beta
C^{2}+9c\right) }$. Notice that in this case one constant remains free. The
integration of (\ref{po}) yields in this case a trigonometric solution,
which, using the same Miura transformation (\ref{Miura}), may also be
converted into a solution of the KdV-equations.

\subsubsection{$\mathcal{H}_{4}^{-}$}

\begin{figure}[h!]
\centering  \includegraphics[width=7.0cm]{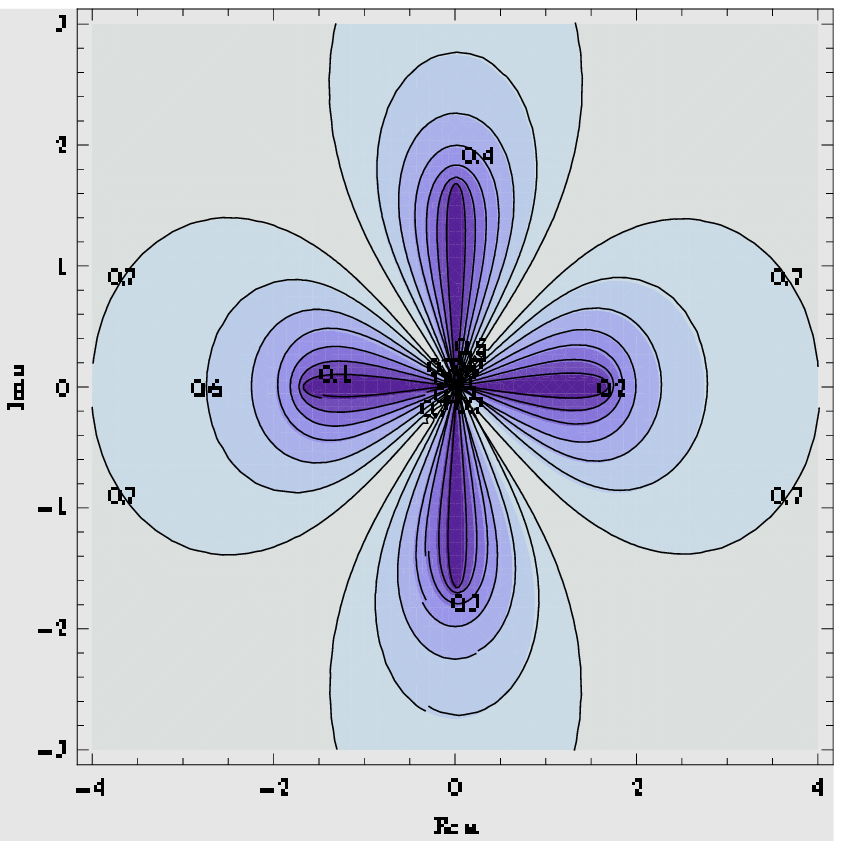} %
\includegraphics[width=7.0cm]{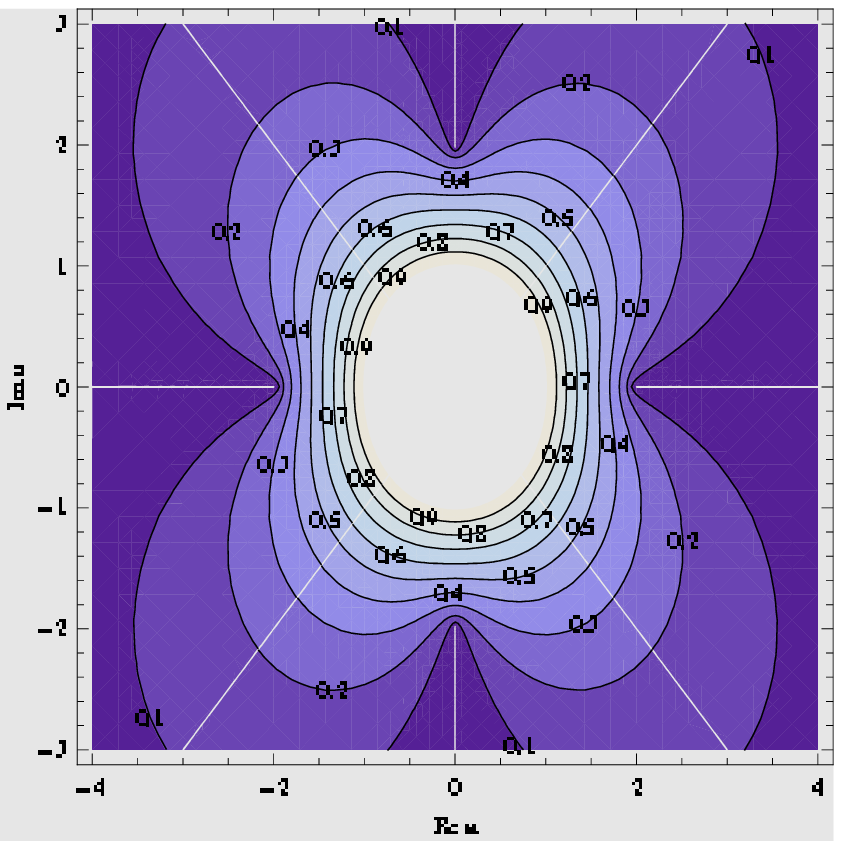}
\caption{$\mathcal{PT}$-symmetric solutions for $\mathcal{H}_{4}^{-}:$ (a)
Star node at the origin for $c=1$, $\protect\beta =2$, $\protect\gamma =1$
and $B=(15/2)^{1/4}$; (b) centre at the origin for $c=1$, $\protect\beta =1$%
, $\protect\gamma =-1$ and $B=(15/2)^{1/4}$.}
\label{fig22}
\end{figure}

In this case the polynomial of the right hand side of (\ref{po}) is of sixth
order. We present here just one very symmetric solution by assuming a
factorization of the form $Q(u)=u^{2}(u^{2}-B^{2})(u^{2}-C^{2})$, which is
possible with the simple choice%
\begin{equation}
\kappa _{1}=\kappa _{2}=0,\qquad B=iC\qquad \text{and\qquad }C^{4}=\frac{15c%
}{\beta }.
\end{equation}%
Thus we have made contact with the solution (\ref{closed}). Parameterizing $%
B=r_{B}e^{i\theta _{B}}$ and $\lambda =r_{\lambda }e^{i\theta _{\lambda }}$
the eigenvalues of the Jacobian when linearized about $u=0$ are easily
computed to%
\begin{equation}
j_{1}=\pm i\sqrt{r_{\lambda }}r_{B}^{2}\exp \left[ \frac{i}{2}(4\theta
_{B}+\theta _{\lambda })\right] \quad \text{and\quad }j_{2}=\mp i\sqrt{%
r_{\lambda }}r_{B}^{2}\exp \left[ -\frac{i}{2}(4\theta _{B}+\theta _{\lambda
})\right] .  \label{bj}
\end{equation}%
This means for the $\mathcal{PT}$-symmetric solution we always obtain two
real degenerate eigenvalues and therefore a star node at $u=0$. For the
positive square root in (\ref{po}) with $B=\pm (15c/\beta )^{1/4}$ or with $%
B=\pm i(15c/\beta )^{1/4}$ the node is stable or unstable, respectively. The
stability property is reversed for the negative square root. Taking the
branch cuts at $\func{Im}u=0$, $\func{Re}u=0$, $\func{Im}u=\func{Re}u$ and $%
\func{Im}u=-\func{Re}u$ into account we obtain closed curves. The four zeros
of $Q(u)$, i.e. $\pm (15c/\beta )^{1/4}$, $\pm i(15c/\beta )^{1/4}\,$, are
surrounded by the trajectories. All these features can be seen in figure \ref%
{fig22}a. We may also tune the coupling constants in such a way that $%
4\theta _{B}+\theta _{\lambda }=0$, which by (\ref{bj}) implies that the
fixed point at the origin should become a centre. Indeed, for a specific
choice we observe this in figure \ref{fig22}b. Apart from the origin, the
remaining four fixed points are now situated outside of the trajectories.

\begin{figure}[h!]
\centering  \includegraphics[width=7.0cm]{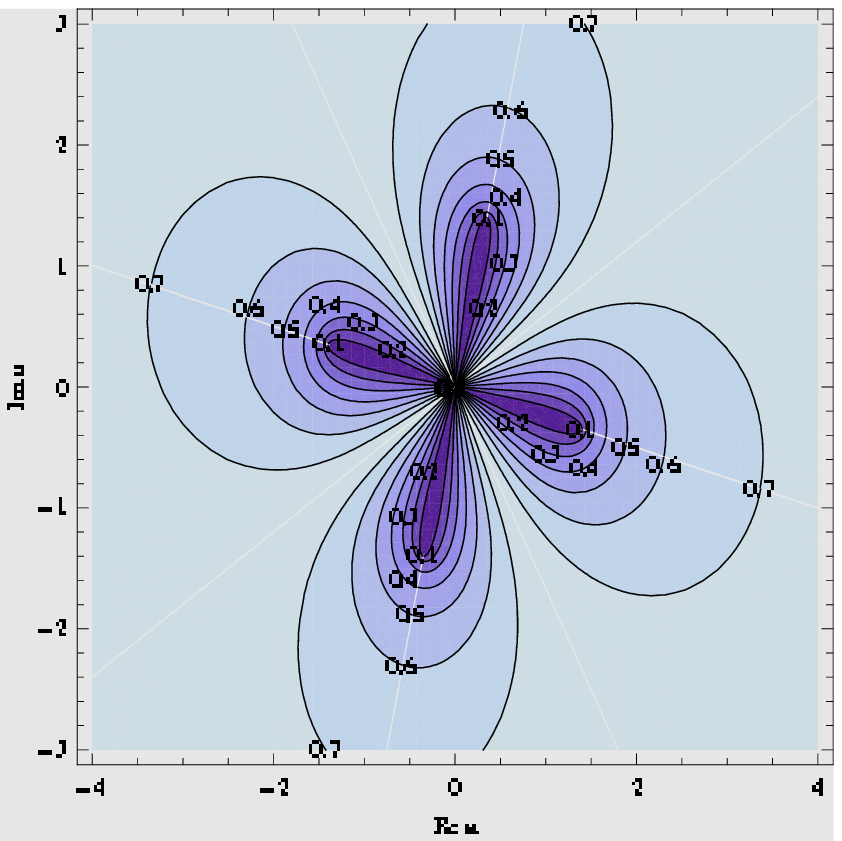} %
\includegraphics[width=7.0cm]{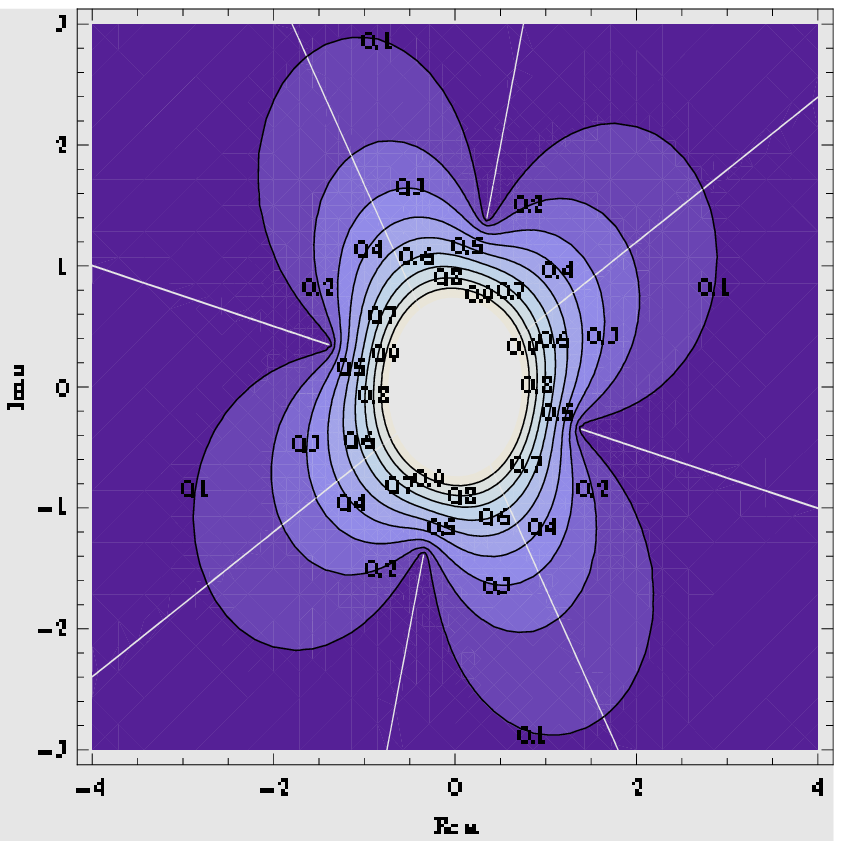}
\caption{Broken $\mathcal{PT}$-symmetric solution for $\mathcal{H}_{4}^{-}$:
(a) Star node at the origin for $c=1$, $\protect\beta =2+i3$, $\protect%
\gamma =1$ and $B=(15/2+i3)^{1/4}$; (b) centre at the origin for $c=1$, $%
\protect\beta =2+i3$, $\protect\gamma =-1$ and $B=(30/13-i45/13)^{1/4}$.}
\label{fig23}
\end{figure}

Since we have no free parameter left in our factorization this solution can
not be broken spontaneously. A complete breaking is carried out by
complexifying $\beta $ or $\gamma $. When choosing $\func{Im}\beta \neq 0$
we can still achieve in (\ref{bj}) that $4\theta _{B}+\theta _{\lambda }=\pm
\pi $, such that the star node nature of the fixed point is preserved
despite the fact that the $\mathcal{PT}$-symmetry is lost. For a particular
choice this behaviour is depicted in figure \ref{fig23}a. More surprising is
the fact that we can also achieve that $4\theta _{B}+\theta _{\lambda }=0$
in the broken case. This means the eigenvalues in (\ref{bj}) are purely
imaginary and the fixed point at the origin is a centre. We depict this
possibility in figure \ref{fig23}b. This means we have closed trajectories
even in the $\mathcal{PT}$-symmetrically broken case.

\begin{figure}[h!]
\centering  \includegraphics[width=7.0cm]{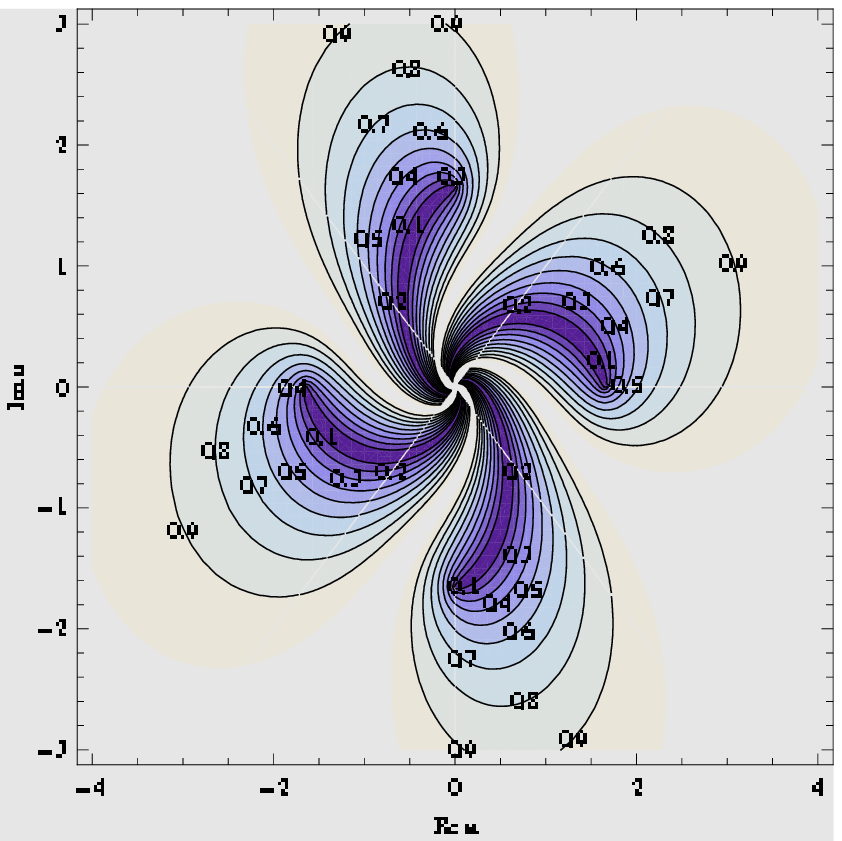} %
\includegraphics[width=7.0cm]{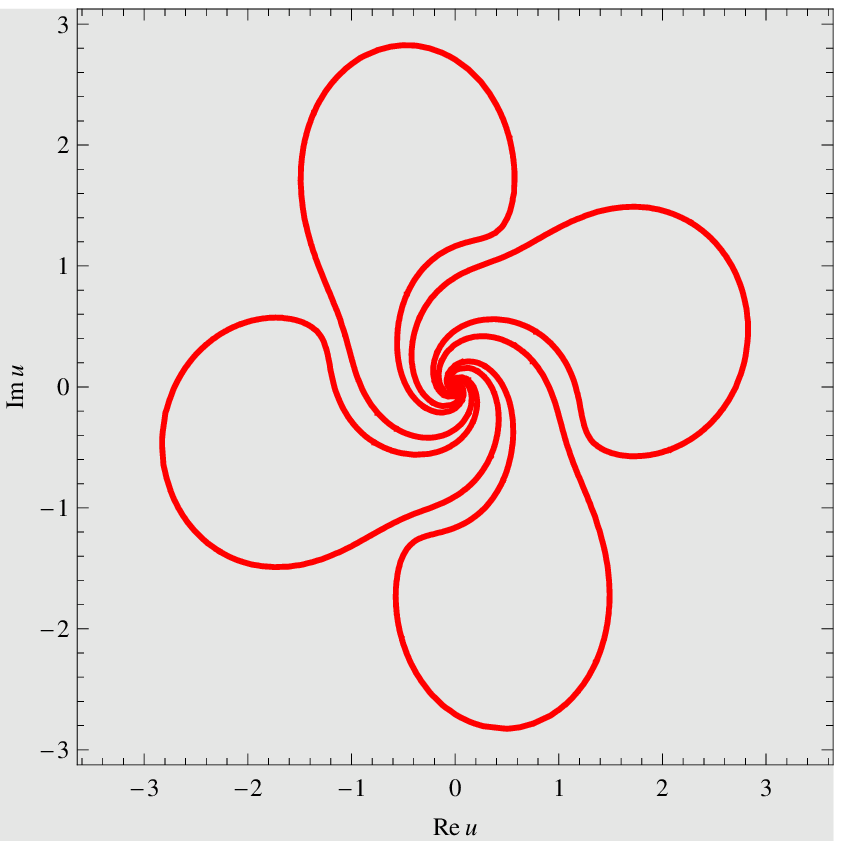}
\caption{Broken $\mathcal{PT}$-symmetric solution for $\mathcal{H}_{4}^{-}$:
(a) focus at the origin for $c=1$, $\protect\beta =2$, $\protect\gamma =1+i3$
and $B=(15/2)^{1/4}$; (b) focus at the origin for $\func{Im}\protect\zeta %
_{0}=0.8$, $c=1$, $\protect\beta =2+i3$, $\protect\gamma =-1+i2$ and $%
B=(30/13-i45/13)^{1/4}$.}
\label{figb23}
\end{figure}

In contrast for $\func{Im}\gamma \neq 0$ the eigenvalues $j_{i}$ will become
complex and the fixed point at the origin turns into an unstable or stable
focus. Taking the branch cut structure into account we observed this
behaviour in figure \ref{figb23}a for $\beta \in \mathbb{R}$, $\gamma \in 
\mathbb{C}$ and in figure \ref{figb23}b for $\beta \in \mathbb{C}$, $\gamma
\in \mathbb{C}$.

Once again we observed that when identifying a coupling constant as a
bifurcation parameter, the origin undergoes a Hopf bifurcation.

Making once more the identification $u\rightarrow x$ and $\zeta \rightarrow
t $ for our traveling wave equation together with the identification 
\begin{equation}
\kappa _{1}=0,\quad \kappa _{2}=\gamma E,\quad \text{and\quad }\beta =\gamma
g(1+\varepsilon )(2+\varepsilon )
\end{equation}%
equation (\ref{po}) converts into the classical deformation of the harmonic
oscillator%
\begin{equation}
H=E=\frac{1}{2}p^{2}-\frac{c}{2\gamma }x^{2}+gx^{2}(ix)^{\varepsilon }.
\end{equation}%
Setting furthermore $c=0$, which in our setting corresponds to a static
solution, we obtain precisely the potential treated in the seminal paper by
Bender and Boettcher \cite{Bender:1998ke}.

\section{Complex coupled nonlinear wave equation of Ito type}

It is known for some time that the Korteweg-deVries field $u(x,t)$ may be
coupled to a nonlinear field $v(x,t)$ without destroying the integrability
of the newly constructed system. This coupling is carried out by the
introduction of an additional dispersion term involving the second field $%
v(x,t)$ and the assumption of a specific form of the evolution equation for
this field. The coupled system describing this type of scenario acquires the
general form%
\begin{eqnarray}
u_{t}+\alpha vv_{x}+\beta uu_{x}+\gamma u_{xxx} &=&0,\quad \ \ \ \ \ \ \ \ \
\alpha ,\beta ,\gamma \in \mathbb{C},  \label{e1} \\
v_{t}+\delta (uv)_{x}+\phi v_{xxx} &=&0,\quad \ \ \ \ \ \ \ \ \ \delta ,\phi
\in \mathbb{C}.  \label{e2}
\end{eqnarray}%
Imposing the constraints $\beta =3\gamma $, $\phi =1$, $\delta =3$ on the
constants and leaving $\alpha $, $\gamma $ free, the Hirota method can be
applied to establish that the system (\ref{e1}) and (\ref{e2}) possesses
N-soliton solutions \cite{Hirota:1981wb}. When selecting in addition $\gamma
=-1/2$ as in \cite{Satsuma:1982pc} or $\alpha =-2$, $\beta =-6$, $\gamma =-1$%
, $\delta =-2$ and $\phi =0$ as in \cite{Ito:1982vp}, it was established in
both cases that the system possesses infinitely many charges. For $\phi =0$
the system (\ref{e1}) and (\ref{e2}) was shown to possess soliton solutions
of cusp type \cite{Kawa}. Notice also that when complexifying the KdV-field
in (\ref{KdV}) as $u_{\text{KdV}}\rightarrow u+iv$ we obtain the system (\ref%
{e1}) and (\ref{e2}) for the special choice $\alpha =-\beta _{\text{KdV}}$, $%
\beta =\beta _{\text{KdV}}$, $\gamma =\gamma _{\text{KdV}}$, $\delta =\beta
_{\text{KdV}}$ and $\phi =\gamma _{\text{KdV}}$. In the following we will
mainly discuss the case\ $\phi =0$, such that even the real case of (\ref{e1}%
) and (\ref{e2}) is distinct from the complexified version of the KdV
equation.

It is straightforward to verify that for the choice $\delta =\alpha $ the
system of equations (\ref{e1}) and (\ref{e2}) results from a variation of a
Hamiltonian whose density is given by%
\begin{equation}
\mathcal{H}_{I}=-\frac{\alpha }{2}uv^{2}-\frac{\beta }{6}u^{3}+\frac{\gamma 
}{2}u_{x}^{2}+\frac{\phi }{2}v_{x}^{2},  \label{H}
\end{equation}%
when using the variational derivative and time evolution in the standard way%
\begin{equation}
w_{t}=\frac{\partial }{\partial x}\left( \sum\nolimits_{n=0}^{\infty
}(-1)^{n}\frac{d^{n}}{dx^{n}}\frac{\partial \mathcal{H}_{I}}{\partial w_{nx}}%
\right) _{x}\quad \ \ \ \ \text{for\quad }w=u,v.  \label{var}
\end{equation}%
The Hamiltonian resulting from the density (\ref{H}) is manifestly $\mathcal{%
PT}$-symmetric as it remains invariant under a simultaneous parity
transformation and time reversal which may be realised in four alternative
ways as%
\begin{eqnarray}
\mathcal{PT}_{++} &:&x\mapsto -x,t\mapsto -t,i\mapsto -i,u\mapsto u,v\mapsto
v\quad \ \ \ \ \ \ \text{for }\alpha ,\beta ,\gamma ,\phi \in \mathbb{R},
\label{p++} \\
\mathcal{PT}_{+-} &:&x\mapsto -x,t\mapsto -t,i\mapsto -i,u\mapsto u,v\mapsto
-v\quad \ \ \ \ \text{for }\alpha ,\beta ,\gamma ,\phi \in \mathbb{R}, \\
\mathcal{PT}_{-+} &:&x\mapsto -x,t\mapsto -t,i\mapsto -i,u\mapsto
-u,v\mapsto v\quad \ \ \ \ \text{for }i\alpha ,i\beta ,\gamma ,\phi \in 
\mathbb{R}, \\
\mathcal{PT}_{--} &:&x\mapsto -x,t\mapsto -t,i\mapsto -i,u\mapsto
-u,v\mapsto -v\quad \ \ \text{for }i\alpha ,i\beta ,\gamma ,\phi \in \mathbb{%
R},  \label{p--}
\end{eqnarray}%
depending on whether we choose the fields $u,v$ to be $\mathcal{PT}$%
-symmetric or anti-symmetric. All possibilities ensure that $\mathcal{PT}%
_{ij}:\mathcal{H}_{I}\mapsto $ $\mathcal{H}_{I}$ holds with $i,j\in \{+,-\}$%
. First we will exploit these possibilities to explain the reality of the
energies and in section 5 we use them to define new physically feasible
models.

\subsection{$\mathcal{PT}$-symmetric, spontaneously broken and broken
solutions}

As usual in this context we assume that the fields acquire the form of a
traveling wave $u(x,t)=u(\zeta )$ and $v(x,t)=v(\zeta )$ with $\zeta =x-ct$.
In \cite{Bijantravel} it was shown for the case $\phi =0$ that the
possibility to have only one field to be a traveling wave and not the other
is inconsistent. We extrapolate here without rigorous proof that this
assumption can be made without loss of generality even in other cases. Let
us briefly recall how these equations may be solved in a systematic way. To
begin with we integrate (\ref{e1}) and (\ref{e2}) to%
\begin{eqnarray}
-cu+\frac{\alpha }{2}v^{2}+\frac{\beta }{2}u^{2}+\gamma u_{\zeta \zeta }
&=&\kappa _{1},  \label{1} \\
-cv+\alpha uv &=&\kappa _{2},  \label{2}
\end{eqnarray}%
with integration constants $\kappa _{1},\kappa _{2}$. In the following we
always exclude the case $\kappa _{2}=0$ as this implies the vanishing of the
new field $v=0$, which means a reduction to the deformed KdV equation, or a
constant KdV-field $u=\alpha /c$. Multiplying (\ref{1}) by $u_{\zeta }$,
using (\ref{2}) to replace $v$ by $u$ in (\ref{1}), we can integrate once
more and obtain%
\begin{equation}
u_{\zeta }^{2}=\frac{2}{\gamma }\left( \kappa _{3}+\kappa _{1}u+\frac{c}{2}%
u^{2}-\frac{\beta }{6}u^{3}+\frac{\kappa _{2}^{2}}{2}\frac{1}{\alpha u-c}%
\right) .  \label{ux}
\end{equation}%
This equation is difficult to solve directly, but following \cite{Kawa} we
trade the $u$ field for the $v$ field with%
\begin{equation}
u=\frac{c}{\alpha }+\frac{\kappa _{2}}{\alpha v}\qquad \text{and\qquad }%
u_{\zeta }=-\frac{\kappa _{2}}{\alpha v^{2}}v_{\zeta }  \label{uv}
\end{equation}%
and obtain%
\begin{equation}
v_{\zeta }^{2}=-\frac{v}{3\alpha \gamma \kappa _{2}^{2}}\sum%
\limits_{k=0}^{4}a_{k}v^{k},  \label{vvv}
\end{equation}%
where%
\begin{equation}
\begin{array}{ll}
a_{0}=\beta \kappa _{2}^{3},\quad a_{1}=3c\kappa _{2}^{2}(\beta -\alpha ), & 
a_{2}=3\kappa _{2}(\beta c^{2}-2\alpha c^{2}-2\alpha ^{2}\kappa _{1}), \\ 
a_{3}=c^{3}(\beta -3\alpha )-6\alpha ^{2}(c\kappa _{1}+\alpha \kappa
_{3}),\quad \quad & a_{4}=-3\alpha ^{3}\kappa _{2}.%
\end{array}%
\end{equation}%
At first sight this looks even less encouraging than (\ref{ux}). However,
now the right hand side is a polynomial and in case we can factorize the sum 
$\sum\nolimits_{k=0}^{4}a_{k}v^{k}$ into some convenient form we may be able
to integrate (\ref{vvv}) similarly as in the previous section. Up to one
integration the solution is therefore%
\begin{equation}
\pm \sqrt{\lambda }\left( \zeta -\zeta _{0}\right) =\int dv~\frac{1}{\sqrt{%
R(v)}},  \label{sol}
\end{equation}%
with $\lambda =\alpha ^{2}/\gamma \kappa _{2}$ and $\lambda R(v)$
corresponding to the right hand side of (\ref{vvv}). We note that not all
conceivable assumptions for the sum will lead to meaningful or nontrivial
solutions. Taking for instance $R(v)=v^{4}(A+v)$ or $R(v)=v^{3}(A+v)(B+v)$
with some unknown constants $A$ and $B$ leads in both cases to $\kappa
_{2}=0 $ or $\beta =0$, which we exclude for the above mentioned reasons.

When demanding vanishing asymptotic boundary conditions for the $u$-field
and its derivatives, the relations (\ref{1}) and (\ref{ux}) imply that we
satisfy the additional constraints%
\begin{equation}
\lim_{\zeta \rightarrow \pm \infty }u,u_{\zeta },u_{\zeta \zeta }=0\quad
\Rightarrow \quad \kappa _{1}=\frac{\alpha \kappa _{2}^{2}}{2c^{2}}%
,~~~\kappa _{3}=\frac{\kappa _{2}^{2}}{2c},~~~\ ~~\lim_{\zeta \rightarrow
\pm \infty }v=-\frac{\kappa _{2}}{c}.  \label{abound}
\end{equation}

\subsubsection{Type I solutions}

The simplest possible factorization for the sum in (\ref{vvv}) is $%
R(v)=v(v-A)^{4}$, which holds up to the constraints%
\begin{eqnarray}
\kappa _{1} &=&\frac{c^{2}\left( 2\beta \alpha -9\alpha ^{2}-\beta
^{2}\right) }{16\alpha ^{2}\beta },\quad \kappa _{2}=\frac{3\sqrt{3}%
c^{2}(\alpha -\beta )^{2}}{16\alpha ^{3/2}(-\beta )^{3/2}},\quad \\
\kappa _{3} &=&\frac{c^{3}(3\alpha -\beta )\left( 9\alpha ^{2}-6\beta \alpha
+5\beta ^{2}\right) }{96\alpha ^{3}\beta ^{2}}\qquad \text{and\qquad }A=%
\frac{\sqrt{3}c(\beta -\alpha )}{4\alpha ^{3/2}\sqrt{-\beta }}.  \notag
\end{eqnarray}%
For a given specific model, i.e.~fixed $\alpha ,\beta ,\gamma ,\phi $, this
means that all remaining free parameters are fixed. Vanishing asymptotic
boundary conditions for $u$ require by the first two equations in (\ref%
{abound}) only \emph{one} further relation $\beta =-3\alpha $, despite the
fact that one has to solve \emph{two} constraining equations. Solving (\ref%
{sol}) for the given factorization yields%
\begin{equation}
\zeta -\zeta _{0}=\pm \frac{1}{A\sqrt{\lambda }}\left[ \frac{\func{arctanh}%
\left( \frac{\sqrt{v}}{\sqrt{A}}\right) }{\sqrt{A}}+\frac{\sqrt{v}}{A-v}%
\right] .  \label{v}
\end{equation}%
First we focus on the real solution, which is obtained from (\ref{v}) for $%
\lambda >0$ and $0\,<v<A$. A further solution is obtained when replacing in (%
\ref{v}) the $\func{arctanh}$ by $\func{arcoth}$, which produces a real
solution for $0\,<A<v$. Again we are faced with the problem that we are not
able to extract from (\ref{v}) the function $v(\zeta )$. Nonetheless, we can
re-write (\ref{v}) as%
\begin{equation}
v(\zeta )=A+\frac{\sqrt{v(\zeta )}\sqrt{A}}{\func{arctanh}\left( \frac{\sqrt{%
v(\zeta )}}{\sqrt{A}}\right) \pm A^{3/2}\sqrt{\lambda }(\zeta -\zeta _{0})}
\label{vb}
\end{equation}%
similarly as done in \cite{Kawa}. This is of course still not $v(\zeta )$,
but this form is very useful to extract various types of information in an
analytical manner. Simply by considering the functions on the right hand
side of (\ref{vb}), we observe the asymptotic behaviour $\lim\nolimits_{%
\zeta \rightarrow \pm \infty }v(\zeta )=A$. In principle this is already
sufficient to extract the qualitative behaviour of $v(\zeta )$, since we
also have assumed a simple factorization for the derivative of $v$ in (\ref%
{vvv}). We can, however, be more precise because the formulation (\ref{vb})
is also ideally suited to be solved numerically. In principle this
possibility will become more powerful when considering more complicated
scenarios. However, in these cases one encounters also more occurrences of
the field $v$ and it is not always obvious for which one the equation should
be solved.

\begin{figure}[h]
\centering\includegraphics[width=7.0cm]{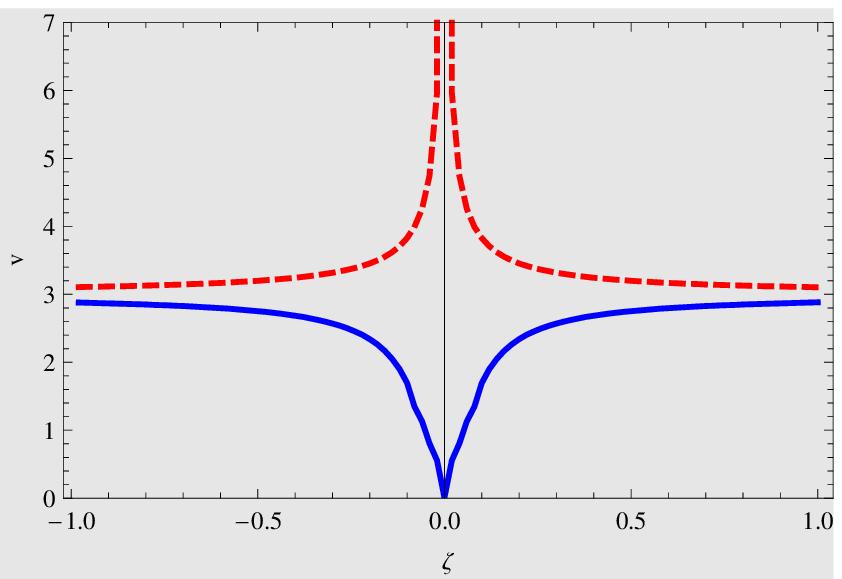} %
\includegraphics[width=7.0cm]{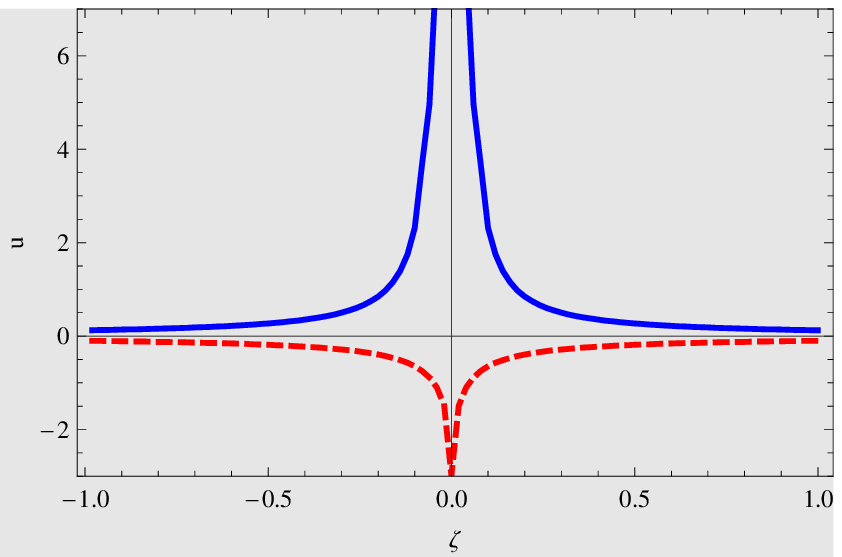}
\caption{Cusp solutions with asymptotically vanishing boundary conditions
for the $u$-field of the Ito type equation with $A=3$, $c=1$, $\protect%
\alpha =-1/3$, $\protect\beta =1$, $\protect\gamma =-1/27^{2}$ and $\protect%
\zeta _{0}=0$: (a) $v$-field; (b) $u$-field.}
\label{fig29}
\end{figure}

For the purpose of a numerical study we discretise the equations as $%
v_{n+1}=F(v_{n})$ and subsequently solve them iteratively for given values
of $\zeta $. The fixed points $v_{f}(\zeta )$ of such discretised equations
are known to be stable if and only if $\left\vert F^{\prime }\left[
v_{f}(\zeta )\right] \right\vert <1$. We use this criterium to facilitate
the numerical investigations. Concretely we solve the recursive equation%
\begin{equation}
v_{n+1}(\zeta )=A+\frac{\sqrt{v_{n}(\zeta )}\sqrt{A}}{\func{arctanh}\left( 
\frac{\sqrt{v_{n}(\zeta )}}{\sqrt{A}}\right) \pm A^{3/2}\sqrt{\lambda }%
(\zeta -\zeta _{0})},
\end{equation}%
which converges very rapidly to a precision of $\sim 10^{-5\text{ }}$%
typically already for less than $150$ iterations. We proceed similarly for
the solution when $0\,<A<v$. In figure \ref{fig29} we depict two types of
cusp solutions obtained in this manner.

In principle we could also proceed in this manner when taking complex
initial conditions, but it is simpler to produce the contour plot in the way
outlined at the beginning of section 3.1. For the same values of the
parameters as in figure \ref{fig29} we depict our results in figure \ref%
{fig30}, indicating as before the imaginary parts of $\zeta _{0}$ on some
particular trajectories.

\begin{figure}[h]
\centering  \includegraphics[width=4.9cm]{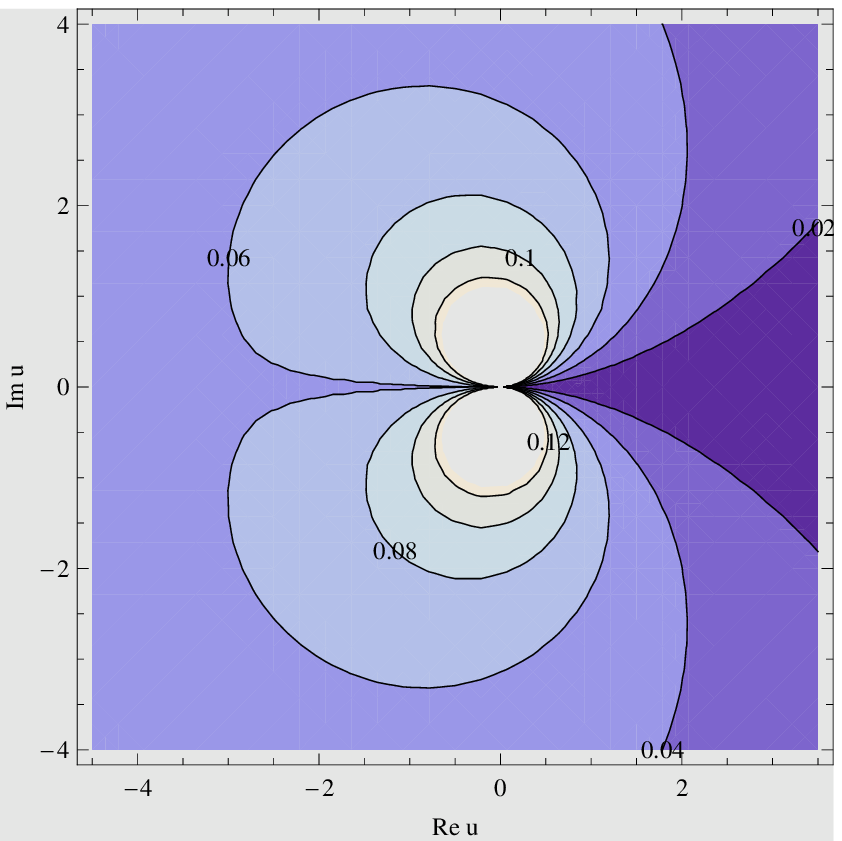} %
\includegraphics[width=4.9cm]{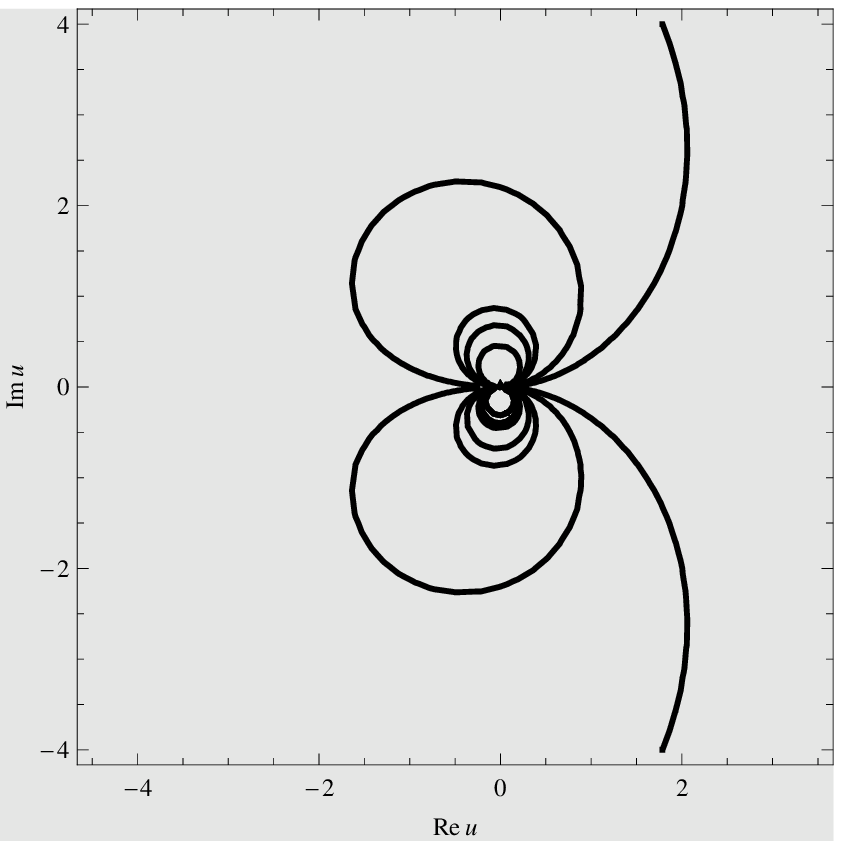} %
\includegraphics[width=4.9cm]{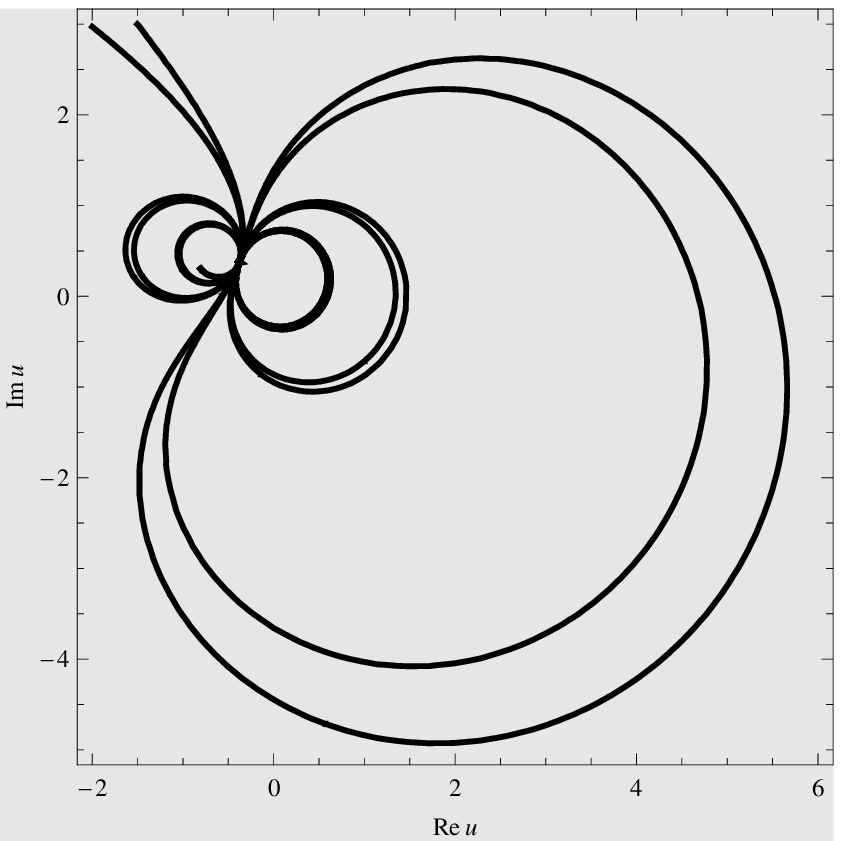} %
\includegraphics[width=4.9cm]{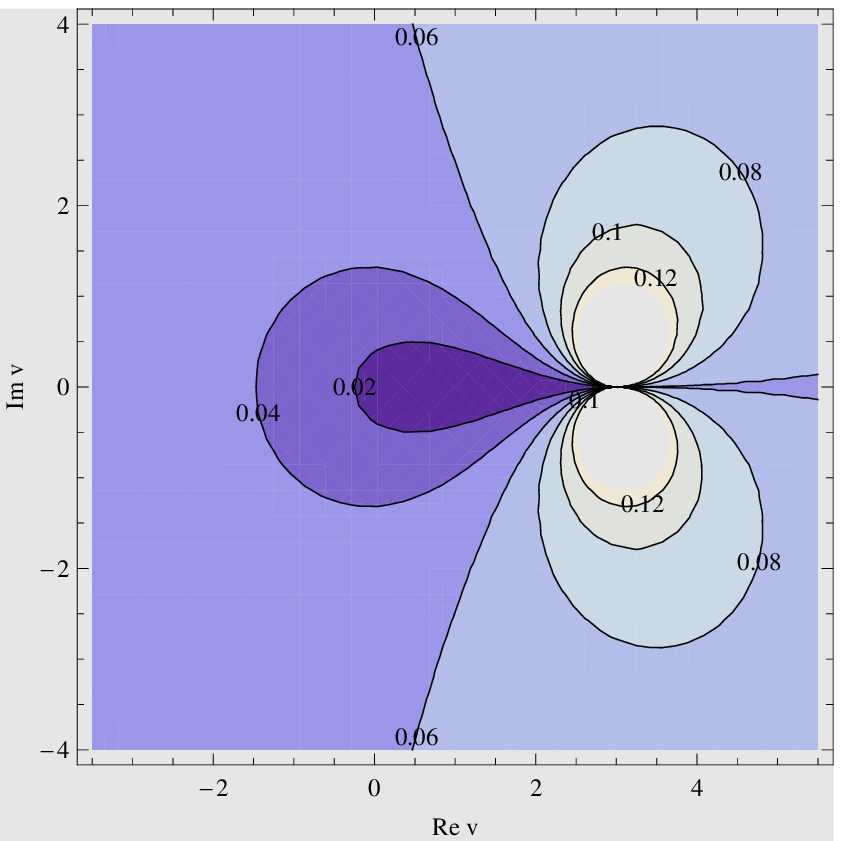} %
\includegraphics[width=4.9cm]{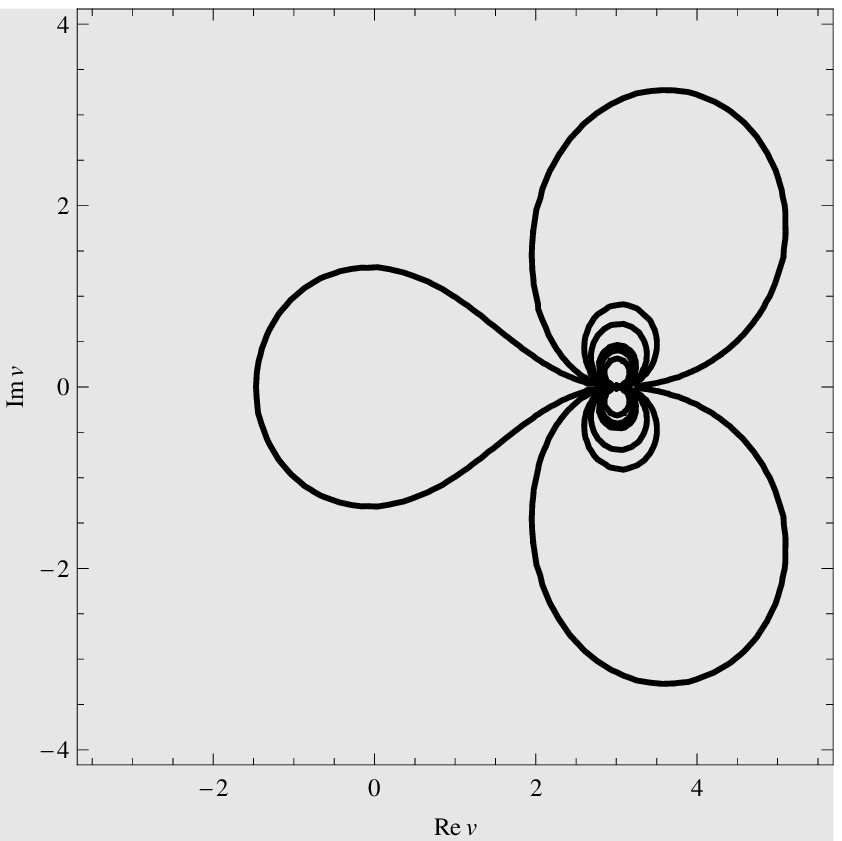} %
\includegraphics[width=4.9cm]{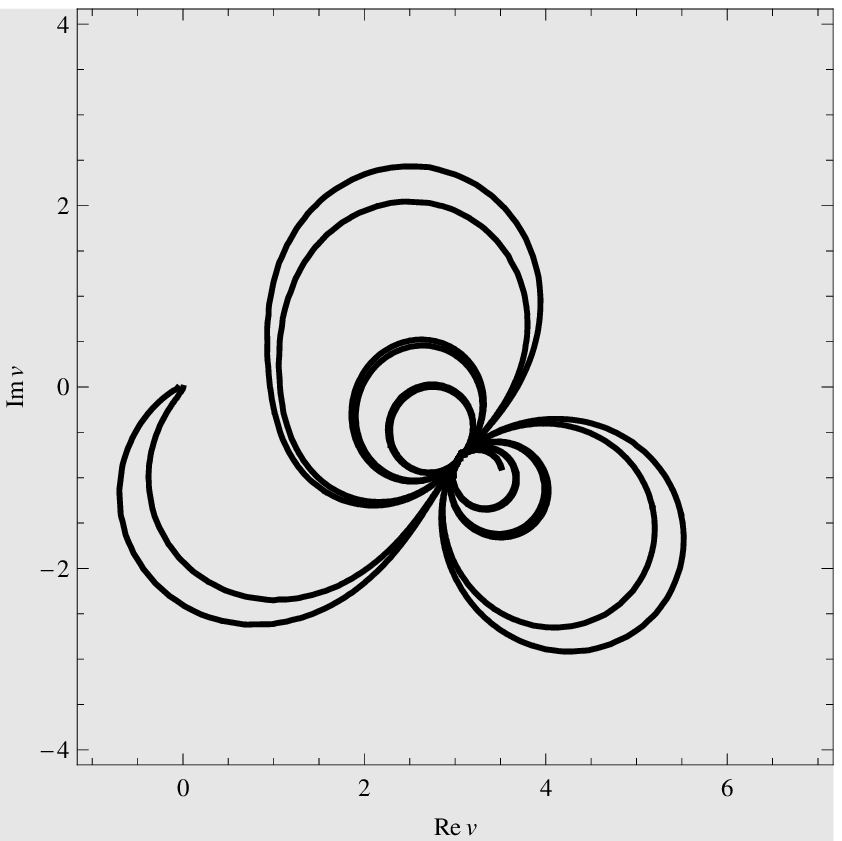}
\caption{Complex type I solutions with asymptotically vanishing boundary
conditions for the $u$-field and corresponding $v$-field of the Ito type
equation: (a) $\mathcal{PT}$-symmetric case with $A=3$, $c=1$, $\protect%
\alpha =-1/3$, $\protect\beta =1$ and $\protect\gamma =-1/27^{2}$; (b) same
values as in panel (a) for a single trajectory with $\protect\zeta _{0}=0.04$%
; (c) broken $\mathcal{PT}$-symmetric case with $A\approx 3.054-0.783i$, $%
c=1 $, $\protect\alpha =-1/3$, $\protect\beta =1-i$ and $\protect\gamma %
=-1/27^{2}-i/2$.}
\label{fig30}
\end{figure}

As expected from (\ref{vb}) we observe in figures \ref{fig30} that the
complex solutions tend to the same asymptotic value as the real ones. The
point $A$ is an unstable focus in the $\mathcal{PT}$-symmetric and its
broken version. In panel (b) more Riemann sheets are taken into account,
revealing more substructure compared to panel (a). In the $\mathcal{PT}$%
-symmetric case we also observe the crucial feature that $v^{\ast }(\zeta
)=v(-\zeta )$ and $u^{\ast }(\zeta )=u(-\zeta )$, which guarantees the
reality of the energy as defined by the expression in (\ref{Energy}).

\subsubsection{Type II solutions}

Next we introduce an additional parameter $B$ and assume the factorization
of the form $R(v)=v(v-A)^{2}(v-B)^{2}$, which imposes the four constraining
equations%
\begin{eqnarray}
\quad \kappa _{1} &=&\frac{(\beta -2\alpha )c^{2}}{2\alpha ^{2}}+\frac{%
\left( A^{2}+4AB+B^{2}\right) \alpha }{2},\quad  \label{c11} \\
\kappa _{2} &=&\frac{2AB(A+B)\alpha ^{3}}{c(\beta -\alpha )},
\end{eqnarray}%
\begin{eqnarray}
\kappa _{3} &=&\frac{2AB(A+B)^{2}\alpha ^{3}}{c(\alpha -\beta )}+\frac{%
c^{3}(3\alpha -2\beta )}{6\alpha ^{3}}-\frac{c\left( A^{2}+4AB+B^{2}\right) 
}{2}, \\
A+B &=&\pm \frac{\sqrt{3}c(\alpha -\beta )}{2\alpha ^{3/2}\sqrt{-\beta }}.
\label{c44}
\end{eqnarray}%
This means our five constants $\kappa _{1},\kappa _{2},\kappa _{3},A,B$ are
constrained by four equations, such that one of them remains free and thus
allows us to adjust for some desired boundary conditions. We exclude the
trivial solutions $A=B=A+B=0$ as they all lead to $\kappa _{2}=0$. Solving (%
\ref{sol}) for this factorization gives%
\begin{equation}
\zeta -\zeta _{0}=\pm \frac{2}{(A-B)\sqrt{\lambda }}\left[ \frac{\func{%
arctanh}\left( \frac{\sqrt{v}}{\sqrt{B}}\right) }{\sqrt{B}}-\frac{\func{%
arctanh}\left( \frac{\sqrt{v}}{\sqrt{A}}\right) }{\sqrt{A}}\right] ,
\label{g}
\end{equation}%
which reduces to (\ref{v}) in the limit $B\rightarrow A$. Arguing as in the
previous case, the real solutions are obtained from (\ref{v}) for $\lambda
>0 $ and $0\,<v<A<B$. For other configurations of the ordering we replace in
(\ref{g}) the $\func{arctanh}$ by $\func{arcoth}$ when the argument becomes
greater than one. For the same reasons as in the previous section we isolate 
$v$ from this equation and obtain%
\begin{equation}
v(\zeta )=A\tanh ^{2}\left[ \frac{1}{2}\sqrt{\lambda }\sqrt{A}(B-A)(\zeta
-\zeta _{0})+\sqrt{\frac{A}{B}}\func{arctanh}\left( \frac{\sqrt{v(\zeta )}}{%
\sqrt{B}}\right) \right] .  \label{xxx}
\end{equation}%
Simply by considering the functions on the right hand side of (\ref{xxx}),
we observe the asymptotic behaviour $\lim\nolimits_{\zeta \rightarrow \pm
\infty }v(\zeta )=A$. In this case the vanishing asymptotic boundary
conditions for $u$ can be implemented without additional constraints on the
model defining parameters $\alpha ,\beta ,\gamma ,\phi $. For other
orderings of $v,A,B$ we may also obtain $\lim\nolimits_{\zeta \rightarrow
\pm \infty }v(\zeta )=B$. When $\lim\nolimits_{\zeta \rightarrow \pm \infty
}v(\zeta )=A$ all constraints in (\ref{abound}) and (\ref{c11})-(\ref{c44})
are satisfied with the choice \ 
\begin{equation}
A=\frac{c^{2}(\beta -3\alpha )}{2\sqrt{3}\alpha ^{3/2}\sqrt{-\beta }}~\quad 
\text{and\quad }B=-\frac{c\sqrt{-\beta }}{\sqrt{3}\alpha ^{3/2}},
\label{const}
\end{equation}%
whereas when $\lim\nolimits_{\zeta \rightarrow \pm \infty }v(\zeta )=B$ we
need to exchange $A$ and $B$ in (\ref{const}). For the solution reported in 
\cite{Kawa} the possible values for $\alpha ,\beta ,\gamma $ were restricted
because the value of $B$ was pre-selected, but as we have shown here this is
not necessary.

\begin{figure}[h]
\centering  \includegraphics[width=7.0cm]{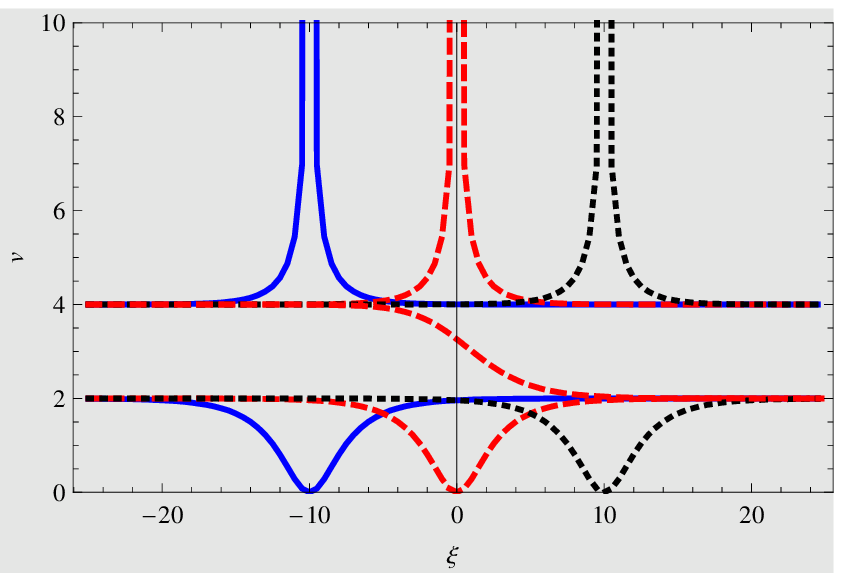} %
\includegraphics[width=7.0cm]{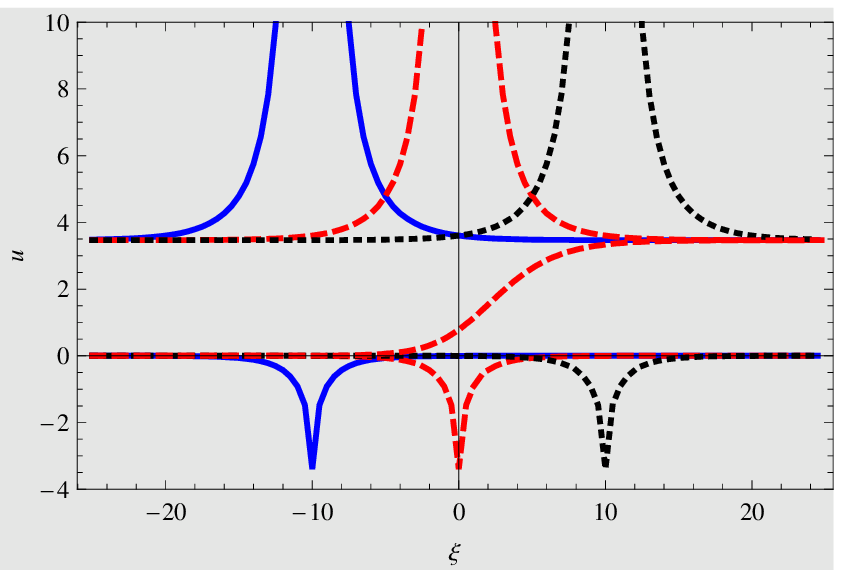}
\caption{Soliton, kink and cusp type solutions for the Ito type equations
with $\protect\alpha =-\protect\beta =1/(2\protect\sqrt{3})$, $\protect%
\gamma =1$, $c=-1$, $A=2$ and $B=4$. The initial values are takes to be $%
\protect\zeta _{0}=-10$ solid (blue), $\protect\zeta _{0}=0$ dashed (red)
and $\protect\zeta _{0}=-10$ dotted (black). (a) $v$-field; (b) $u$-field.}
\label{fig32}
\end{figure}

Choosing vanishing boundary conditions we depict the real solutions in
figure \ref{fig32} for the different regimes and real initial values $\zeta
_{0}$.

Our findings disagree slightly with those in \cite{Kawa}, where the case $%
0<v<|A|<|B|$ was reported to be of cusp type, whereas we observe that it is
of a standard soliton nature. Our numerical findings are in agreement with
the factorization of the right hand side of equation (\ref{vvv}), which
implies that $v_{\zeta }(0)=0$ and not infinity as needed for a cusp
solution. We also find a new kink type solution in the region $0<|A|<v<|B|$
not reported by Kawamoto. In the regions $0<v<A<B$ and $0<A<B<v$ we observe
explicitly the $\mathcal{PT}$-symmetry $\zeta \rightarrow -\zeta +2\zeta
_{0} $, $v\rightarrow v$ and $u\rightarrow u$.

Next we investigate some complex solutions by taking the initial values $%
\zeta _{0}$ to be purely imaginary. For this type of the factorization the
linearisation around the point $A$ and $B$ is straightforward as the square
root in (\ref{sol}) can be taken. Parameterizing $A=r_{A}e^{i\theta _{A}}$, $%
B=r_{B}e^{i\theta _{B}}$ and $\lambda =r_{\lambda }e^{i\theta _{\lambda }}$
the eigenvalues of the Jacobian when linearized about $v=A$ are computed to%
\begin{eqnarray}
j_{k} &=&\pm \sqrt{r_{A}r_{\lambda }}\left[ \cos \left( \frac{3\theta _{A}}{2%
}+\frac{\theta _{\lambda }}{2}\right) r_{A}-\cos \left( \frac{\theta _{A}}{2}%
+\theta _{B}+\frac{\theta _{\lambda }}{2}\right) r_{B}\right]  \label{jk} \\
&&+i(-1)^{k}\sqrt{r_{A}r_{\lambda }}\left[ \sin \left( \frac{3\theta _{A}}{2}%
+\frac{\theta _{\lambda }}{2}\right) r_{A}-\sin \left( \frac{\theta _{A}}{2}%
+\theta _{B}+\frac{\theta _{\lambda }}{2}\right) r_{B}\right]  \notag
\end{eqnarray}%
for $k=1,2$. For the linearisation around $v=B$ we obtain (\ref{jk}) with $%
A\leftrightarrow B$.

\begin{figure}[h]
\centering  \includegraphics[width=7.0cm]{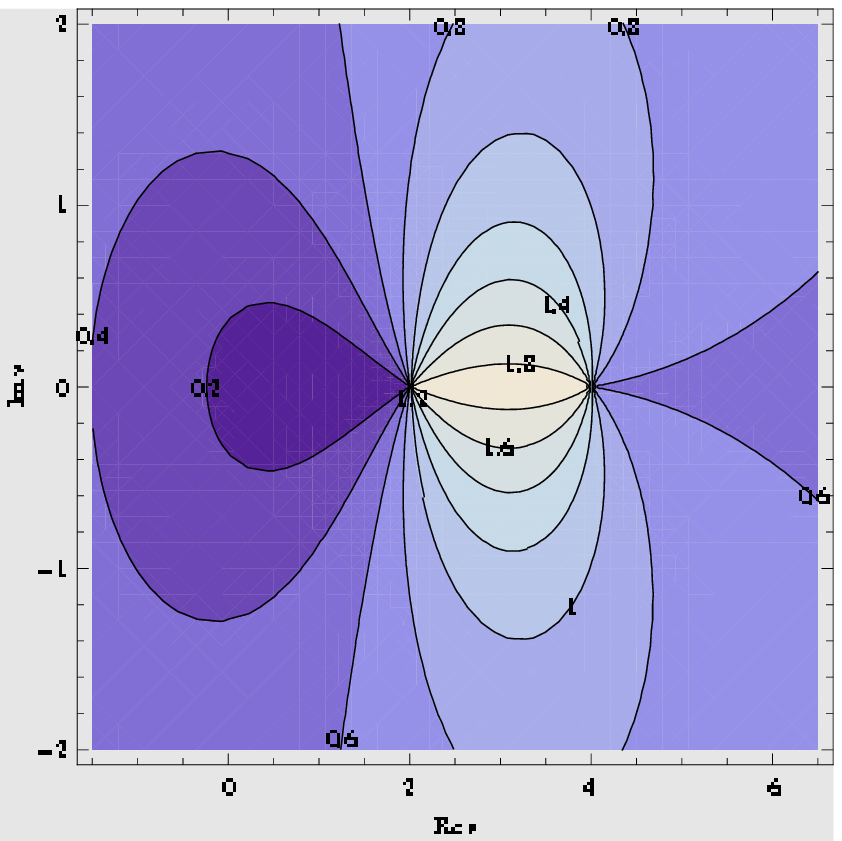} %
\includegraphics[width=7.0cm]{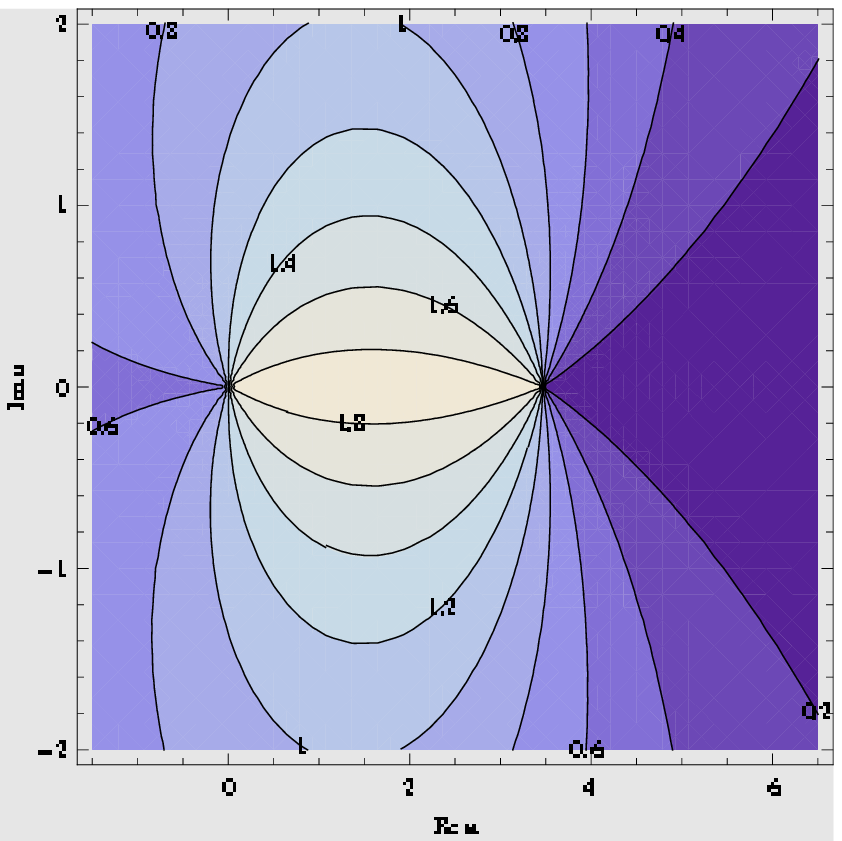}
\caption{Complex asymptotically constant type II trajectories for the $%
\mathcal{PT}$-symmetric Ito type system for purely complex initial values $%
\protect\zeta _{0}$ with $\protect\alpha =-\protect\beta =1/(2\protect\sqrt{3%
})$, $\protect\gamma =1$, $c=-1$, $A=2$ and $B=4$. (a) $v$-field; (b) $u$%
-field.}
\label{fig33}
\end{figure}

By tuning our free parameters we can produce any desired characteristic
behaviour for the fixed points at $A$ and $B$. For instance in the $\mathcal{%
PT}$-symmetric setting for $A,B\in \mathbb{R}^{+}$ we always obtain two real
degenerate eigenvalues when $\lambda \in \mathbb{R}^{+}$ and therefore star
nodes at $v=A$ and $v=B$. In figure \ref{fig33} we report an example of this
type with $j_{1}=j_{2}=\mp 1/\sqrt{6}$.

We also observe in figure \ref{fig33}a that the complex trajectories
surround the real solution with the asymptotic point or points in common.
For instance, the trajectory with $\func{Im}\zeta _{0}=1$ corresponds to a
complexified version of a real soliton solution in the regime $0<v<A<B$ with
asymptotic behaviour $\lim\nolimits_{\zeta \rightarrow \pm \infty }v(\zeta
)=A$. In the $u$-plane the real solution becomes a cusp solution running off
to infinity, whereas the complex solutions close. We may also identify
complexified versions of the kink solutions in the regime $0<A<v<B$, such as
for instance the trajectory with $\func{Im}\zeta _{0}=5$ in the $v$-plane
with asymptotic behaviour $\lim\nolimits_{\zeta \rightarrow +\infty }v(\zeta
)=A$ and $\lim\nolimits_{\zeta \rightarrow -\infty }v(\zeta )=B$. In the $u$%
-plane the qualitative behaviour remains the same with asymptotic behaviour $%
\lim\nolimits_{\zeta \rightarrow -\infty }u(\zeta )=0$ and $%
\lim\nolimits_{\zeta \rightarrow \infty }u(\zeta )=-c/\alpha $. These
features may also be derived analytically from (\ref{xxx}). The reality of
the energy is once more guaranteed be the symmetry $v^{\ast }(\zeta
)=v(-\zeta )$ and $u^{\ast }(\zeta )=u(-\zeta )$.

\begin{figure}[h!]
\centering  \includegraphics[width=7.0cm]{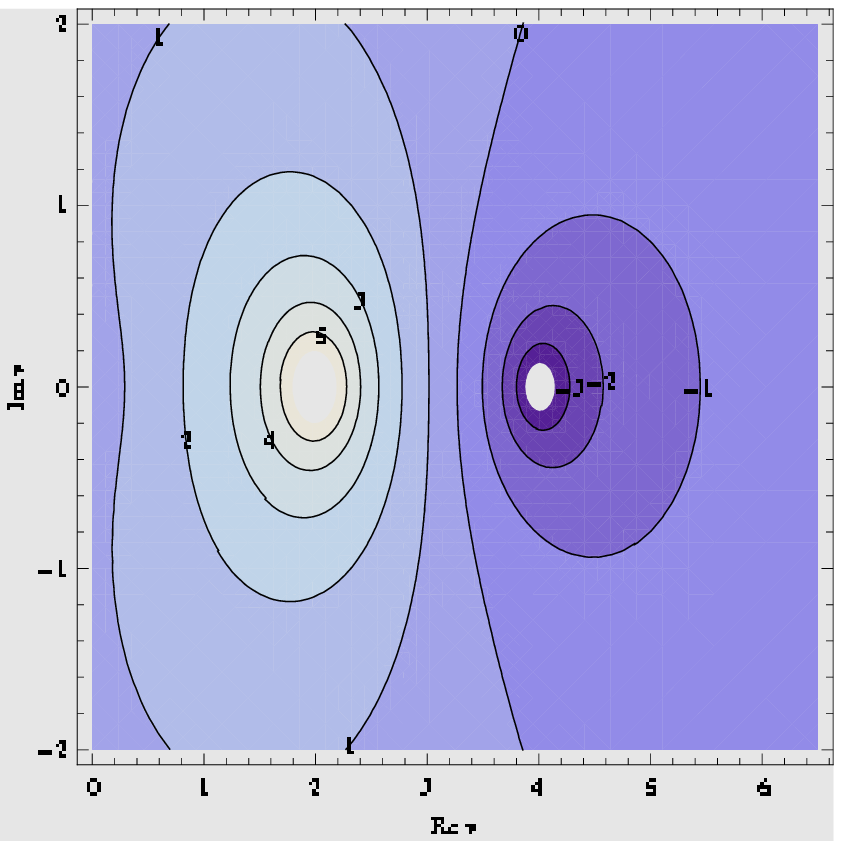} %
\includegraphics[width=7.0cm]{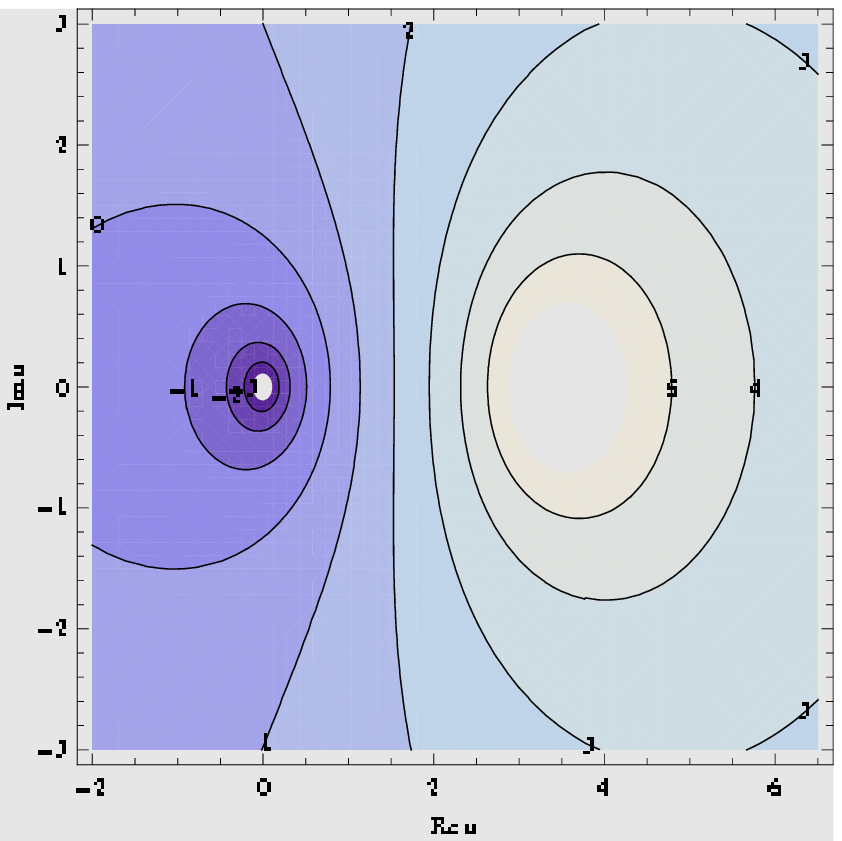}
\caption{Complex periodic trajectories for type II $\mathcal{PT}$-symmetric
Ito type system for purely complex initial values $\protect\zeta _{0}$ with $%
\protect\alpha =-\protect\beta =1/(2\protect\sqrt{3})$, $\protect\gamma =-1$%
, $c=-1$, $A=2$ and $B=4$. (a) $v$-field; (b) $u$-field.}
\label{fig39}
\end{figure}

\begin{figure}[h!]
\centering  \includegraphics[width=7.0cm]{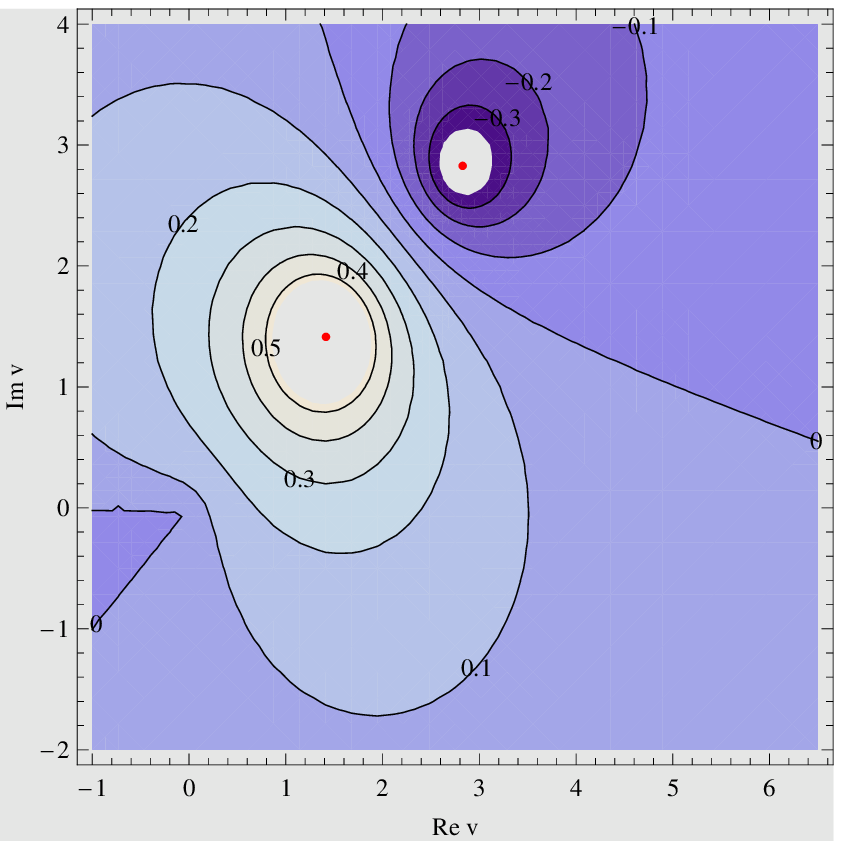} %
\includegraphics[width=7.0cm]{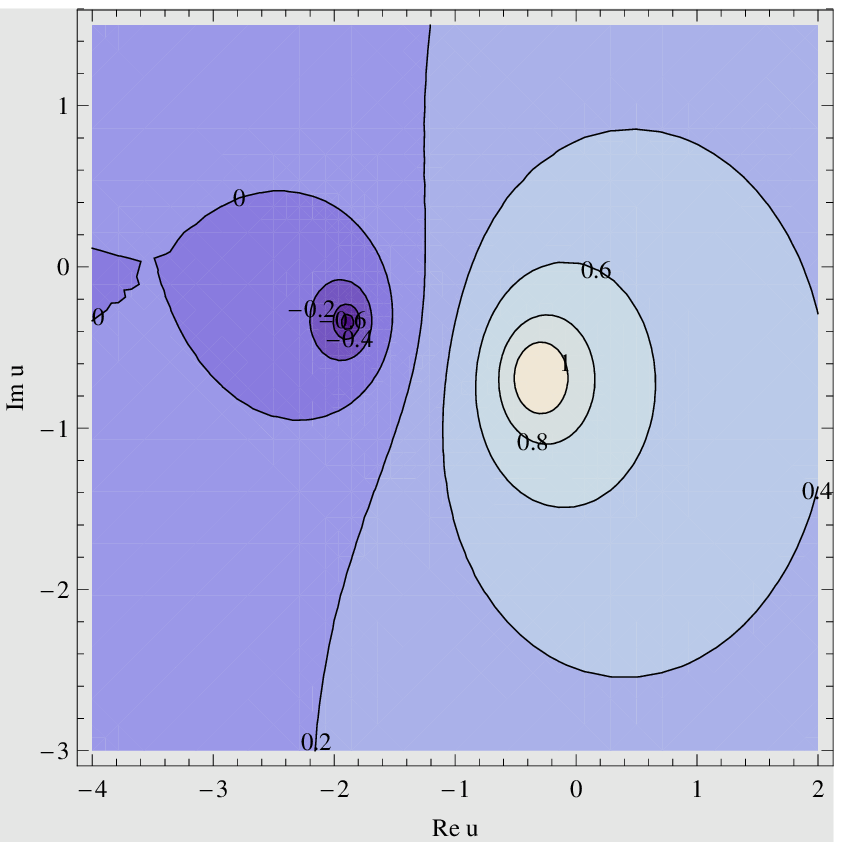}
\caption{Complex periodic trajectories for type II broken $\mathcal{PT}$%
-symmetric Ito type system for purely complex initial values $\protect\zeta %
_{0}$ with $\protect\alpha =1/(2\protect\sqrt{3})$, $\protect\beta =\frac{%
(1-2i)+2\protect\sqrt{-1-i}}{2\protect\sqrt{3}}$, $\protect\gamma =\frac{1}{%
48}\left( -i+\protect\sqrt{-1-i}\right) $ $c=-1$, $A=(1+i)\protect\sqrt{2}$
and $B=(2+2i)\protect\sqrt{2}$. (a) $v$-field; (b) $u$-field.}
\label{2period}
\end{figure}

Similarly as for the trigonometric solution of the KdV equation, by changing
the parameters we can predict some periodic solutions. In the $\mathcal{PT}$%
-symmetric setting this is achieved for $A,B\in \mathbb{R}^{+}$ when taking $%
\lambda \in \mathbb{R}^{-}$, as in this case the two eigenvalues become
purely complex and therefore the fixed point becomes a centre. For this type
of solution this can be achieved either for the point $v=A$ or $v=B$. We
depict an example of such solutions in figure \ref{fig39} with eigenvalues $%
j_{1/2}=\pm i1/\sqrt{6}$. In the $v$-plane we observe that the complex
trajectories encircle the points $A$ or $B$ whereas in the $u$-plane the
trajectories surround the points $-c/\alpha $ and $0$ for $\func{Im}\zeta
_{0}>0$ or $\func{Im}\zeta _{0}<0$, respectively.

Again we observe the symmetry relations $v^{\ast }(\zeta )=v(-\zeta )$ and $%
u^{\ast }(\zeta )=u(-\zeta )$ ensuring the reality of the energy, but as for
the trigonometric solution of the KdV equation we may compute the energy in
this case explicitly. Taking the trajectories surrounding $A$ in the $v$%
-plane to be the contour $\Gamma $ we compute%
\begin{eqnarray}
E_{T_{A}} &=&\oint\nolimits_{\Gamma }\mathcal{H}\left[ v(\zeta )\right] 
\frac{dv}{v_{\zeta }}=\oint\nolimits_{\Gamma }\frac{\mathcal{H}\left[ v%
\right] }{\sqrt{\lambda }\sqrt{v}(v-A)(v-B)}dv  \label{E1} \\
&=&-\pi \frac{\sqrt{-\gamma \kappa _{2}}}{\alpha \sqrt{A}(A-B)}\left[
cA^{2}+\kappa _{2}A+\frac{\beta }{3}\left( \frac{c}{\alpha }+\frac{\kappa
_{2}}{\alpha A}\right) ^{3}\right] .  \label{E2}
\end{eqnarray}%
For the trajectories surrounding $B$ in the $v$-plane we obtain in a similar
way $E_{T_{B}}=E_{T_{A}}(A\leftrightarrow B)$.

Considering the amount of free parameters we have at our disposal, the
expression for (\ref{jk}) also suggests that we will be able to generate any
type of fixed point even for the broken $\mathcal{PT}$-symmetric scenario.
Most unexpected is probably that we may even generate periodic orbits in
that case. For this to happen we require $3\theta _{A}+\theta _{\lambda
}=\pi $ and $\theta _{A}+2\theta _{B}+$ $\theta _{\lambda }=\pi $ to hold. A
solution to these equations, leading to the eigenvalues $j_{1/2}=\pm i2\sqrt{%
2}$, is presented in figure \ref{2period}.

\begin{figure}[h!]
\centering  \includegraphics[width=7.0cm]{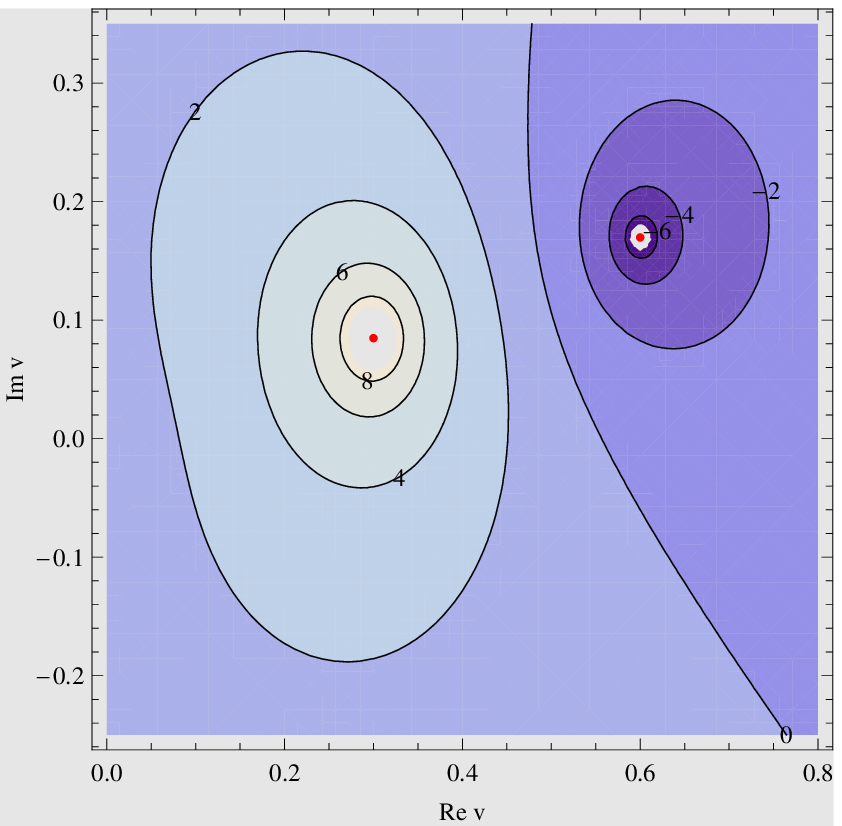} %
\includegraphics[width=7.0cm]{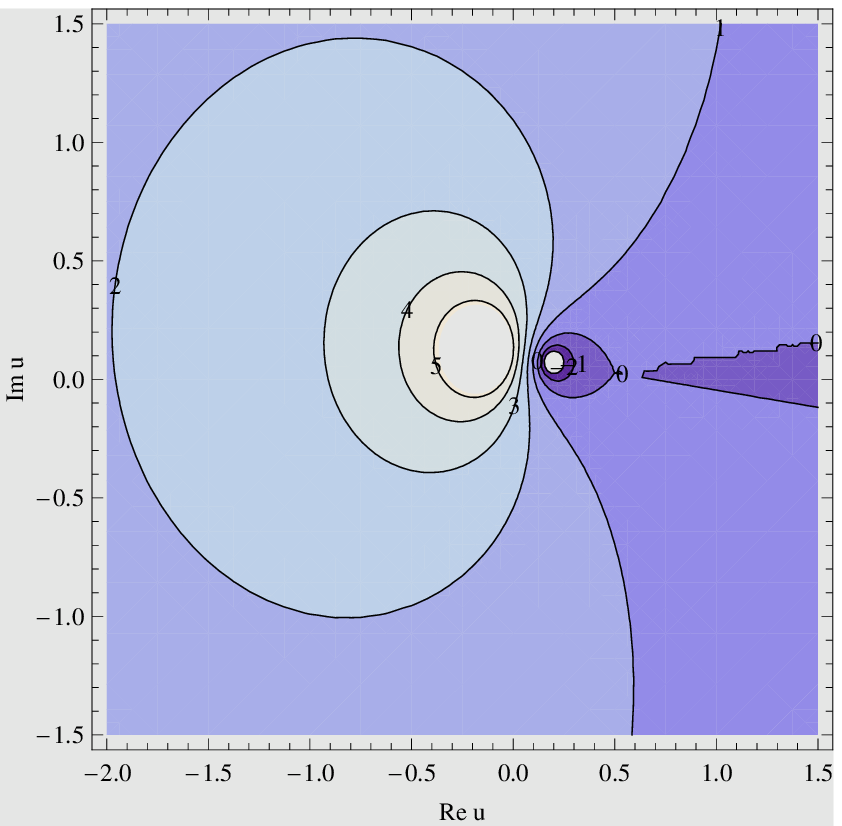}
\caption{Complex periodic trajectories for type II broken $\mathcal{PT}$%
-symmetric Ito type system with real energy $E_{T_{A}}\approx -0.4275$ for
purely complex initial values $\protect\zeta _{0}$ for the parameter choice
in (\protect\ref{pr1}) and (\protect\ref{pr2}) with $r_{\protect\alpha }=2$
and $\protect\mu =2$; (a) $v$-field; (b) $u$-field.}
\label{fig36}
\end{figure}

As already seen for the complex KdV-system in section 2.1.2, we can also
break the $\mathcal{PT}$-symmetry in a more controlled way and restore the
reality of the energy given by (\ref{E2}). Making for instance the parameter
choice%
\begin{eqnarray}
\alpha &=&r_{\alpha }e^{-\frac{i}{2}\arctan \left( \frac{7}{4\sqrt{2}}%
\right) },\quad \quad \beta =3r_{\alpha }e^{i\frac{\pi }{4}},\quad \quad
\gamma =\mu \frac{\beta -\alpha }{\alpha },\quad \quad c=-1,  \label{pr1} \\
A &=&\frac{c(\beta -\alpha )}{2\sqrt{3}\alpha ^{3/2}\sqrt{-\beta }},\quad
\quad B=2A,  \label{pr2}
\end{eqnarray}%
with $\mu \in \mathbb{R}^{+}$ and $r_{\alpha }\in \mathbb{R}$ being
unconstrained constants, we obtain for the energy of a periodic trajectory
around the point $A$ the real expression 
\begin{equation}
E_{T_{A}}=-\frac{\pi }{3r_{\alpha }^{2}}\sqrt{\frac{5\mu }{3}}.
\end{equation}%
Notice that in this case we have two free parameters available allowing to
tune the real energy together with the model, despite the fact that any of
these the Hamiltonians is neither Hermitian nor $\mathcal{PT}$-symmetric. We
depict an example for such type of trajectory in figure \ref{fig36}. We
notice that the trajectories are qualitatively the same as those with
complex energies depicted for instance in figure \ref{2period}.

\subsubsection{Type III solutions}

Next we assume that $R(v)=v(v-A)^{2}(B+v)(C+v)$, which is achievable upon
the constraints%
\begin{eqnarray}
\kappa _{1} &=&\frac{(\beta -2\alpha )c^{2}+[A^{2}+2A(B+C)+BC]\alpha ^{3}}{%
2\alpha ^{2}},  \notag \\
\kappa _{2} &=&\frac{A[2BC+A(B+C)]\alpha ^{3}}{c(\beta -\alpha )}, \\
\kappa _{3} &=&\frac{[c^{2}-[A^{2}+2(B+C)A+BC]\alpha ^{2}]c}{2\alpha ^{2}}-%
\frac{A(2A+B+C)[2BC+A(B+C)]\alpha ^{3}}{2(\beta -\alpha )c}-\frac{\beta c^{3}%
}{3\alpha ^{3}},  \notag
\end{eqnarray}%
\begin{equation}
A=-\frac{\sqrt{3}\sqrt{-\beta \alpha ^{3}(\alpha -\beta )^{2}c^{2}BC(B+C)^{2}%
}}{\beta \alpha ^{3}(B+C)^{2}}-\frac{2BC}{B+C}.  \notag
\end{equation}%
Now we have two free parameters at our disposal. In this case the solution
of (\ref{sol}) leads to%
\begin{equation}
\zeta -\zeta _{0}=\pm \frac{2}{A\sqrt{B}\sqrt{\lambda }}\frac{1}{v}\left\{ F%
\left[ \arcsin \left( \frac{\sqrt{B}}{\sqrt{v}}\right) |\frac{C}{B}\right]
+\Pi \left[ \frac{A}{B};-\arcsin \left( \frac{\sqrt{B}}{\sqrt{v}}\right) |%
\frac{C}{B}\right] \right\} ,
\end{equation}%
with $F\left[ \phi |m\right] $ denoting an elliptic function of the first
and $\Pi \left[ n;\phi |m\right] $ an incomplete elliptic function.

\begin{figure}[h!]
\centering\includegraphics[width=7.0cm]{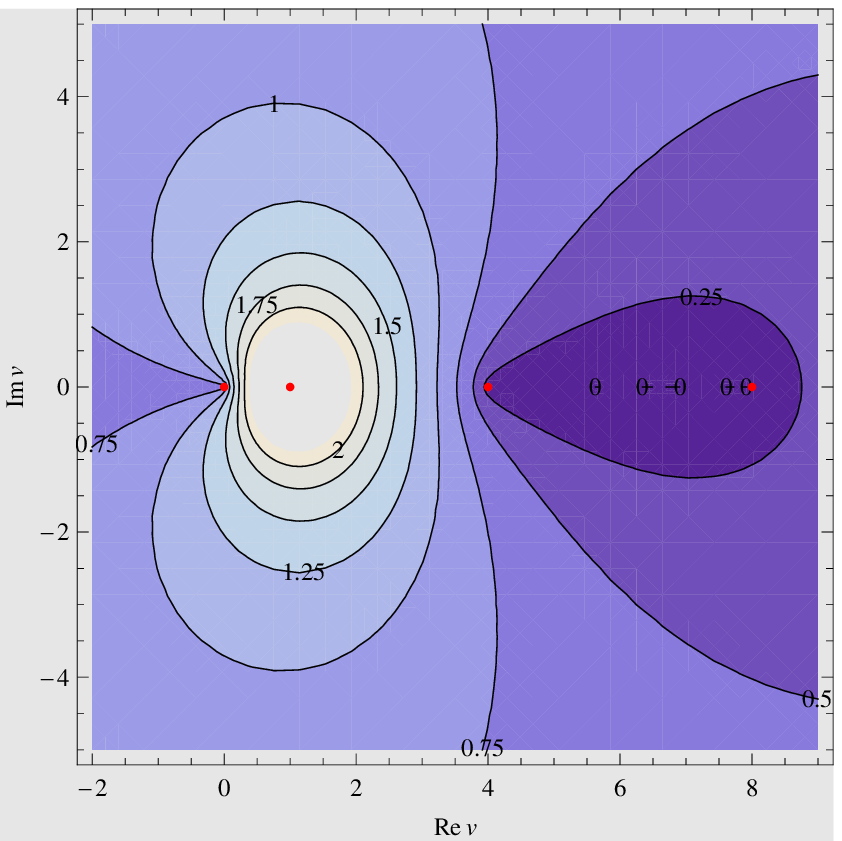} %
\includegraphics[width=7.0cm]{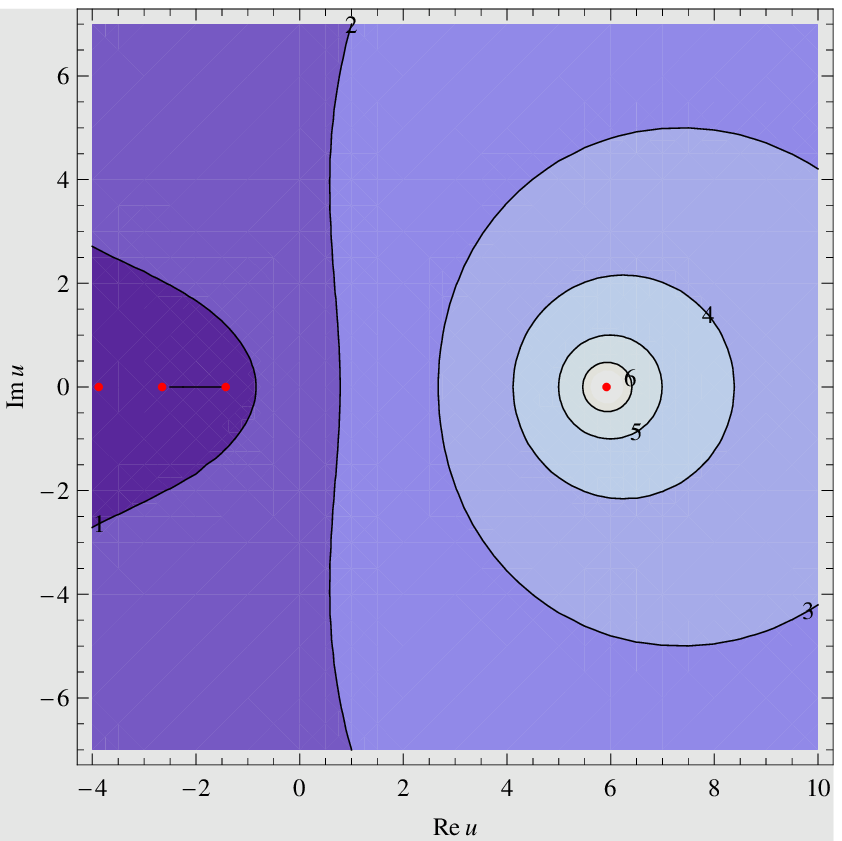}
\caption{Complex periodic trajectories for the type III $\mathcal{PT}$%
-symmetric Ito type system for purely complex initial values $\protect\zeta %
_{0}$ with $\protect\alpha =-\protect\beta =2\protect\sqrt{6}/19$, $\protect%
\gamma =-1$, $c=-1$, $A=1$, $B=4$ and $C=8$.}
\label{fig40}
\end{figure}

Choosing $B$ and $C$ conveniently we compute some trajectories similarly as
in the previous subsection and depict our results in figure \ref{fig40}.

We identify some trajectories surrounding the point $A$ in the $v$-plane for
which we compute the energy as 
\begin{eqnarray}
E_{T_{A}} &=&\oint\nolimits_{\Gamma }\mathcal{H}\left[ v(\zeta )\right] 
\frac{dv}{v_{\zeta }}=\oint\nolimits_{\Gamma }\frac{\mathcal{H}\left[ v%
\right] }{\sqrt{\lambda }\sqrt{(v-B)(v-C)v}(v-A)}dv \\
&=&\pi \frac{\sqrt{-\gamma \kappa _{2}}}{\alpha \sqrt{(A-B)(A-C)A}}\left[
cA^{2}+\kappa _{2}A+\frac{\beta }{3}\left( \frac{c}{\alpha }+\frac{\kappa
_{2}}{\alpha A}\right) ^{3}\right] .
\end{eqnarray}%
As seen for the type II solutions we may also in this case break the $%
\mathcal{PT}$-symmetry and still obtain periodic solutions. Moreover, we can
break the symmetry further, that is also on the level of the Hamiltonian,
and still render the expression for $E_{T_{A}}$ real. Since in this case we
have even an additional parameter $C$ at our disposal it is conceivable that
we might even have a free variable left in the expression for the real
energy. We leave this question for a future investigation.

An example for broken $\mathcal{PT}$-symmetry is depicted in figure \ref%
{fig99}.

\begin{figure}[h!]
\centering\includegraphics[width=7.0cm]{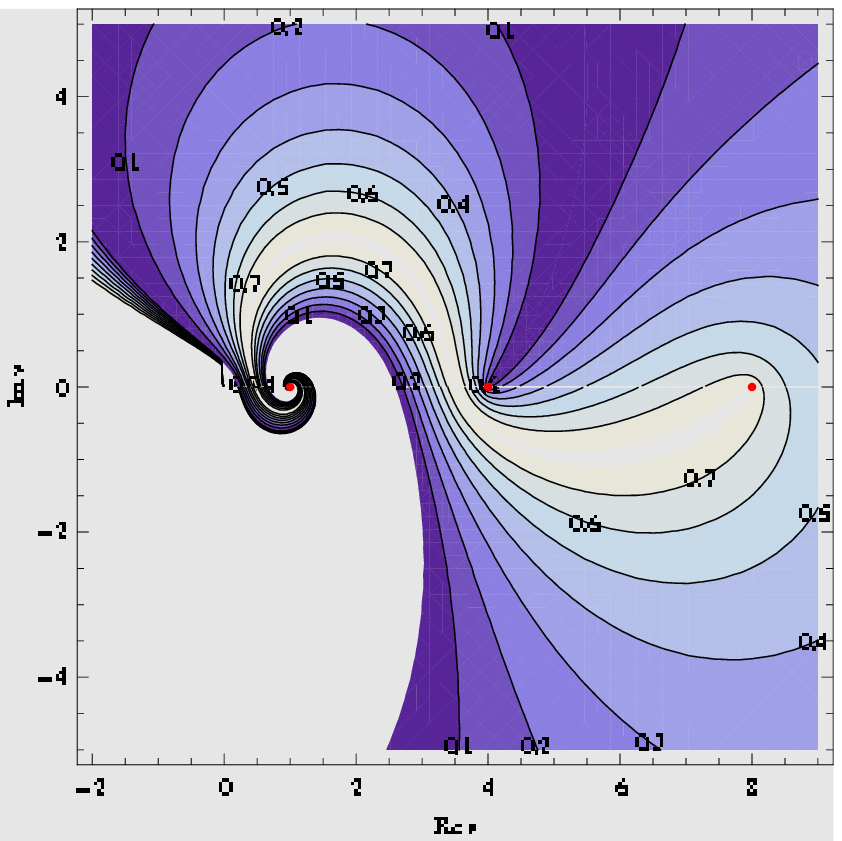} %
\includegraphics[width=7.0cm]{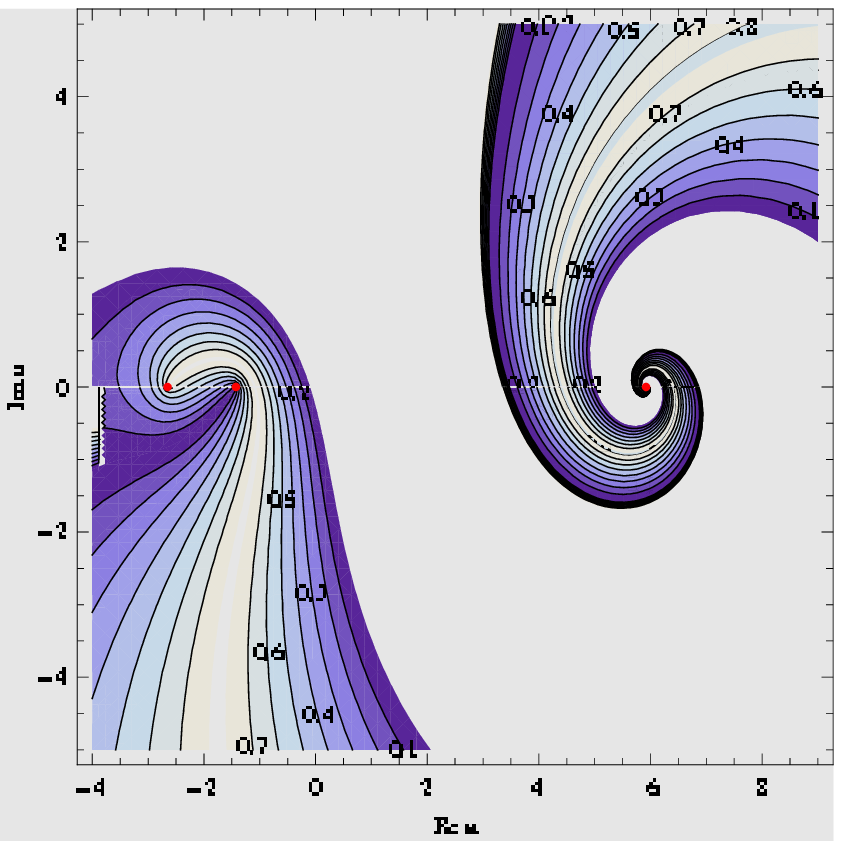}
\caption{Complex periodic trajectories for the type III with broken $%
\mathcal{PT}$-symmetry for the Ito type system for purely complex initial
values $\protect\zeta _{0}$ with $\protect\alpha =-\protect\beta =2\protect%
\sqrt{6}/19$, $\protect\gamma =-1+i$, $c=-1$, $A=1$, $B=4$ and $C=8$.}
\label{fig99}
\end{figure}

\newpage

\section{$\mathcal{PT}$-symmetric deformations of the Ito type equations}

Since the Hamiltonian $\mathcal{H}_{I}$ admits the four different types of
antilinear symmetries (\ref{p++})-(\ref{p--}), we have now the options to
apply the deformation maps $\delta _{\varepsilon }^{+}$ and $\delta
_{\varepsilon }^{-}$ introduced in (\ref{delta}) to the fields $u$ and $v$
in different combinations. Accordingly we define four different $\mathcal{PT}
$-symmetric models with two deformation parameters $\varepsilon ,\mu $
suitable normalized by the following Hamiltonian densities%
\begin{eqnarray}
\mathcal{H}_{\varepsilon ,\mu }^{++} &=&-\frac{\alpha }{2}uv^{2}-\frac{\beta 
}{6}u^{3}-\frac{\gamma }{1+\varepsilon }(iu_{x})^{\varepsilon +1}-\frac{\phi 
}{1+\mu }(iv_{x})^{\mu +1},\quad  \label{Hplus} \\
\mathcal{H}_{\varepsilon ,\mu }^{+-} &=&\frac{\alpha }{1+\mu }u(iv)^{\mu +1}-%
\frac{\beta }{6}u^{3}-\frac{\gamma }{1+\varepsilon }(iu_{x})^{\varepsilon
+1}+\frac{\phi }{2}v_{x}^{2}, \\
\mathcal{H}_{\varepsilon ,\mu }^{-+} &=&-\frac{\alpha }{2}uv^{2}-\frac{%
i\beta }{(1+\varepsilon )(2+\varepsilon )}(iu)^{2+\varepsilon }+\frac{\gamma 
}{2}u_{x}^{2}-\frac{\phi }{1+\mu }(iv_{x})^{\mu +1}, \\
\mathcal{H}_{\varepsilon ,\mu }^{--} &=&\frac{\alpha }{1+\mu }u(iv)^{\mu +1}-%
\frac{i\beta }{(1+\varepsilon )(2+\varepsilon )}(iu)^{2+\varepsilon }+\frac{%
\gamma }{2}u_{x}^{2}+\frac{\phi }{2}v_{x}^{2}.\quad
\end{eqnarray}%
By construction we have $\mathcal{PT}_{ij}:\mathcal{H}_{\varepsilon ,\mu
}^{ij}\mapsto $ $\mathcal{H}_{\varepsilon ,\mu }^{ij}$ with $i,j\in \{+,-\}$
and $\lim_{\varepsilon ,\mu \rightarrow 1}\mathcal{H}_{\varepsilon ,\mu
}^{ij}=\mathcal{H}_{I}$. The corresponding equations of motion resulting
from (\ref{var}) are%
\begin{equation}
\begin{array}{rr}
\mathcal{H}_{\varepsilon ,\mu }^{++}:u_{t}+\alpha vv_{x}+\beta uu_{x}+\gamma
u_{xxx,\varepsilon }=0, & ~~\mathcal{H}_{\varepsilon ,\mu
}^{+-}:u_{t}+\alpha v_{\mu }v_{x}+\beta uu_{x}+\gamma u_{xxx,\varepsilon }=0,
\\ 
v_{t}+\alpha (uv)_{x}+\phi v_{xxx,\mu }=0, & v_{t}+\alpha (uv_{\mu
})_{x}+\phi v_{xxx}=0,%
\end{array}
\label{eq1}
\end{equation}%
\begin{equation}
\begin{array}{rr}
\mathcal{H}_{\varepsilon ,\mu }^{-+}:u_{t}+\alpha vv_{x}+\beta
u_{\varepsilon }u_{x}+\gamma u_{xxx}=0, & ~~\mathcal{H}_{\varepsilon ,\mu
}^{--}:u_{t}+\alpha v_{\mu }v_{x}+\beta u_{\varepsilon }u_{x}+\gamma
u_{xxx}=0, \\ 
v_{t}+\alpha (uv)_{x}+\phi v_{xxx,\mu }=0, & v_{t}+\alpha (uv_{\mu
})_{x}+\phi v_{xxx}=0.%
\end{array}
\label{eq2}
\end{equation}%
Naturally there exist also possibilities to construct non-Hamiltonian
deformations. Noting for instance that the first equation related to $%
\mathcal{H}_{\varepsilon ,\mu }^{++}$ is also invariant under $\mathcal{PT}%
_{+-}$ we may combine it with the second equation resulting from $\mathcal{H}%
_{\varepsilon ,\mu }^{+-}$ and define the $\mathcal{PT}_{+-}$-invariant
system%
\begin{equation}
\begin{array}{r}
u_{t}+\alpha vv_{x}+\beta uu_{x}+\gamma u_{xxx,\varepsilon }=0, \\ 
v_{t}+\alpha (uv_{\mu })_{x}+\phi v_{xxx}=0.%
\end{array}
\label{nonH}
\end{equation}

We have now numerous new physically feasible theories to be investigated.
Here we will only present few examples. Having two parameters at our
disposal allows to obtain more analytic expressions for the solutions.
Technically we have two problems to overcome. First of all we have to
decouple the equations and subsequently carry out all the integrations. In
order to achieve the first aim we focus mainly on the case $\phi =0$ as this
will allow to express $u$ in terms of $v$ in a simple manner. This choice
will eliminate the deformation term in $\mathcal{H}_{\varepsilon ,\mu }^{++}$
and we therefore concentrate here on $\mathcal{H}_{\varepsilon ,\mu }^{+-}$
to study the interplay between the two parameters.

\subsection{The model $\mathcal{H}_{\protect\varepsilon ,\protect\mu }^{+-}(%
\protect\alpha ,\protect\beta ,\protect\gamma ,u,v)$}

As a further simplification we set the constant $\kappa _{2}$ to zero, which
results in the first integration of the equation for $v_{t}$. We may then
integrate again and obtain a decoupled equation solely involving the field $%
u $%
\begin{equation}
u_{\zeta }^{\varepsilon +1}=i^{1-\varepsilon }\frac{\varepsilon +1}{\gamma
\varepsilon }\left[ \kappa _{3}+\kappa _{1}u+\frac{c}{2}u^{2}-\frac{\beta }{6%
}u^{3}+\frac{\alpha }{2}\frac{1-\mu }{1+\mu }\left( \frac{\alpha }{c}\right)
^{\frac{1+\mu }{1-\mu }}u^{\frac{2}{1-\mu }}\right] ,  \label{ue}
\end{equation}%
with the two fields related as%
\begin{equation}
u=\frac{c}{\alpha }(iv)^{1-\mu }.  \label{uvv}
\end{equation}%
To be able to carry out the final integration in (\ref{ue}) we would like
the right hand side to acquire the form of a factorizable polynomial in $u$.
This is possible for several specific choices and with the additional choice
of $\varepsilon $ we may even construct analytic solutions.

$\mathcal{H}_{\varepsilon ,1/3}^{+-}(\alpha ,\beta ,\gamma ,u,v)$

This case is entirely reducible to the deformation of the KdV $\mathcal{H}%
_{\varepsilon }^{+}(\beta ,\gamma ,u)$ when noticing that%
\begin{equation}
\mathcal{H}_{\varepsilon }^{+}\left( \beta -\frac{3}{2}\frac{\alpha ^{3}}{%
c^{2}},\gamma \frac{2\varepsilon }{1+\varepsilon }i^{\varepsilon
-1},u\right) =\mathcal{H}_{\varepsilon ,1/3}^{+-}\left[ \alpha ,\beta
,\gamma ,u,\left( \frac{\alpha }{\beta }u\right) ^{3/2}\right] .
\end{equation}%
where we used the relation between $u$ and $v$ as specified in (\ref{uvv})
for $\mu =1/3$.

$\mathcal{H}_{-1/2,1/2}^{+-}$

\begin{figure}[h!]
\centering  
\includegraphics[width=7.0cm]{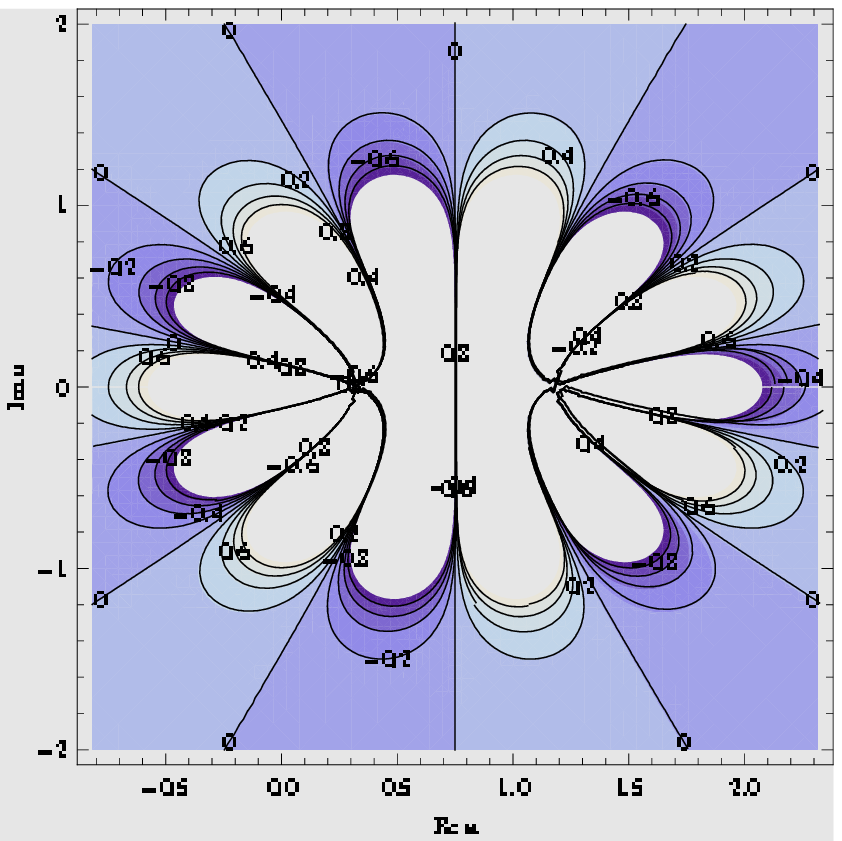} 
\includegraphics[width=7.0cm]{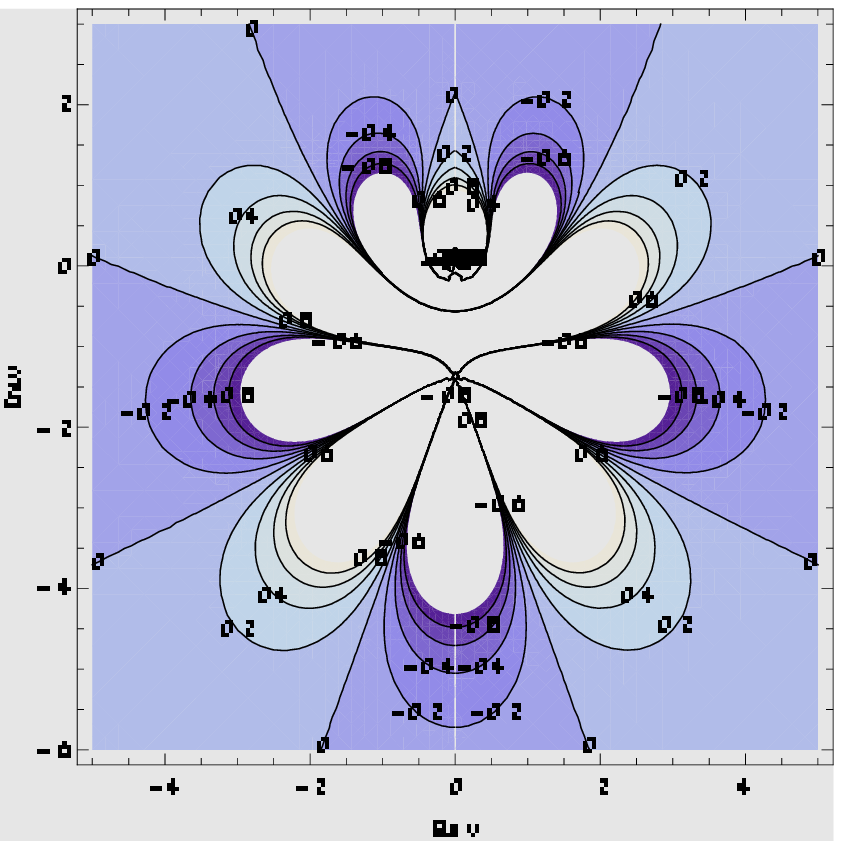}
\caption{Complex periodic trajectories for the $\mathcal{PT}_{+-}$-symmetric
system $\mathcal{H}_{-1/2,1/2}^{+-}$ for purely complex initial values $%
\protect\zeta _{0}$ with $\protect\alpha =1$, $\protect\beta =3$,$\protect%
\gamma =1$, $c=1$, $A=(3-\protect\sqrt{3})/4$ and $B=(3-\protect\sqrt{3})/4.$%
}
\label{fig41}
\end{figure}

The choice $\mu =1/2$ converts the right hand side of (\ref{ue}) into a
fourth order polynomial. Let us next assume the factorization is of the form 
$\lambda (u-A)^{2}(u-B)^{2}$, which is indeed possible if the following
constraints hold%
\begin{eqnarray}
\lambda &=&\frac{e^{-i\pi /4}\alpha ^{4}}{6c^{3}\gamma },\quad \kappa _{1}=%
\frac{c^{6}\beta ^{3}-12c^{4}\alpha ^{4}\beta }{48\alpha ^{8}},\quad \kappa
_{3}=\frac{c^{5}\left( c^{2}\beta ^{2}-12\alpha ^{4}\right) ^{2}}{384\alpha
^{12}}, \\
A &=&\frac{c^{3}\alpha ^{4}\beta \pm \sqrt{3}\sqrt{c^{6}\alpha ^{8}\beta
^{2}-8c^{4}\alpha ^{12}}}{4\alpha ^{8}},\quad B=\frac{c^{3}\beta \alpha
^{4}\mp \sqrt{3}\sqrt{c^{6}\alpha ^{8}\beta ^{2}-8c^{4}\alpha ^{12}}}{%
4\alpha ^{8}}.
\end{eqnarray}%
Notice that all the free parameters are fixed in this case, as a consequence
of having already pre-selected $\kappa _{2}=0$. We analyse this model in
figure \ref{fig41}.

We observe that the $PT_{+-}$-symmetry manifests itself now through\ $%
v^{\ast }(\zeta )=-v(-\zeta )$ and $u^{\ast }(\zeta )=u(-\zeta )$.
Furthermore we recognise that the asymptotic limits of the $u$-field are $A$
and $B$.

Of course there are plenty more models one may explore.

\subsection{The non-Hamiltonian deformation (\protect\ref{nonH})}

Let us briefly comment on one of the possibilities to construct deformations
of non-Hamiltonian systems (\ref{nonH}). We consider again the case $\phi =0$
and $\kappa _{2}=0$ and integrate the first equation in (\ref{nonH}) for $%
\varepsilon =1$ and $\mu \neq 3$ 
\begin{equation}
u_{\zeta }^{2}=\frac{2}{\gamma }\left[ \kappa _{3}+\kappa _{1}u+\frac{c}{2}%
u^{2}-\frac{\beta }{6}u^{3}+\frac{\alpha }{2}(\frac{c}{\alpha })^{\frac{2}{%
\mu -1}}\frac{1-\mu }{3-\mu }u^{\frac{3-\mu }{1-\mu }}\right] .  \label{44}
\end{equation}%
The $v$-field is related to the $u$-field via (\ref{uvv}). The boundary
conditions have to be treated by distinguishing different case. For
instance, when $(3-\mu )/(1-\mu )\geq 0$ the vanishing asymptotic boundary
conditions for $u$ and its derivatives demand that $\kappa _{1}=\kappa
_{3}=0 $.

The simplest case to consider is $\mu =-1$, which is just the KdV-system
with a re-defined speed of the wave $c\rightarrow c+\alpha ^{2}/c$.

A further simple example is to take $\mu =1/3$ for which we can find an
explicit solution $u(\zeta )$. The right hand side of (\ref{44}) can be
factorized into $\lambda (u-A)^{2}(u-B)^{2}$ with%
\begin{eqnarray}
\lambda &=&\frac{c^{3}}{4\alpha ^{2}\gamma },\quad \kappa _{1}=\frac{\alpha
^{4}\beta ^{3}-9c^{4}\alpha ^{2}\beta }{27c^{6}}\quad \kappa _{3}=\frac{%
\left( \alpha ^{3}\beta ^{2}-9c^{4}\alpha \right) ^{2}}{162c^{9}}, \\
A &=&\frac{c^{3}\alpha ^{2}\beta -\sqrt{3}\sqrt{c^{6}\alpha ^{4}\beta
^{2}-6c^{10}\alpha ^{2}}}{3c^{6}},\quad B=\frac{\alpha ^{2}\beta c^{3}+\sqrt{%
3}\sqrt{c^{6}\alpha ^{4}\beta ^{2}-6c^{10}\alpha ^{2}}}{3c^{6}}.
\end{eqnarray}%
for which (\ref{44}) can be integrated further and solved for $u$. We find 
\begin{equation}
u(\zeta )=\frac{Be^{\pm A(\zeta -\zeta _{0})\sqrt{\lambda }}-Ae^{\pm B(\zeta
-\zeta _{0})\sqrt{\lambda }}}{e^{\pm A(\zeta -\zeta _{0})\sqrt{\lambda }%
}-e^{\pm B(\zeta -\zeta _{0})\sqrt{\lambda }}}\quad \text{and\quad }v(\zeta
)=-i\left( \frac{\alpha u(\zeta )}{c}\right) ^{3/2}.
\end{equation}%
The system is easily linearized and qualitatively we find a similar
behaviour as for the Hamiltonian systems, which we will, however, not
present here in more detail. Thus, even though it is less clear how the $%
\mathcal{PT}$-symmetry can be utilized, we find that they are not
fundamentally different from the Hamiltonian systems.

\section{Conclusions}

The main focus of this paper was to investigate the effects of $\mathcal{PT}$%
-symmetry and its breaking in complex nonlinear wave equations. In general
we find that unlike as claimed for orbits in quantum mechanical one particle
models \cite{Nana,Bender:2006tz,Bender:2008fr,Bender:2010eg}, trajectories
in the complex plane of the field of a $\mathcal{PT}$-symmetric nonlinear
system are not fundamentally distinct from those associated to systems with
spontaneously or completely broken $\mathcal{PT}$-symmetry. Just from the
general type of trajectory one can not conclude which type of setting one is
considering. Of course invoking the information on the symmetry one can
identify on this basis the $\mathcal{PT}$-symmetric case over the broken
ones. However, based on this criterium the spontaneously broken cases are
indistinguishable from the completely broken ones. This behaviour extends to
the type of fixed points one may find. We observed that essentially all
types of fixed points, except saddle points, may occur, irrespective of the $%
\mathcal{PT}$-symmetry properties of the model. When identifying and varying
certain constants of the model as bifurcation parameter the fixed point may
undergo a Hopf bifurcation.

It appears that the entire phase space is filled out in the broken case,
thus suggesting a chaotic behaviour. However, from the Poincar\'{e}%
-Bendixson theorem\footnote{\textbf{Poincar\'{e}-Bendixson Theorem: }\emph{%
Let }$\varphi _{t}$\emph{\ be a flow for a two dimensional dynamical systems
and let }$\mathcal{D}$\emph{\ be a closed, bounded and connected set }$%
\mathcal{D}\in \mathbb{R}^{2}$\emph{, such that }$\varphi _{t}(\mathcal{D}%
)\subset \mathcal{D}$\emph{\ for all time. Furthermore }$\mathcal{D}$\emph{\
does not contain any fixed point. Then there exists at least one limit cycle
in }$\mathcal{D}$\emph{.\smallskip }} we know that for a closed bounded and
connected region in two dimensions this is impossible to occur.

The nature of the fixed points also indicates that the models are not
Hamiltonian in the real and imaginary part of $u$, as only saddle points and
centres would emerge in that case, whereas we found different types of fixed
points and the absence of saddle points.

With regard to the energies we confirmed that fully $\mathcal{PT}$-symmetric
systems have real energies, which could be calculated explicitly in many
cases. When breaking this symmetry spontaneously by complexifying some free
parameters in the solutions we obtained complex energies. As expected, the
models related to the complex conjugates of these parameters have complex
conjugate energies. More surprising are the findings obtained in various
models that one may regain the reality of the energy by breaking the
symmetry further. For the type III solutions of the Ito type systems it is
conceivable the one might have free parameters in the expressions for the
energy even when fixing the model. We conjecture that these models possess a
different kind of antilinear symmetry yet to be identified.

We found that complex $\mathcal{PT}$-symmetric soliton solutions behave
similarly as their real counterparts, albeit in the complex plane. The
one-solitons travel in the complex plane while maintaining their overall
shape and the two-soliton solution can be associated to two separate
one-soliton solutions in the distant past and future. However, when the $%
\mathcal{PT}$-symmetry is broken the nature of the one-soliton solution
changes into a breather, which only regains its shape after it has traveled
a certain time and distance. The two-soliton solutions for this case can
also be separated into these breathers in the distant past and future. The
energy of the two-soliton solution was found to be the sum of the energies
of the constituent one-soliton solutions in all scenarios. It would be
interesting to investigate these features also for N-soliton solutions for N%
\TEXTsymbol{>}2 and different types of systems, such as the Ito type.

Besides guaranteeing the reality of the energy, $\mathcal{PT}$-symmetry was
noted to be useful for various other reasons. We found that the symmetry
allows for a natural "$\varepsilon $-prescription" to facilitate the
computation of the energies in the complex plane by means of the Cauchy
theorem. Identifying the different types of $\mathcal{PT}$-symmetries it
also allows to formulate new physically feasible models with real energies.
This procedure constitutes a natural and more general framework for some
models which have already been known before, such as the generalized
KdV-equations with the modified one as a special case.

Since many of the models discussed here are integrable, it is worth
mentioning that most of the analysis carried out here for the Hamiltonian
may also be performed for other conserved quantities of the same order in
the fields having the same type of symmetry property. For some of the
charges this is in fact not the case. Although in many deformed cases the
question of integrability has not yet been answered decisively, the
statement also holds for the lowest charges which are certain to exist.

The main difference between the deformed models and their undeformed
counterparts is in general the occurrence of more and more Riemann sheets
with increasing integer value for $\varepsilon $. We showed that in many
cases non-integer values and even negative values for $\varepsilon $ give
rise to interesting solvable models. The overall structure of the
trajectories and the nature of the fixed points is not fundamentally
different, except that they usual extend over several Riemann sheets.
Clearly we only presented here a limited number of solutions and many cases
still need to be explored, especially for the second type of deformation for
which even the factorization of the $P(u)$-function remains an open issue.

When comparing the Ito type system with the KdV system we note that the
former is not simply an add on to the latter, but gives rise to more complex
structures. We found even for the undeformed case some new solutions such as
those of kink type hitherto not reported. We also pointed out some minor
discrepancies when compared to the literature. With regard to the
deformations, which are all entirely new proposals, the conclusions are
similar as for the deformed KdV-case. The major difference is that the
interplay between the two deformation parameters allows for even more
possibilities. Also in this case many possibilities remain unexplored.

As indicated at the end of section 2.1.3 and 3.1.3 we can obtain simple
quantum mechanical systems as special cases from our analysis. For instance,
the archetype deformation of the harmonic oscillator Hamiltonian $%
H=p^{2}+x^{2}(ix)^{\varepsilon }$ can be obtained in various ways via the
identification for the traveling wave $u\rightarrow x$ and $\zeta
\rightarrow t$. As discussed, the deformed models $\mathcal{H}_{\varepsilon
}^{-}$ yield precisely the quantum mechanical model with potential $V(x)=$ $%
x^{2}(ix)^{\varepsilon }$, of which some cases were studied in \cite%
{Anderson}. From the models presented here there are more possibilities to
arrive at such potential, as for instance also the non-Hamiltonian models (%
\ref{nonH}) give rise to such type of potentials as may be seen from (\ref%
{44}). Exploiting these observations allows to obtain many solutions and
properties of these systems easily from our analysis overlooked up to now.
Obviously, one may also construct new interesting quantum mechanical models
in this way which have not been investigated so far.

In addition, this opens up immediately the more general question of studying
properties of these systems as continuous functions of $\varepsilon $,
rather than selecting just certain specific values as in this manuscript.
Building for instance on the analogy with the potential systems should
certainly reveal fundamentally different kinds of behaviour in some regions,
such that for instance $\mathcal{H}_{\varepsilon }^{-}$ will probably have a
qualitatively different behaviour for negative values of $\varepsilon $. We
leave these type of questions for future investigations.

Clearly it would be very interesting to extend these type of analysis to
other non-linear field equations such as Burgers, Bussinesque, KP,
generalized shallow water equations, extended KdV equations with compacton
solution, etc.

\bigskip

\textbf{Acknowledgments:} AF would like to thank the UGC Special Assistance
Programme in the Applied Mathematics Department of the University of
Calcutta and S.N. Bose National Centre for Basic Sciences for providing
infrastructure and financial support. AC is supported by a City University
Research Fellowship.

\appendix

\section{The ten similarity classes for $J$}

For convenience we recall here the ten different similarity classes
characterizing the fixed points of a two dimensional linear system (\ref{lin}%
) by the eigenvalues $j_{1}$ and $j_{2}$ of the Jacobian matrix $J$ at the
fixed point.

\begin{center}
$%
\begin{array}{|l|l|l|}
j_{i}\in \mathbb{R} & j_{1}>j_{2}>0~~~~~ & \text{unstable node} \\ \hline
& j_{2}<j_{1}<0 & \text{stable node} \\ \hline
& j_{2}<0<j_{1} & \text{saddle point} \\ \hline
j_{1}=j_{2}\text{, diagonal }J & j_{i}>0 & \text{unstable star node} \\ 
\hline
& j_{i}<0 & \text{stable star node} \\ \hline
j_{1}=j_{2}\text{, nondiagonal }J~~~~ & j_{i}>0 & \text{unstable improper
node} \\ \hline
& j_{i}<0 & \text{stable improper node} \\ \hline
j_{i}\in \mathbb{C} & \func{Re}j_{i}>0 & \text{unstable focus} \\ \hline
& \func{Re}j_{i}=0 & \text{centre} \\ \hline
& \func{Re}j_{i}<0 & \text{stable focus}%
\end{array}%
$
\end{center}

\noindent {\small Table 1: Nature of a fixed point as classified by the
eigenvalues of the Jacobian.}


\end{document}